\documentclass[preprint, 3p, twocolumn]{elsarticle}

\usepackage[utf8]{inputenc}
\usepackage[cmex10]{amsmath}		% Math
\usepackage{amssymb}				% Symbols of the number sets (e.g., \mathbb{N})
\usepackage{siunitx}				% Typeset SI units
\usepackage[capitalise]{cleveref}	% \cref{} -> Automatically write Fig., Eq.,
    \crefformat{equation}{(#2#1#3)}		% Redefine Equation labels to be typeset as "(1)", instead of "Eq. (1)".

\biboptions{compress}

\usepackage{url}		% Better handling of URL links

\usepackage{graphicx}	% Images

\usepackage[caption=false,font=footnotesize]{subfig}

\usepackage{tikz}
\usetikzlibrary{arrows}
\usetikzlibrary{shapes}

\newcommand{\mymk}[1]{
  \tikz[baseline=(char.base)]\node[anchor=south west, draw,rectangle, rounded corners, inner sep=0.1pt, minimum size=3.5mm,
    text height=2mm](char){\ensuremath{#1}} ;}

\newcommand*\circled[1]{\tikz[baseline=(char.base)]{
            \node[shape=circle,draw,inner sep=0.1pt] (char) {#1};}}
\makeatother

\usepackage[]{algorithm, algpseudocode}
\makeatletter
\renewcommand{\ALG@beginalgorithmic}{\small}

\newcommand{\settest}[1]{\renewcommand{\max}{maximize}}

\graphicspath{{Figures/}}

\begin{document}

\begin{frontmatter}

\title{Joint Traffic-Aware UAV Placement and Predictive Routing for Aerial Networks}

\author{Eduardo Nuno Almeida\corref{cor1}}
\ead{eduardo.n.almeida@inesctec.pt}

\author{André Coelho}
\ead{andre.f.coelho@inesctec.pt}

\author{José Ruela}
\ead{jruela@inesctec.pt}

\author{Rui Campos}
\ead{rui.l.campos@inesctec.pt}

\author{Manuel Ricardo}
\ead{mricardo@inesctec.pt}

\address{INESC TEC and Faculdade de Engenharia, Universidade do Porto, Portugal}

\cortext[cor1]{Corresponding author at INESC TEC and Faculdade de Engenharia, Universidade do Porto, Campus da FEUP, Rua Dr. Roberto Frias, 4200-465 Porto, Portugal}

%\journal{Ad Hoc Networks, Elsevier}

%%% \maketitle  % IEEEtran only

%%%%%%%%%%%%%%%%%%%%%%%%%%%%%%%%%%%%%%%%%%%%%%%%%%%%%%%%%%%%%%%%%
% ABSTRACT AND KEYWORDS
%%%%%%%%%%%%%%%%%%%%%%%%%%%%%%%%%%%%%%%%%%%%%%%%%%%%%%%%%%%%%%%%%
\begin{abstract}

Aerial networks, composed of Unmanned Aerial Vehicles (UAVs) acting as Wi-Fi access points or cellular base stations, are emerging as an interesting solution to provide on-demand wireless connectivity to users, when there is no network infrastructure available, or to enhance the network capacity. This article proposes a traffic-aware topology control solution for aerial networks that holistically combines the placement of UAVs with a predictive and centralized routing protocol. The synergy created by the combination of the UAV placement and routing solutions allows the aerial network to seamlessly update its topology according to the users' traffic demand, whilst minimizing the disruption caused by the movement of the UAVs. As a result, the Quality of Service (QoS) provided to the users is improved. The components of the proposed solution are described and evaluated individually in this article by means of simulation and an experimental testbed. The results show that all the components improve the QoS provided to the users when compared to the corresponding baseline solutions.

\end{abstract}


\end{frontmatter}

%%%%%%%%%%%%%%%%%%%%%%%%%%%%%%%%%%%%%%%%%%%%%%%%%%%%%%%%%%%%%%%%%
% INTRODUCTION
%%%%%%%%%%%%%%%%%%%%%%%%%%%%%%%%%%%%%%%%%%%%%%%%%%%%%%%%%%%%%%%%%
\section{Introduction} \label{Introduction-Section}

%%% CONTEXT
In recent years, the need for broadband wireless connectivity has been steadily increasing. From online video streaming to remote vehicle piloting, new applications require reliable wireless links with high throughput and low delay. Additionally, some scenarios pose additional challenges when planning the network. In emergency scenarios, such as forest fires and earthquakes, groups of first-responders are distributed throughout a large area and need to communicate among themselves and with a remote command center \cite{zhao2019uav}. In addition to traditional voice and text services, these communications may also include broadband services requiring the exchange of multimedia content. In some circumstances, existing networks might not be able to provide reliable and broadband wireless connectivity due to failures of the base stations or lack thereof. Other scenarios that exacerbate the challenges of network planning are Temporary Crowded Events (TCEs), such as music festivals and outdoor festivities \cite{almeida2018traffic}. TCEs are characterized by a high density of users that are concentrated in predefined areas for short periods of time and generate significant and variable traffic, which is influenced by the event dynamics.

In order to satisfy the Quality of Service (QoS) requirements in these scenarios, novel network architectures are being considered. An interesting solution relies on the use of Unmanned Aerial Vehicles (UAVs) acting as aerial Wi-Fi Access Points (APs) or cellular base stations, forming aerial wireless networks \cite{zeng2016wireless, yanmaz2018drone}. Due to the mobility of the UAVs and the ability to position them in the 3D space, aerial networks can quickly adapt to the dynamic conditions of the environment and users' traffic demand. Thus, aerial networks are excellent solutions to provide on-demand wireless connectivity when there is no network infrastructure available or to enhance the capacity of existing networks with the deployment of temporary additional aerial base stations.

\begin{figure}
    \centering
    \includegraphics[width=1\linewidth]{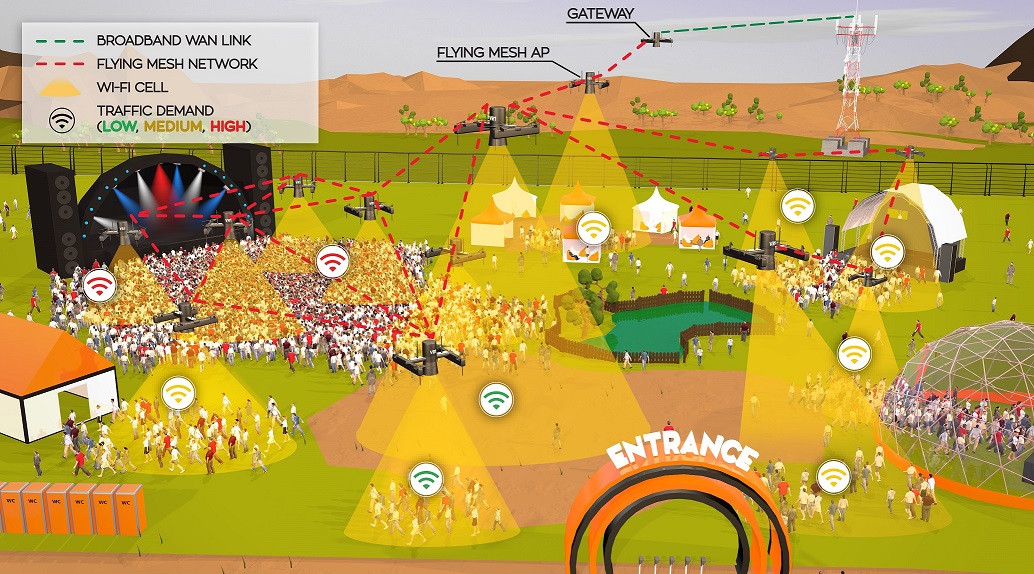}
    \caption{Traffic-Aware Multi-Tier Flying Network (TMFN) providing always-on broadband Internet connectivity to the users attending a music festival.}
    \label{TMFN-Figure: TMFN}
\end{figure}

%%% OUR PREVIOUS WORK
In \cite{almeida2018traffic} we proposed an aerial network architecture named \textbf{Traffic-Aware Multi-Tier Flying Network} (TMFN), which is illustrated in \cref{TMFN-Figure: TMFN}. The TMFN is composed of Flying Mesh Access Points (FMAPs) and Gateway (GW) UAVs, which are organized in a two-tier multi-hop architecture. The access tier consists of FMAPs, which are rotary-wing UAVs acting as aerial Wi-Fi APs that form small cells to serve the users on the ground. The backhaul tier is composed of Gateway UAVs that forward traffic from the FMAPs to the Internet. The TMFN can be dynamically repositioned and reconfigured according to the users' traffic demand, in order to improve the overall provided QoS.

To control the TMFN topology, we proposed a \textbf{Network Planning (NetPlan)} algorithm in \cite{almeida2018traffic}. The NetPlan algorithm determines the horizontal positions and Wi-Fi cell ranges of the hovering FMAPs in order to improve the TMFN's aggregate throughput. To this end, the NetPlan algorithm positions the FMAPs closer to the users generating more traffic with shorter Wi-Fi cells, with the remaining FMAPs being distributed throughout the coverage area. As a follow-up of this work, we proposed the \textbf{RedeFINE routing solution} \cite{coelho2018redefine, coelho2019routing}. RedeFINE is a predictive centralized routing protocol for high-capacity multi-hop aerial networks, which is able to determine, in advance, the forwarding tables of the FMAPs and the time instants they shall be updated in order to minimize communications disruptions. By assuming that the future trajectories of the FMAPs are known, RedeFINE is able to predict the time instants when each FMAP should update its forwarding table so that the overall network throughput is maximized. This is achieved by eliminating the process of neighbor discovery and the time wasted in updating the forwarding table in traditional routing solutions, where nodes typically recover from link failures after they occur and are detected. Finally, we proposed a \textbf{gateway UAV placement (GWP)} algorithm \cite{coelho2019traffic}. The GWP algorithm takes advantage of the knowledge of the FMAPs' future positions and offered traffic to determine the position of the gateway UAVs in order to enable communications paths with high capacity. Although the NetPlan, the RedeFINE and the GWP solutions were developed independently, they were designed to be used simultaneously. To the best of our knowledge, there are no solutions combining the UAV placement, routing and gateway placement problems in a single solution for aerial networks.

%%%%% FIGURE %%%%%
%%%%%%%%%%%%%%%%%%

%%% ARTICLE CONTRIBUTIONS
In this article, we propose a novel solution resulting from the holistic combination of the NetPlan algorithm \cite{almeida2018traffic}, the RedeFINE routing solution \cite{coelho2018redefine}, including an inter-flow interference routing metric \cite{coelho2019routing}, and the GWP algorithm \cite{coelho2019traffic} applied to the TMFN. The synergy created by the integration of the three components allows the TMFN to improve the QoS provided to the users whilst minimizing the communications disruption within the TMFN's backhaul tier. The proposed solution determines the updated TMFN topology periodically with each update cycle having the following sequence of operations. First, the NetPlan algorithm determines the updated positions and Wi-Fi cell ranges of the FMAPs for the following cycle based on the users' positions and their offered traffic. Then, the RedeFINE routing solution determines the optimal forwarding tables for the FMAPs and the time instants they shall be updated, considering their future trajectories calculated through the initial and final positions of the FMAPs. Finally, the GWP algorithm determines the optimal position of the GWs considering the FMAPs' future positions and offered traffic. Although the three components were designed to be used simultaneously, they are evaluated individually using the ns-3 simulator \cite{ns3Simulator}. This allows the evaluation of the corresponding component without the interference from the remaining ones. The NetPlan algorithm is also evaluated in an experimental testbed. In this process, we evaluate the air-to-ground and ground-to-air channel propagation models in the testbed for UAVs hovering at low altitudes in an open-air environment. This allows the verification of the theoretical models proposed in the literature \cite{khuwaja2018survey}. Moreover, using the experimental channel models, instead of the theoretical ones, in the ns-3 simulations allows a more accurate reproduction of the testbed conditions.

The contributions of this article are three-fold:

\begin{itemize}
    \item A novel solution resulting from the holistic combination of the NetPlan, RedeFINE and GWP solutions for aerial networks, which improves the network performance and minimizes the communications disruptions;

    \item Evaluation of the NetPlan algorithm in an experimental testbed;

    \item Experimental evaluation of the air-to-ground and ground-to-air channel propagation models for UAVs hovering at low altitudes in an open-air environment.
\end{itemize}

%%% ARTICLE STRUCTURE
The rest of this article is organized as follows.
\cref{RelatedWork-Section} discusses the related work.
\cref{TMFN-Section} presents the TMFN network architecture and the holistic solution proposed in this article.
\cref{NetplanAlgorithm-Section} explains the NetPlan algorithm.
\cref{NetplanAlgorithmEvaluation-Section} contains the evaluation of the NetPlan algorithm.
\cref{RedeFINE-Section} describes the RedeFINE routing protocol.
\cref{RedeFINEEvaluation-Section} contains the evaluation of RedeFINE.
\cref{GWPAlgorithm-Section} presents the GWP algorithm.
\cref{GWPAlgorithmEvaluation-Section} discusses the evaluation of the GWP algorithm.
\cref{Conclusions-Section} draws the conclusions and future work.

%%%%%%%%%%%%%%%%%%%%%%%%%%%%%%%%%%%%%%%%%%%%%%%%%%%%%%%%%%%%%%%%%
% RELATED WORK
%%%%%%%%%%%%%%%%%%%%%%%%%%%%%%%%%%%%%%%%%%%%%%%%%%%%%%%%%%%%%%%%%
\section{Related Work} \label{RelatedWork-Section}

%%% UAV PLACEMENT ALGORITHMS
Following the emergence of aerial networks, UAV placement algorithms have been proposed to determine the positions of the UAVs that maximize a given objective function \cite{cicek2019uav, yanmaz2018drone}. In \cite{galkin2018backhaul, mozaffari2015drone, kalantari2017backhaul}, the authors propose solutions to maximize the area and number of users served by the aerial network. The determination of the UAV positions that maximize the QoS / Quality of Experience (QoE) provided to the users are the main objective of the works proposed in \cite{he2019resource, zeng2016throughput, li2017optimal, alzenad20183d}. In addition to QoS, other objectives may also be considered when designing UAV placement algorithms, including the minimization of the UAV's transmission power \cite{alzenad20173Dplacement} or the support of first-responders in emergency scenarios \cite{zhao2019uav}.

Recently, UAV placement algorithms based on Machine Learning (ML) or Reinforcement Learning (RL) techniques have been proposed \cite{wang2018machine, jiang2017machine}. The use of these artificial intelligence techniques has been proven to achieve similar or even better results than deterministic alternatives, due to the ability to automatically extract and learn the most relevant features that influence the decision-making process without human intervention. The Q-Learning technique is explored in \cite{ghanavi2018efficient, colonnese2019qsquare, liu2018deployment} in order to determine the positions of the UAVs that maximize the QoS / QoE provided to the users. The solutions presented in \cite{munaye2019uav, liu2019trajectory, balakrishnan2019deep} take advantage of deep learning techniques to design UAV placement algorithms that maximize the QoS provided to the users, by controlling the position, Tx power and OFDMA resource scheduling of the UAVs. A deep RL placement algorithm, based on Echo State Networks (ESNs), is shown in \cite{challita2018deep} to determine the trajectories of multiple UAVs, so that the interference caused in the ground network and the wireless transmission latency are minimized. ESNs are also explored in \cite{chen2017caching}, in which an algorithm for cache-enabled UAVs is proposed to determine the trajectories and content to cache at each UAV, in order to maximize the QoE provided to the users and minimize the Tx power of the on-board base stations. Moreover, an ML framework is presented in \cite{zhang2018machine} to predict congestion and traffic demand surges in cellular networks, which is used to position UAVs to enhance the capacity of local networks using minimal Tx and UAV movement power.

Overall, despite the results achieved by the UAV placement algorithms, most of them do not consider nor differentiate the users' traffic demands and optimize the SNR for all users independently of their traffic demand. Therefore, the network capacity provided by the UAVs may not be fully exploited by inactive users, who are not offering or receiving traffic.

%%% ROUTING PROTOCOLS
Most of the state of the art routing solutions for aerial networks were built upon the protocols employed in Mobile Ad-hoc Networks (MANETs) and Vehicular Ad-hoc Network (VANETs) \cite{jiang2018, lakew2020}. In particular, some predictive approaches have been proposed in \cite{li2017, gankhuyag2017, song2018, khaledi2018, sliwa2019}. Predictive solutions usually consider the positions of the UAVs over time, which are inferred based on their speed, moving direction, and predetermined mobility models \cite{arafat2019}. In \cite{xianfengLi2017}, a routing protocol based on the Ad-Hoc On-Demand Distance Vector (AODV) protocol \cite{perkins2003} is proposed. It uses Global Positioning System (GPS) information and employs mobility prediction to estimate the stability of the links, aiming to minimize the delay of the routing discovery process, typical in reactive routing protocols \cite{jiang2018, sahingoz2014, tareque2015}. A similar approach is proposed in \cite{rosati2016}, where the Optimized Link State Routing (OLSR) protocol \cite{clausen2003} is improved to predict topology changes and react before link disruptions occur. Nevertheless, the overhead inherent to proactive protocols is not addressed \cite{bekmezci2013, sahingoz2014}. Moreover, both solutions have specific hardware requirements, in order to determine the location of the UAVs with high accuracy in short time intervals. In \cite{barritt2017}, a Software-Defined Networking (SDN) routing approach that anticipates topology changes is proposed; however, to the best of our knowledge, its performance evaluation has not yet been presented and the optimal placement of the UAVs is not explored.

Overall, predictive routing solutions for aerial networks employ the distributed routing paradigm; hence, the UAVs need to exchange probe packets, which may introduce high overhead and may not scale for large networks. Moreover, in aerial networks, UAVs typically need to be placed close to each other, in order to ensure high-capacity air-to-air links. This leads to interference between concurrent flows, which is not a local concept, since it depends on all the interfering nodes along a path. Therefore, a solution that performs routing decisions considering a holistic and centralized view of the network is worthy to be considered.

%%% GATEWAY PLACEMENT
GW placement in wireless networks is a commonly treated problem in the literature. Over the years, different studies have been carried out \cite{maolin2009gateways, seyedzadegan2013zero, targon2010joint, aoun2006gateway, jahanshahi2019gateway}. However, most of them aim at minimizing the number of GWs while optimizing their placement, in order to meet some QoS metrics, including throughput and delay, and reducing the energy consumption. In \cite{muthaiah2008single}, the authors show how the GW placement and the transmission power affect the network throughput. However, they do not consider the traffic demand of nodes. Similarly, the work presented in \cite{oueis2019core} aims at determining the optimal placement for an Evolved Packet Core (EPC), amongst a set of BSs in a self-deployed cellular network. Nevertheless, they do not have control over the mobility of the EPC nodes and assume that the nodes have the same traffic demand. In \cite{larsen2017}, the authors show how the placement of a UAV acting as network relay between two ground nodes affects the throughput achieved; however, this study is only valid for a pair of ground nodes. A model-free approach to find the optimal positions of a relay UAV is presented in \cite{zhong2019}; its main drawback is the time required to converge to the optimal position.

%%% CONCLUSION
Overall, state of the art work has been focused on the UAV placement, routing and GW placement problems for aerial networks. However, to the best of our knowledge, no solution proposes a holistic combination of the UAV placement, predictive routing and GW placement, leveraging the knowledge and ability to incorporate the future decisions of all components when determining the future TMFN topology, in order to improve the QoS provided to the users whilst minimizing the communications disruption.

%%%%%%%%%%%%%%%%%%%%%%%%%%%%%%%%%%%%%%%%%%%%%%%%%%%%%%%%%%%%%%%%%
% TRAFFIC-AWARE MULTI-TIER FLYING NETWORK
%%%%%%%%%%%%%%%%%%%%%%%%%%%%%%%%%%%%%%%%%%%%%%%%%%%%%%%%%%%%%%%%%
\section{Traffic-Aware Multi-Tier Flying Network} \label{TMFN-Section}

The TMFN, originally proposed in \cite{almeida2018traffic} and illustrated in \cref{TMFN-Figure: TMFN}, consists of a multi-tier aerial network of FMAPs and Gateway UAVs which dynamically reconfigures its topology according to the users' traffic demand, in order to improve the QoS provided to users on the ground. FMAPs and Gateway UAVs are organized in a multi-hop network architecture that is able to cover large areas and provide on-demand wireless connectivity. The first tier (access network) is composed of FMAPs, which are rotary-wing UAVs acting as Wi-Fi APs; they form high-capacity small cells that can be dynamically configured and positioned according to the variable traffic demand of the moving users. FMAPs are able to continuously detect and seek the users that generate more traffic to provide them more bandwidth, so that the aggregate throughput is improved. The second tier (backhaul network) is composed of Gateway UAVs, which are responsible for forwarding traffic to the Internet using dedicated broadband wireless links. Gateway UAVs are dynamically positioned according to the FMAPs' positions and offered traffic, in order to maximize the throughput forwarded to the Internet. Due to the multi-hop architecture, FMAPs can act as relays between gateway UAVs and other FMAPs.

%%%%% FIGURE %%%%%
\begin{figure}
    \centering
    \includegraphics[width=1\linewidth]{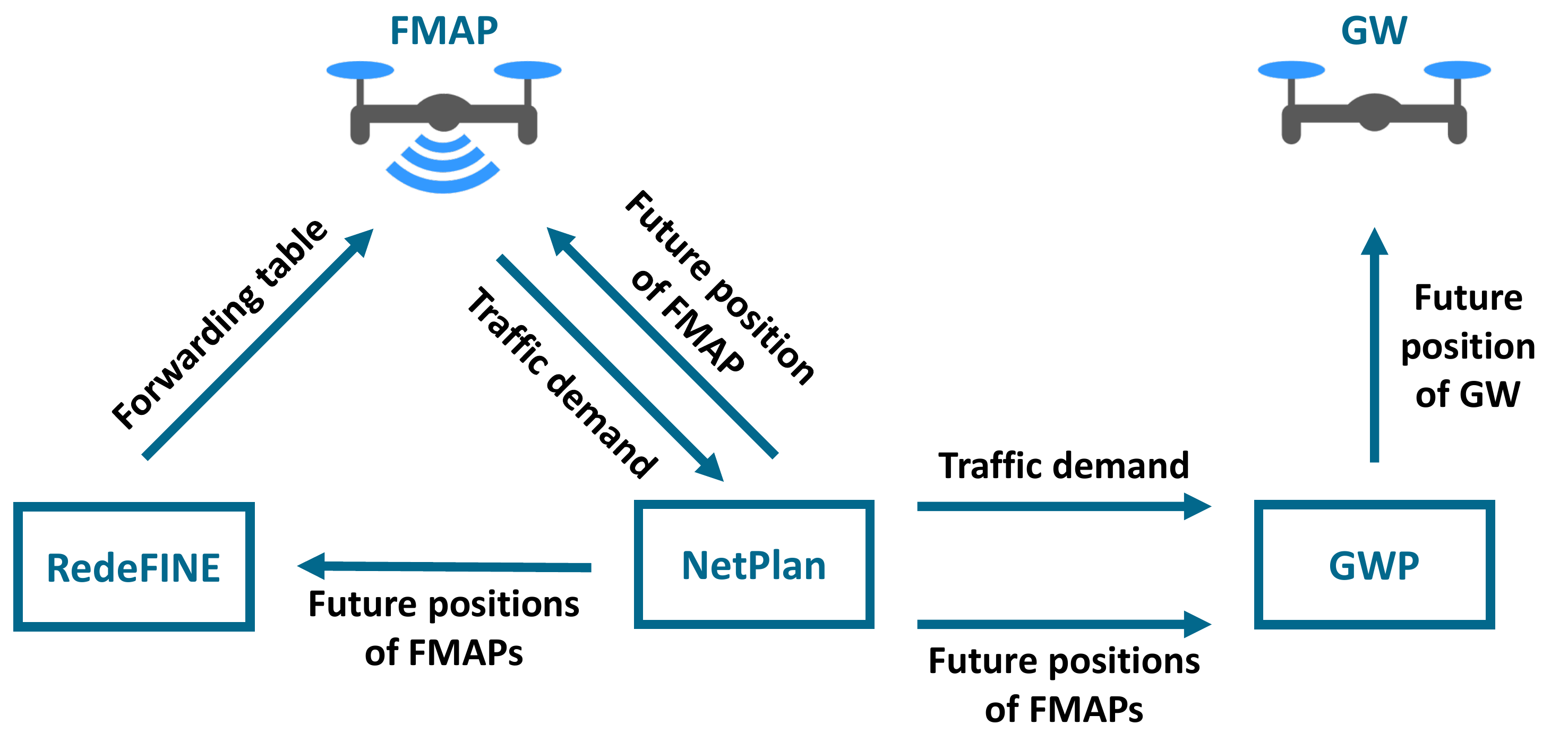}
    \caption{Proposed TMFN solution, illustrating the NetPlan, the RedeFINE and the GWP algorithms and their interactions.}
    \label{TMFN-Figure: TMFN components}
\end{figure}
%%%%%%%%%%%%%%%%%%

The TMFN is controlled by three components working cooperatively: i) the NetPlan algorithm; ii) the RedeFINE routing protocol; and iii) the GWP algorithm. The interactions between the three components are explained in \cref{TMFN-Figure: TMFN components}. By sharing the decisions taken among its components, the TMFN is able to quickly adapt not only to the variable users' traffic demand, but also to the movement of the UAVs themselves. To allow sharing information in advance, the components run simultaneously in a central station.

The integrated proposed solution determines the updated TMFN topology periodically with each update cycle having the following sequence of operations. The first step is to take a snapshot of the TMFN and the users. Then, all components consider this snapshot and the decisions taken by the remaining components as their inputs and determine the final updated TMFN topology as follows. First, NetPlan determines the updated positions and Wi-Fi cell ranges of the FMAPs according to the users' positions and their offered traffic. Knowing the initial and final positions of the FMAPs, RedeFINE calculates their future trajectories. Using this information, RedeFINE determines the optimal forwarding tables of the FMAPs and the time instants they shall be updated. Lastly, considering the future positions of FMAPs and their offered traffic, GWP determines the optimal position of the GW. When all components are combined, the TMFN is able to seamlessly transition into the updated topology, which provides an improved QoS to the users whilst minimizing the disruption caused by the movement of the UAVs.

%%%%%%%%%%%%%%%%%%%%%%%%%%%%%%%%%%%%%%%%%%%%%%%%%%%%%%%%%%%%%%%%%
% NETWORK PLANNING ALGORITHM
%%%%%%%%%%%%%%%%%%%%%%%%%%%%%%%%%%%%%%%%%%%%%%%%%%%%%%%%%%%%%%%%%
\section{Network Planning Algorithm} \label{NetplanAlgorithm-Section}

The NetPlan algorithm, originally proposed in \cite{almeida2018traffic}, is explained in this section.

\subsection{System Model} \label{NetplanAlgorithm-Section: System model}

\begin{figure}
    \centering
    \includegraphics[width=1\linewidth]{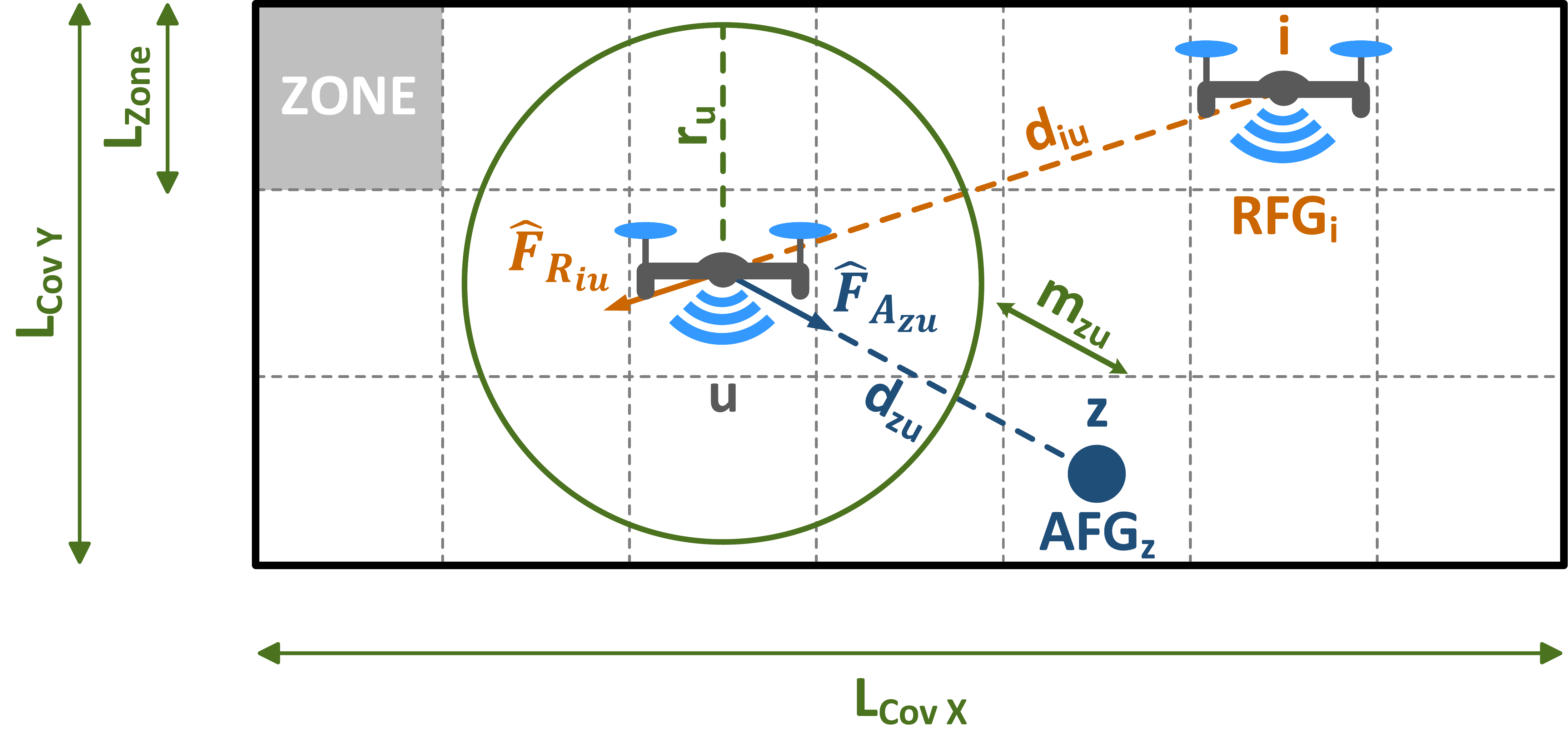}
    \caption{Proposed NetPlan algorithm, representing the map, zones, PFGs, corresponding forces applied on FMAP $ u $, and its Wi-Fi cell.}
    \label{NetplanAlgorithm-Figure: NetPlan algorithm}
\end{figure}

\cref{NetplanAlgorithm-Figure: NetPlan algorithm} represents the model of the system. Let the area to be covered by the TMFN be represented as a rectangle $ L_{\textrm{Cov}_\textrm{X}} \times L_{\textrm{Cov}_\textrm{Y}} $ named \emph{map}. The covered area is further subdivided into smaller fixed-size squares $ L_\textrm{Zone} $, defining \emph{zones}, which represent and aggregate all users on that geographic area. Each zone is identified by its index $ z \in \{ 1, ..., Z \} $.

The TMFN is composed of $ F $ FMAPs, which are identified by index $ f \in \{ 1, ..., F \} $. The FMAPs are hovering at a constant altitude $ H_f $ and may only move in the horizontal plane $ (x, y) $. Each FMAP's Wi-Fi cell operates in a dedicated IEEE 802.11n \SI{20}{MHz} channel in the \SI{5}{GHz} band. The channel model replicates the experimental model evaluated on the field, which is discussed in \cref{NetplanAlgorithmEvaluation-Section: Experimental channel model}. Thus, due to the dominant line-of-sight (LoS) component in the FMAP--FMAP link, this channel is modeled by the Friis path loss model \cite{khuwaja2018survey}. The FMAP--User link is modeled by the Friis path loss and Rician fast-fading characterized by the Rician K-factors $ K_U $ and $ K_D $ for the uplink and downlink directions, respectively.

$ U $ users are positioned throughout the coverage area either generating or receiving traffic from the FMAPs. The users are assumed to be associated to the closest FMAP.

%%%%% FIGURE %%%%%
%%%%%%%%%%%%%%%%%%

\subsection{Overview}

The NetPlan algorithm, illustrated in \cref{NetplanAlgorithm-Figure: NetPlan algorithm}, is inspired in the concepts presented in the Potential Fields (PFs) technique \cite{khatib1986real}. In this sense, PF Generators (PFGs) are virtually deployed on the map, representing users' traffic demand hotspots -- Attractive PF Generators (AFGs) -- and areas already covered by the FMAPs -- Rejective PF Generators (RFGs). Each PFG generates a force field which applies corresponding forces on FMAPs, forcing them to move in the direction of the resulting force applied to them. The FMAPs' Wi-Fi cell ranges are determined directly as a function of the PFGs combined with the current TMFN topology. Hence, by means of the appropriate determination of the PFGs' intensities, locations and corresponding forces, the NetPlan algorithm is able to predominantly position the FMAPs establishing small cells closer to the users generating more traffic (hotspots), while the remaining FMAPs are distributed throughout the map with larger cells, so that the overall map coverage is not compromised.

The NetPlan algorithm runs on a central station that periodically determines the updated positions and Wi-Fi cell ranges of the FMAPs, in which $ T_\textrm{NetPlan} \gg \SI{1}{s} $ is the update period. The updated FMAPs' positions are determined as follows: i) calculate the intensity and location of the PFGs, based on the current TMFN topology and the users' traffic demand; ii) calculate the resulting force applied to the FMAPs; iii) calculate the corresponding displacement vector; and iv) determine the updated position as the sum of the previous coordinates with the displacement vector. The updated FMAPs' Wi-Fi cell ranges are determined directly as a function of the PFGs combined with the current TMFN topology. Finally, the central station transmits the new coordinates and Wi-Fi cell ranges to the FMAPs, which will readjust their positions and configurations accordingly. All of these steps are explained in the following sub-sections.

\subsection{Potential Field Generators}

The calculation of the PFGs' intensity is explained in this sub-section.

\subsubsection{Attractive PFGs} \label{NetplanAlgorithm-Section: Attractive PFGs}

In order to \emph{attract} FMAPs towards high concentrations of generated traffic, an Attractive PFG $ AFG_{z} $ is assigned to each zone $ z $ of the map. Each $ AFG_{z} $ represents the aggregation of all users in zone $ z $ and is located in the center of that zone. Its intensity is defined in \cref{NetplanAlgorithm-Equation: Total AFG}, which includes two components.

%%%%% EQUATION %%%%%
\begin{equation} \label{NetplanAlgorithm-Equation: Total AFG}
AFG_{z} = K_{AFG^{T}} \times AFG^{T}_{z} + K_{AFG^{C}} \times AFG^{C}_{z}
\end{equation}
%%%%%%%%%%%%%%%%%%%%

The $ AFG^{T}_{z} $ component represents the aggregate users' traffic demand in zone $ z $ and is given by ${ AFG^{T}_{z} = K_{AFG^{T}_{T}} \times T_{z} + K_{AFG^{T}_\textrm{Min}} }$, where $ T_{z} $ is the mean aggregate offered throughput of the users in zone $ z $, $ K_{AFG^{T}_{T}} $ is a calibration constant and $ K_{AFG^{T}_\textrm{Min}} $ is the baseline value of $ AFG^{T}_{z} $, ensuring all AFGs have a minimum value so that all zones attract FMAPs and, thus, help maintain general coverage of the map. To limit the value of $ AFG_{z} $, the algorithm considers $ T_{z} \in [0, T_\textrm{Max}] $.

The $ AFG^{C}_{z} $ component is an additional factor that allows zones with insufficient Wi-Fi coverage to attract more FMAPs to that area, thereby ensuring that all zones have proper coverage regardless of the traffic demand. To accomplish this goal, $ AFG^{C}_{z} \in [0, 1] $ and depends on the positions of all FMAPs, as well as their cell ranges. In this sense, ${ m_{zu} = r_{u} - d_{zu} }$ is defined as the distance margin between the edge of the FMAP's Wi-Fi cell $ r_{u} $ (covered in \cref{NetplanAlgorithm-Section: FMAPs WiFi cell range}), and the distance between FMAP $ u $ and the center of zone $ z $. Then, based on the value of ${ m_{z} = \max \{ m_{z1}, ..., m_{zU} \} }$, $ AFG^{C}_{z} $ is defined as a three-branched equation:

\begin{enumerate}
    \item if $ m_{z} \leq 0 $, no FMAP is covering that zone, so $ AFG^{C}_{z} = 1 $ thereby increasing the overall $ AFG_{z} $ intensity;

    \item if $ m_{z} \geq L_\textrm{Zone} $, at least one FMAP is properly covering that zone, hence $ AFG^{C}_{z} = 0 $ and the overall $ AFG_{z} $ is not affected;

    \item if $ 0 < m_{z} < L_\textrm{Zone} $, no FMAP provides sufficient coverage of that zone, thus ${ AFG^{C}_{z} = 1 - m_{z} / L_\textrm{Zone} }$.
\end{enumerate}

Each component is then multiplied by the calibration constants $ K_{AFG^{T}} $ and $ K_{AFG^{C}} $, respectively, to adjust the variation intervals of each component to the characteristics of the map and TMFN.

\subsubsection{Rejective PFGs}

To distribute the FMAPs across the whole map, while focusing on the zones with more traffic demand, a Rejective PFG $ RFG_{u} $ is assigned to each FMAP $ u $. In fact, the FMAPs' RFGs neutralize the AFGs of the zones covered by the FMAP, pushing away nearby FMAPs to cover other areas of the map. The intensity of $ RFG_{u} $ is given by \cref{NetplanAlgorithm-Equation: Total RFG} and revolves around a mean value $ RFG^{C}_{u} $ to which an adjustment value $ RFG^{T}_{u} $ is added. This enables the FMAPs to be predominantly positioned around the zones with higher traffic demand, while ensuring the overall coverage of the map, independently of the exact values of the users' traffic demand.

%%%%% EQUATION %%%%%
\begin{equation} \label{NetplanAlgorithm-Equation: Total RFG}
RFG_{u} = K_{RFG^{T}} \times RFG^{T}_{u} + K_{RFG^{C}} \times RFG^{C}_{u}
\end{equation}
%%%%%%%%%%%%%%%%%%%%

Therefore, the adjustment value is determined as ${ RFG^{T}_{u} = \frac{1}{Z} \sum_{z=1}^{Z} AFG^{T}_{z}  -  \frac{1}{\left| C_{u} \right|} \sum_{z \in C_{u}} AFG^{T}_{z} }$, in which $ C_{u} = \{ z : m_{zu} \geq 0 \} $ is the set of zones $ z $ whose center is located within the Wi-Fi cell covered by FMAP $ u $. If $ C_{u} = \emptyset $, then $ RFG^{T}_{u} = 0 $. The mean value is determined as ${ RFG^{C}_{u} = \frac{1}{Z} \sum_{z=1}^{Z} AFG^{T}_{z} }$. Finally, each component is further multiplied respectively by the calibration constants $ K_{RFG^{T}} $ and $ K_{RFG^{C}} $ to adjust the relative weight of each component as well as the intensity of the final equation to the characteristics of the map and TMFN.

\subsection{Potential Field Generators' Forces}

The calculation of the resulting force applied to each FMAP is explained in this section and illustrated in \cref{NetplanAlgorithm-Figure: NetPlan algorithm}.

\subsubsection{Attractive Forces}

The force $ \vec{F}_{A_{zu}} $ applied by $ AFG_{z} $ to FMAP $ u $ is defined in \cref{NetplanAlgorithm-Equation: F Attractive}. It is directly proportional to $ AFG_{z} $ and the distance between the center of zone $ z $ and FMAP $ u $ ($ d_{zu} $), and has the direction of $ \boldsymbol{\hat{F}_{A_{zu}}} $, which is the unit vector starting at FMAP $ u $ and pointing to the center of zone $ z $. This ensures the FMAP is attracted with greater intensity by the zones located further away and/or with more intensity, so that all zones are properly covered, especially the ones with higher $ AFG_{z} $. The constant $ K_{F_{A}} $ is added to the equation to calibrate the force's intensity.

%%%%% EQUATION %%%%%
\begin{equation} \label{NetplanAlgorithm-Equation: F Attractive}
\vec{F}_{A_{zu}} = \left( K_{F_{A}} \times AFG_{z} \times d_{zu} \right) \times \boldsymbol{\hat{F}_{A_{zu}}}
\end{equation}
%%%%%%%%%%%%%%%%%%%%

\subsubsection{Rejective Forces}

The force $ \vec{F}_{R_{iu}} $ applied by $ RFG_{i} $ to FMAP $ u $ is given by \cref{NetplanAlgorithm-Equation: F Rejective}. It is directly proportional to $ RFG_{u} $, but inversely proportional to the distance between FMAPs $ i $ and $ u $ ($ d_{iu} $), and has the direction of $ \boldsymbol{\hat{F}_{R_{iu}}} $, which is the unit vector starting at FMAP $ u $ and pointing in the opposite direction of FMAP $ i $. This enables FMAPs to move away from nearby and/or intense RFGs, while being mostly unaffected by distant RFGs. As a result, FMAPs will be predominantly positioned around zones with higher traffic demand, while ensuring overall coverage of the map. The constant $ K_{F_{R}} $ is added to the equation to calibrate the force's intensity.

%%%%% EQUATION %%%%%
\begin{equation} \label{NetplanAlgorithm-Equation: F Rejective}
\vec{F}_{R_{iu}} = \left( K_{F_{R}} \times RFG_{i} / d_{iu} \right) \times \boldsymbol{\hat{F}_{R_{iu}}}
\end{equation}
%%%%%%%%%%%%%%%%%%%%

\subsubsection{Resulting Force}

The resulting force $ \vec{F}_{u} $ applied to FMAP $ u $ is given by the sum of all forces applied to it, as defined in \cref{NetplanAlgorithm-Equation: F Resulting}. As a result of this force, an instantaneous acceleration $ \vec{a}_{u} $ is imposed on the FMAP. Assuming the FMAP's position is fixed at the beginning of cycle $ n $, the resulting displacement vector of FMAP $ u $ is given by $ \vec{s}_{u}[n] = K_{s} \times \vec{F}_{u}[n] $, in which $ K_{s} \propto T^2_\textrm{NetPlan} / m $ summarizes the underlying Physics constants as a final calibration constant.

%%%%% EQUATION %%%%%
\begin{equation} \label{NetplanAlgorithm-Equation: F Resulting}
\vec{F}_{u} = \sum_{z=1}^{Z} \vec{F}_{A_{zu}} + \sum_{\substack{i=1 \\ i \neq u}}^{U} \vec{F}_{R_{iu}}
\end{equation}
%%%%%%%%%%%%%%%%%%%%

\subsection{FMAP's Wi-Fi Cell Range} \label{NetplanAlgorithm-Section: FMAPs WiFi cell range}

As discussed in \cref{NetplanAlgorithm-Section: System model}, in order to improve the TMFN's aggregate throughput, the FMAPs closer to the users with higher traffic demand should establish smaller Wi-Fi cells, whereas the remaining FMAPs should establish larger Wi-Fi cells to maintain the overall coverage of the map. To implement this behavior, the FMAP's Wi-Fi cell range is determined according to \cref{NetplanAlgorithm-Equation: FMAPs cell range}. Similar to the FMAP's $ RFG_{u} $, it revolves around a mean value $ R_\textrm{Mean} $ to which an adjustment value $ r_{u_{\Delta}} $ is added. Since the FMAP's $ RFG^{T}_{u} $ already implements this behavior, the adjustment value of the Wi-Fi cell range $ r_{u_{\Delta}} $ is proportional to $ RFG^{T}_{u} $, in which $ K_{r} $ is the calibration constant added to the equation to adjust this value to the characteristics of the TMFN and the event.

%%%%% EQUATION %%%%%
\begin{equation} \label{NetplanAlgorithm-Equation: FMAPs cell range}
r_{u} = R_\textrm{Mean} + K_{r} \times RFG^{T}_{u}
\end{equation}
%%%%%%%%%%%%%%%%%%%%

%%%%%%%%%%%%%%%%%%%%%%%%%%%%%%%%%%%%%%%%%%%%%%%%%%%%%%%%%%%%%%%%%
% EVALUATION OF NETWORK PLANNING ALGORITHM
%%%%%%%%%%%%%%%%%%%%%%%%%%%%%%%%%%%%%%%%%%%%%%%%%%%%%%%%%%%%%%%%%
\section{Evaluation of Network Planning Algorithm} \label{NetplanAlgorithmEvaluation-Section}

%%%%% FIGURE %%%%%
\begin{figure*}[t]
	\centering
	\subfloat[Homogeneous traffic demand.] {
		\includegraphics[width=0.3\linewidth]{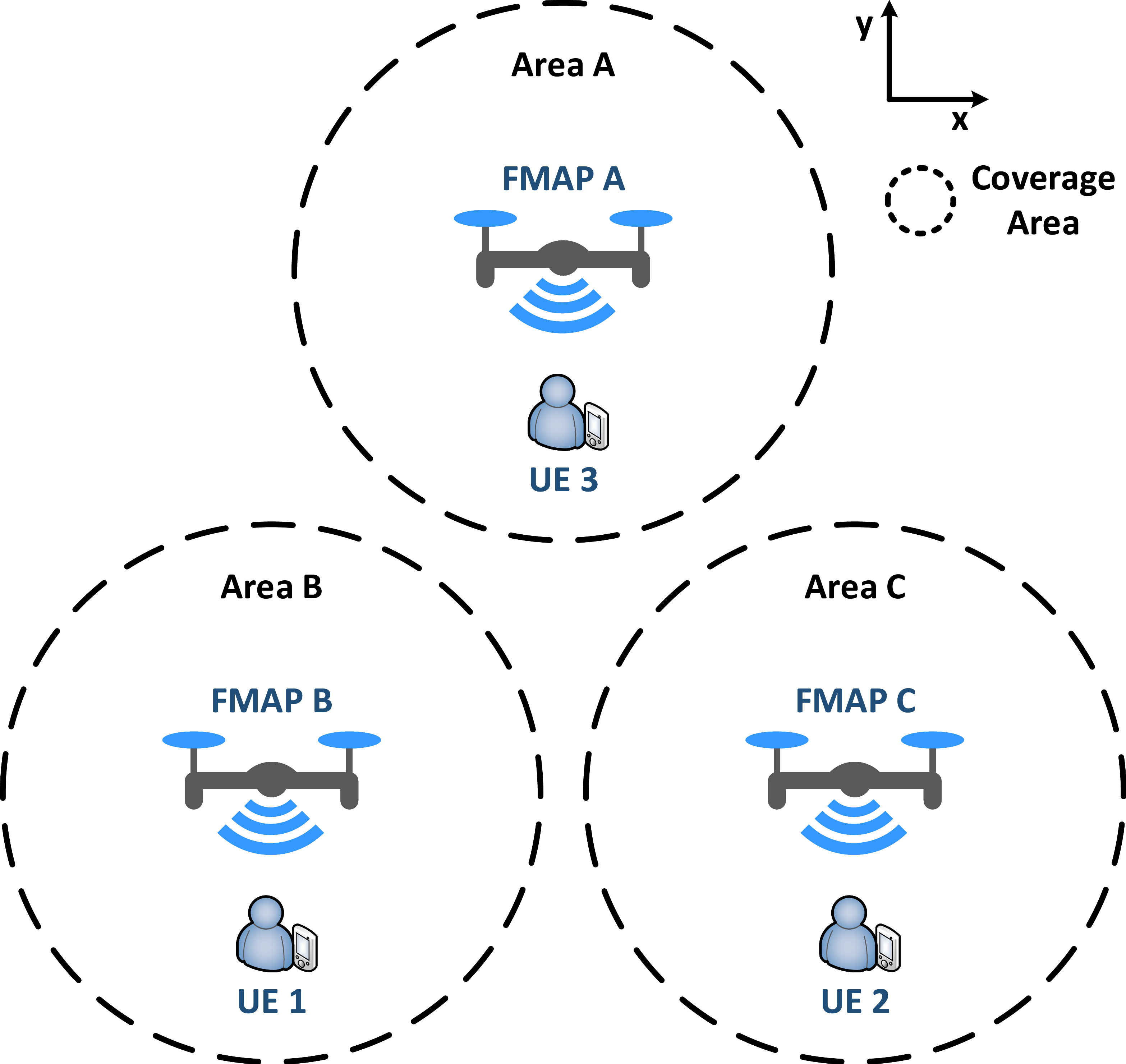}
		\label{NetplanAlgorithmEvaluation-Figure: Homogeneous traffic demand scenario}
	}
	\hfill
	\subfloat[Concentrated traffic demand (without NetPlan).] {
		\includegraphics[width=0.3\linewidth]{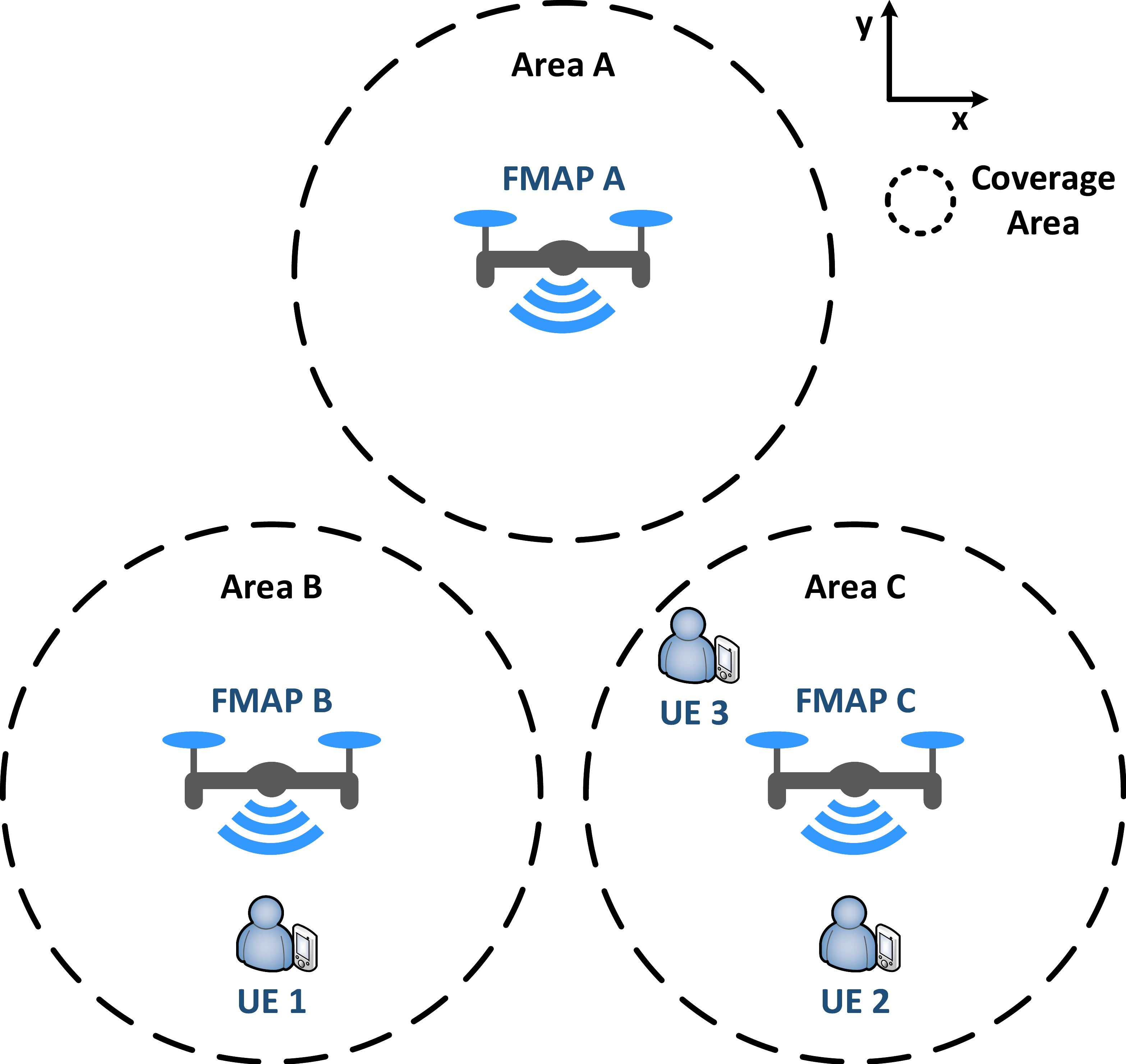}
		\label{NetplanAlgorithmEvaluation-Figure: Concentrated traffic demand scenario without NetPlan}
	}
	\hfill
	\subfloat[Concentrated traffic demand (with NetPlan).] {
		\includegraphics[width=0.3\linewidth]{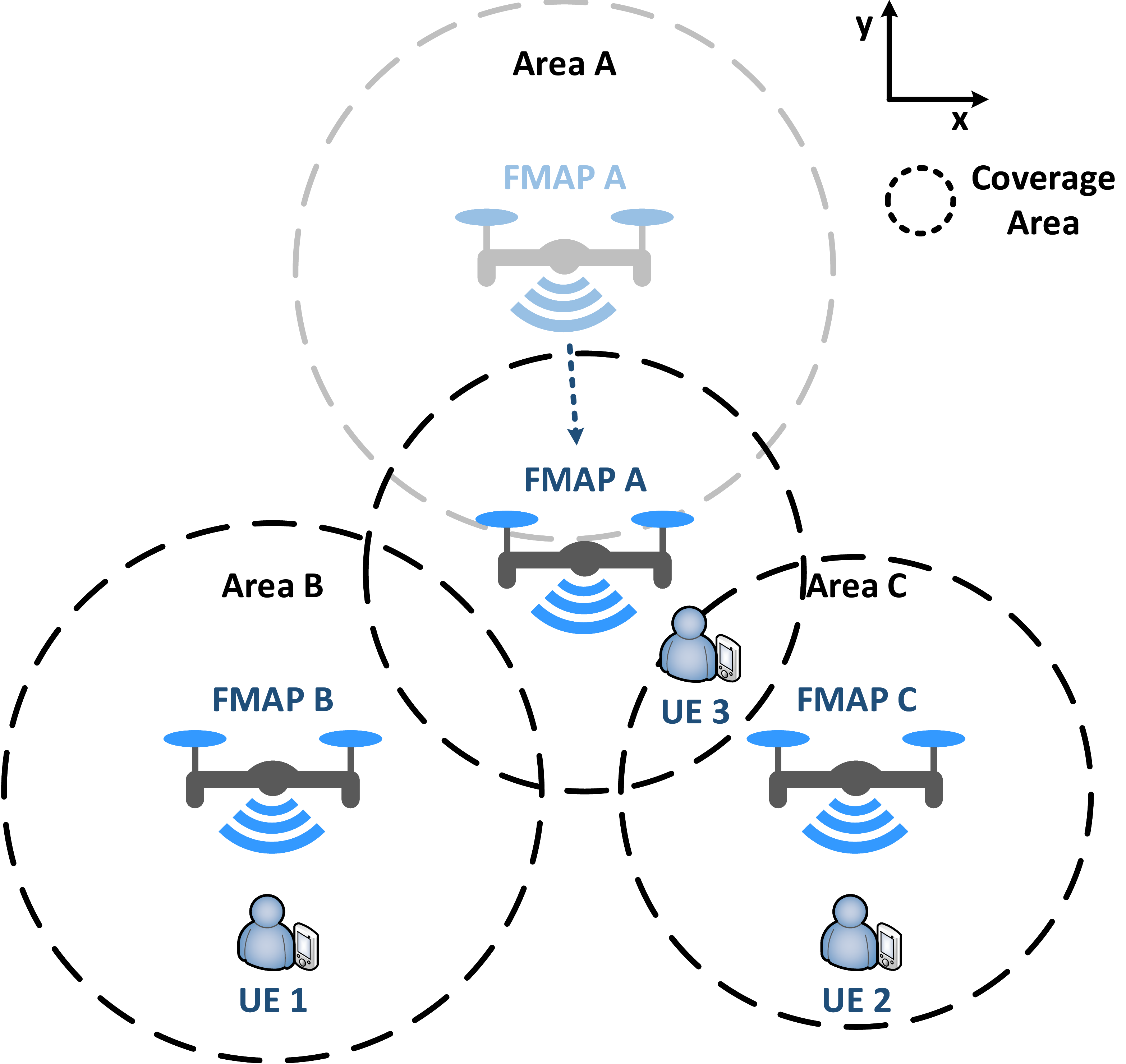}
		\label{NetplanAlgorithmEvaluation-Figure: Concentrated traffic demand scenario with NetPlan}
	}
	\caption{Test scenarios depicting the FMAP positions and the associated UEs.}
	\label{NetplanAlgorithmEvaluation-Figure: Test scenarios}
\end{figure*}
%%%%%%%%%%%%%%%%%%

%%%%% TABLE %%%%%
\begin{table*}[t]
	\centering
	\caption{Association map indicating the FMAP to which a UE is connected.}
	\label{NetplanAlgorithmEvaluation-Table: FMAP-UE association map}
	\begin{tabular}{l l l l}
		\hline % --------------------------------------------------------
		&   UE 1    &  UE 2   & UE 3    \\
		\hline % --------------------------------------------------------
		Homogeneous traffic demand                      &   FMAP B  &   FMAP C  &   FMAP A  \\
		Concentrated traffic demand (without NetPlan)   &   FMAP B  &   FMAP C  &   FMAP C  \\
		Concentrated traffic demand (with NetPlan)      &   FMAP B  &   FMAP C  &   FMAP A  \\
		\hline % --------------------------------------------------------
	\end{tabular}
\end{table*}
%%%%%%%%%%%%%%%%%

The performance of the NetPlan algorithm was evaluated by means of simulations and an experimental testbed. The scenarios defined for the tests are explained in \cref{NetplanAlgorithmEvaluation-Section: Test scenarios}. For each scenario, Matlab simulations of the NetPlan algorithm are presented, enabling the visualization of the final positions of the FMAPs determined by the NetPlan algorithm. In addition, ns-3 simulations \cite{ns3Simulator} were performed to analyze the network performance of the User Equipment (UE) for different offered traffic types and in both the uplink and downlink flow directions.

In terms of experimental results, first the FMAP--UE communications channel was characterized in terms of Signal-to-Noise Ratio (SNR), considering different LoS distances and FMAP altitudes. This enabled the verification of the theoretical models proposed in the literature \cite{khuwaja2018survey}. Moreover, using the experimental channel models in the ns-3 simulations allows a more accurate reproduction of the testbed conditions. Then, the performance of the communications channel, considering the defined scenarios, was assessed with the the Cumulative Distribution Function (CDF) of the throughput and the histogram of the data rates of the packets generated by the FMAPs and the UEs.

%%%%% TABLE %%%%%
\begin{table}
	\centering
	\caption{System and ns-3.29 simulator parameters.}
	\label{NetplanAlgorithm-Table: System and ns3 parameters}
	\begin{tabular}{l l}
		\hline % --------------------------------------------------------
		\multicolumn{2}{c}{Coverage area and TMFN}	\\
		\hline % --------------------------------------------------------
		$ L_{\textrm{Cov}_\textrm{X}}$  &	$ \SI{100}{m} $ \\
		$ L_{\textrm{Cov}_\textrm{Y}}$  &	$ \SI{100}{m} $ \\
		$ L_{\textrm{Zone}} $           &	$ \SI{10}{m} $  \\
		$ Z $                           &	100 zones       \\
		$ F $                           &	3 FMAPs         \\
		$ U $                           &	3 users         \\
		$ H_f $                         &	$ \SI{10}{m} $  \\
		$ K_U $                         &	$ \SI{13}{dB} $ \\
		$ K_D $                         &	$ \SI{40}{dB} $ \\
		\hline % --------------------------------------------------------
		\multicolumn{2}{c}{ns-3.29 simulator parameters}	\\
		\hline % --------------------------------------------------------
		Simulation time         &	(\SI{30}{s} init. +) \SI{20}{s}          \\
		Wi-Fi standard          &	IEEE 802.11n        \\
		Wi-Fi channels          &	\{36, 40, 44\}      \\
		Channel bandwidth       &	\SI{20}{MHz}        \\
		Tx power                &	\SI{0}{dBm}         \\
		Propagation model       &	Friis + Rician      \\
		Application traffic     &	UDP CBR or          \\
		&	TCP \emph{BulkSend} \\
		Packet length           &	\SI{8000}{Bytes}    \\
		MAC queues              &	500 Packets         \\
		MAC auto rate           &	MinstrelHt          \\
		\hline % --------------------------------------------------------
	\end{tabular}
\end{table}
%%%%%%%%%%%%%%%%%

\subsection{Test Scenarios} \label{NetplanAlgorithmEvaluation-Section: Test scenarios}

%%%%% FIGURE %%%%%
\begin{figure*}[t]
	\centering
	\subfloat[SNR measured in the FMAP.] {
		\includegraphics[width=0.47\linewidth]{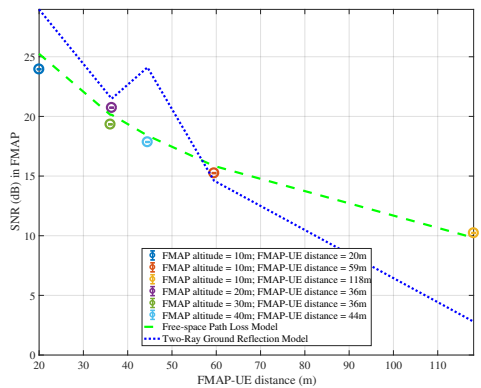}
		\label{NetplanAlgorithmEvaluation-Figure: Experimental SNR FMAP}
	}
	\hfill
	\subfloat[SNR measured in the UE.] {
		\includegraphics[width=0.47\linewidth]{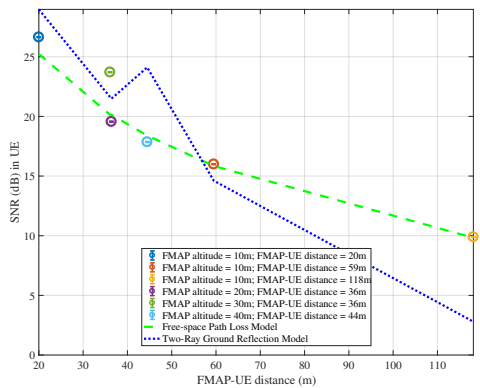}
		\label{NetplanAlgorithmEvaluation-Figure: Experimental SNR UE}
	}
	%%%
	\vfil
	%%%
	\subfloat[SNR PDF measured in the FMAP.] {
		\includegraphics[width=0.47\linewidth]{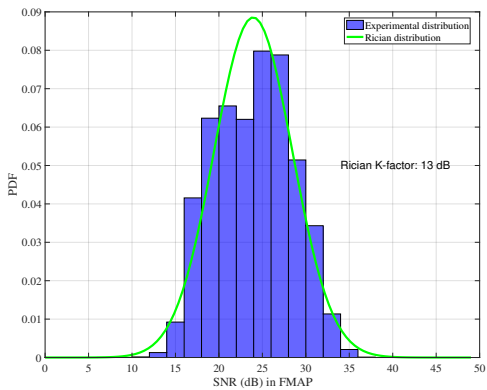}
		\label{NetplanAlgorithmEvaluation-Figure: Experimental SNR PDF FMAP}
	}
	\hfill
	\subfloat[SNR PDF measured in the UE.] {
		\includegraphics[width=0.47\linewidth]{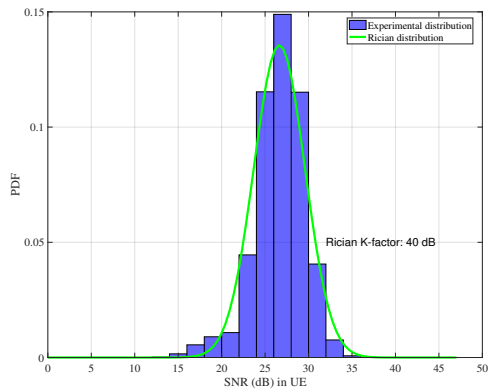}
		\label{NetplanAlgorithmEvaluation-Figure: Experimental SNR PDF UE}
	}
	
	\caption{SNR measured in the FMAP and UE. Subfigures (a) and (b) plot the SNR versus the LoS distance between each other at different altitudes. The free-space path loss and two-ray ground reflection theoretical models are represented by the dashed and dotted lines, respectively. Subfigures (c) and (d) represent the SNR PDF measured in the FMAP placed at \SI{10}{m} altitude, and the UE, considering the LoS distance of approximately \SI{20}{m}.}
	\label{NetplanAlgorithmEvaluation-Figure: Experimental SNR PDF}
\end{figure*}
%%%%%%%%%%%%%%%%%%

The test scenarios aim at evaluating the performance of the NetPlan algorithm in typical networking scenarios. They explore different traffic demands in order to demonstrate the concept and network performance gains of the NetPlan algorithm.

Two scenarios were defined for this evaluation: i) homogeneous traffic demand, which is considered as the baseline; and ii) concentrated traffic demand, which allows the evaluation of the network performance gains. Each scenario is characterized by four factors: i) the UE positions; ii) the UE offered traffic; iii) the FMAP positions; and iv) the FMAP--UE association map, indicating the FMAP to which each UE is connected. The test scenarios are illustrated in \cref{NetplanAlgorithmEvaluation-Figure: Test scenarios} and the FMAP--UE association map is indicated in \cref{NetplanAlgorithmEvaluation-Table: FMAP-UE association map}. Each scenario is further explained in the corresponding subsection.

Three offered traffic types were considered: i) TCP; ii) UDP \SI{25}{Mbit/s} constant bitrate (UDP 25M); and iii) UDP \SI{75}{Mbit/s} constant bitrate (UDP 75M). The TCP traffic flow allows the analysis of the channel in saturation, whereas the results of the UDP 25M and UDP 75M flows allow the analyses of the channel when the offered traffic is low and high, respectively. For each traffic type, both flow directions were simulated: i) UE to FMAP (uplink); and ii) FMAP to UE (downlink).

The ns-3 simulation parameters are shown in \cref{NetplanAlgorithm-Table: System and ns3 parameters}. The ns-3 results were obtained using the \emph{FlowMonitor} \cite{carneiro2009flowmonitor} module, which analyzes the traffic flows at the IP network layer. The experimental results were obtained considering 3 runs for each experiment, under the same conditions.

\subsection{Experimental Channel Model} \label{NetplanAlgorithmEvaluation-Section: Experimental channel model}

The air-to-ground and ground-to-air channel models were evaluated in the experimental setup. The FMAP was hovering at different altitudes and hence different LoS distances to the UE. The SNR received at both the FMAP and the UE was collected using the \emph{horst} software \cite{randolf2017horst}.

The channel model was analyzed in terms of the path loss and fast-fading components. \cref{NetplanAlgorithmEvaluation-Figure: Experimental SNR FMAP,NetplanAlgorithmEvaluation-Figure: Experimental SNR UE} represents the collected SNR values for different FMAP--UE distances. In order to compare the experimental values with the free-space path loss and the two-ray ground reflection models, both are represented in the plots. \cref{NetplanAlgorithmEvaluation-Figure: Experimental SNR PDF FMAP,NetplanAlgorithmEvaluation-Figure: Experimental SNR PDF UE} represents the probability distribution function (PDF) of the experimental SNRs, which was fitted with a Rician distribution.

In terms of the path loss component, it can be concluded that the free-space path loss is the most adequate model of the UE--FMAP link, as suggested in the literature \cite{khuwaja2018survey}. This is an expected conclusion, since there is a dominant LoS component between the communications nodes. Nevertheless, the two-ray ground reflection model, which in addition to the LoS component considers a component reflected on the ground and the effect of the antennas' heights, also provides a close SNR estimation, especially for UE--FMAP distances up to \SI{60}{m} and FMAP's altitudes up to \SI{30}{m}. For the same LoS distance, the angle $ \theta_{LoS} $ between the ground and the LoS ray affected the SNR on the UE and the FMAP differently. When $ \theta_{LoS} $ increased, the SNR measured on the UE increased, whereas the SNR on the FMAP decreased. This can be concluded from \cref{NetplanAlgorithmEvaluation-Figure: Experimental SNR FMAP,NetplanAlgorithmEvaluation-Figure: Experimental SNR UE}, by analyzing the two SNRs measured for the same UE--FMAP LoS distance of \SI{36}{m}, but with FMAP's altitude of \SI{20}{m} or \SI{30}{m}.

Regarding the Rician fast-fading component (K-factor), which represents the ratio of the received power in the dominant component to the non-dominant power, it presents a higher value in the UE, as observed in \cref{NetplanAlgorithmEvaluation-Figure: Experimental SNR PDF FMAP,NetplanAlgorithmEvaluation-Figure: Experimental SNR PDF UE}. The lower values in the FMAP can be justified by the obstructions to the signal caused by the airframe of the UAV and the compensation movements performed by the UAV to maintain its position, which induce changes in the antennas' tilt.

\subsection{Homogeneous Traffic Demand}

This scenario aimed at evaluating the network performance when multiple UEs generating the same amount of traffic were evenly distributed throughout the coverage area. To test this scenario, three UEs were always placed at fixed positions, each one within the coverage area of an FMAP, as illustrated in \cref{NetplanAlgorithmEvaluation-Figure: Homogeneous traffic demand scenario}. Each UE is associated to the closest FMAP, which results in each FMAP having only one UE associated. This is considered as the baseline scenario, since the use of the NetPlan algorithm does not significantly change the positions of the FMAPs when the overall traffic demand is equally distributed throughout the coverage area.

% Given the homogeneity of the traffic demand and the placement of the UEs, the NetPlan algorithm positioned the FMAPs in order to form similar Wi-Fi cells able to provide the same bandwidth to the UEs.

\subsubsection{Simulation Results}

%%%%% FIGURE %%%%%
\begin{figure*}[t]
    \centering

    \subfloat[TCP offered traffic.] {
        \includegraphics[width=0.3\linewidth]{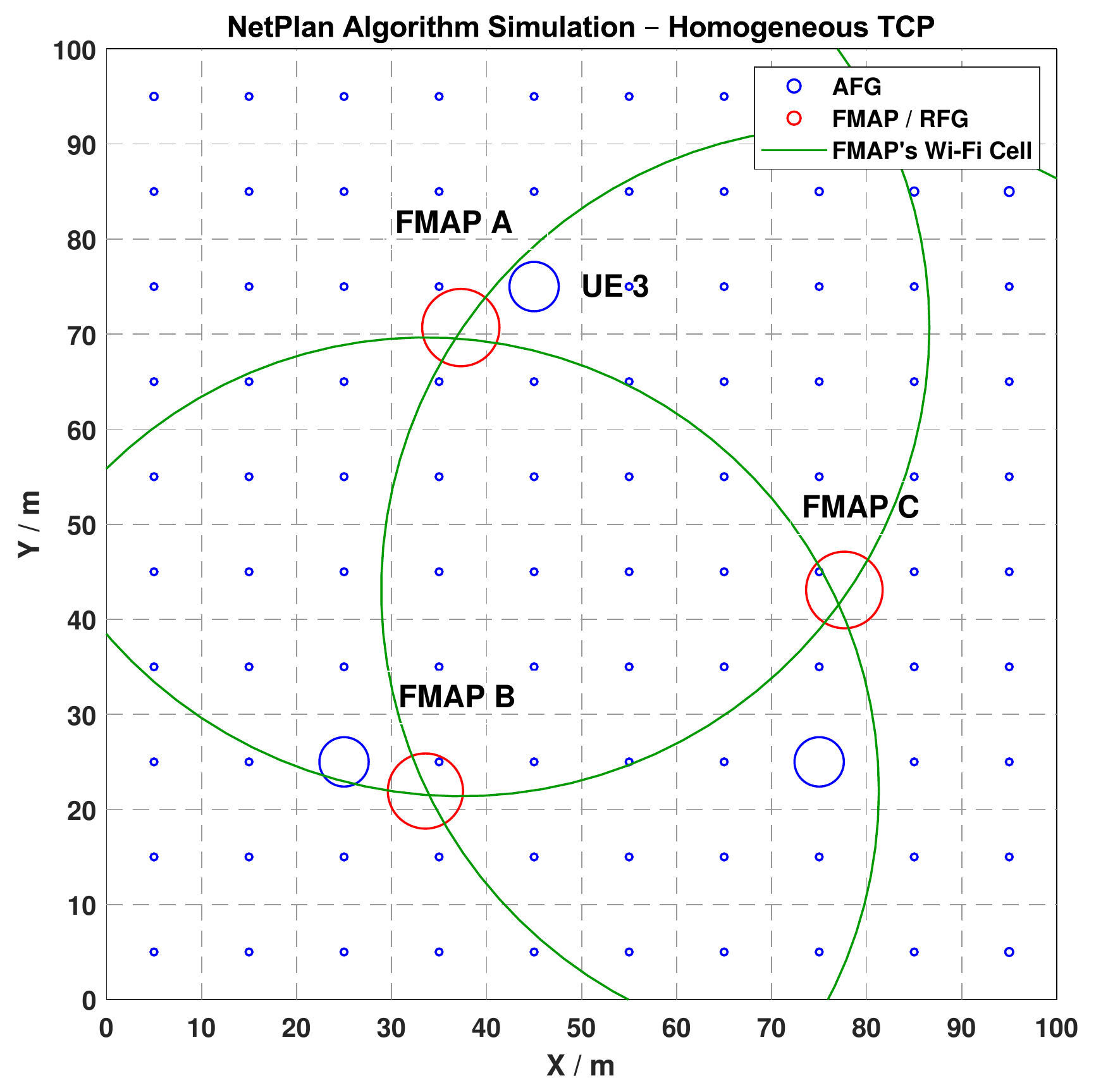}
        \label{NetplanAlgorithmEvaluation-Figure: Homogeneous matlab simulations TCP}
    }
    \hfill
    \subfloat[\SI{25}{Mbit/s} UDP offered traffic.] {
        \includegraphics[width=0.3\linewidth]{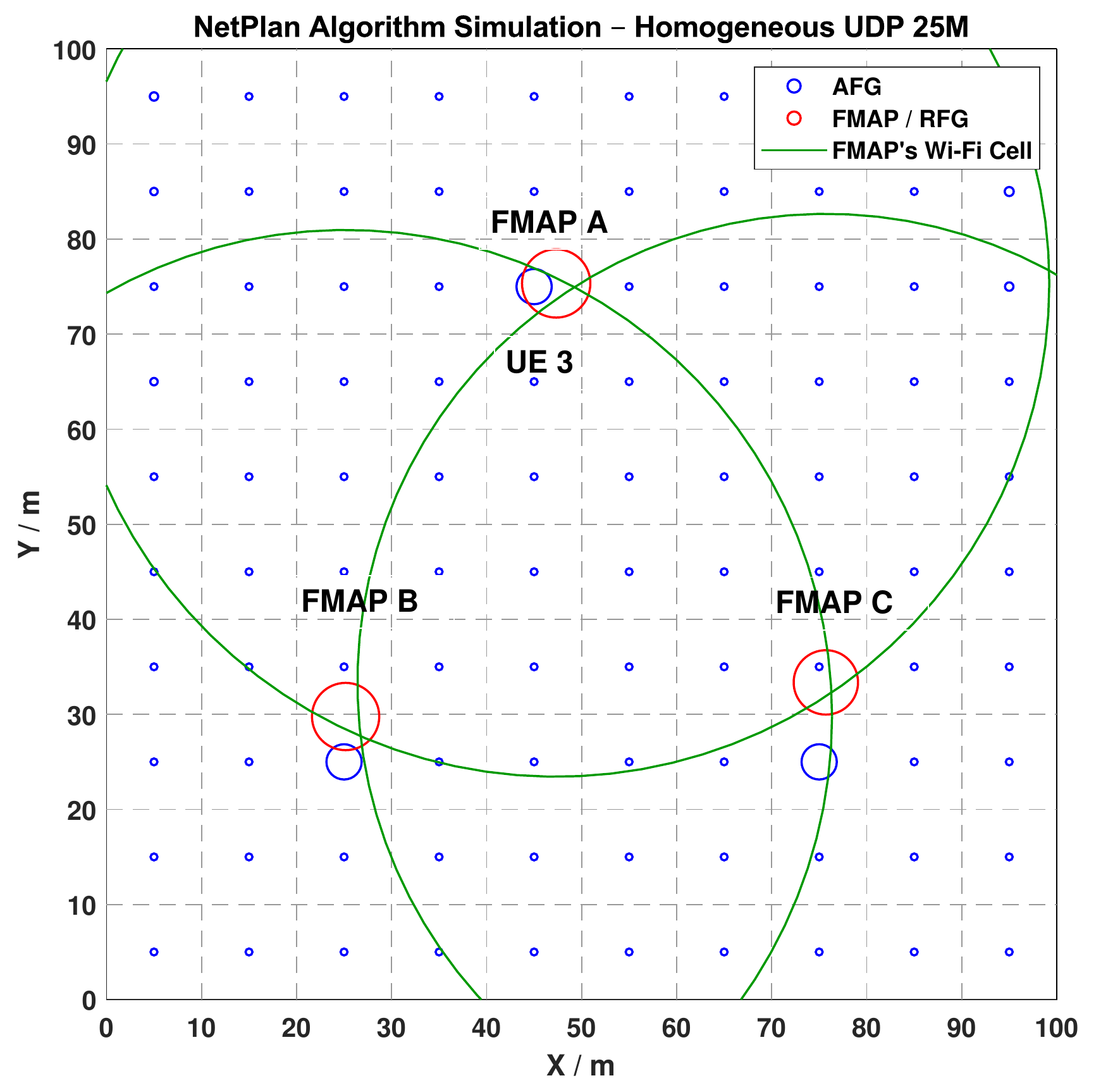}
        \label{NetplanAlgorithmEvaluation-Figure: Homogeneous matlab simulations UDP 25M}
    }
    \hfill
    \subfloat[\SI{75}{Mbit/s} UDP offered traffic.] {
        \includegraphics[width=0.3\linewidth]{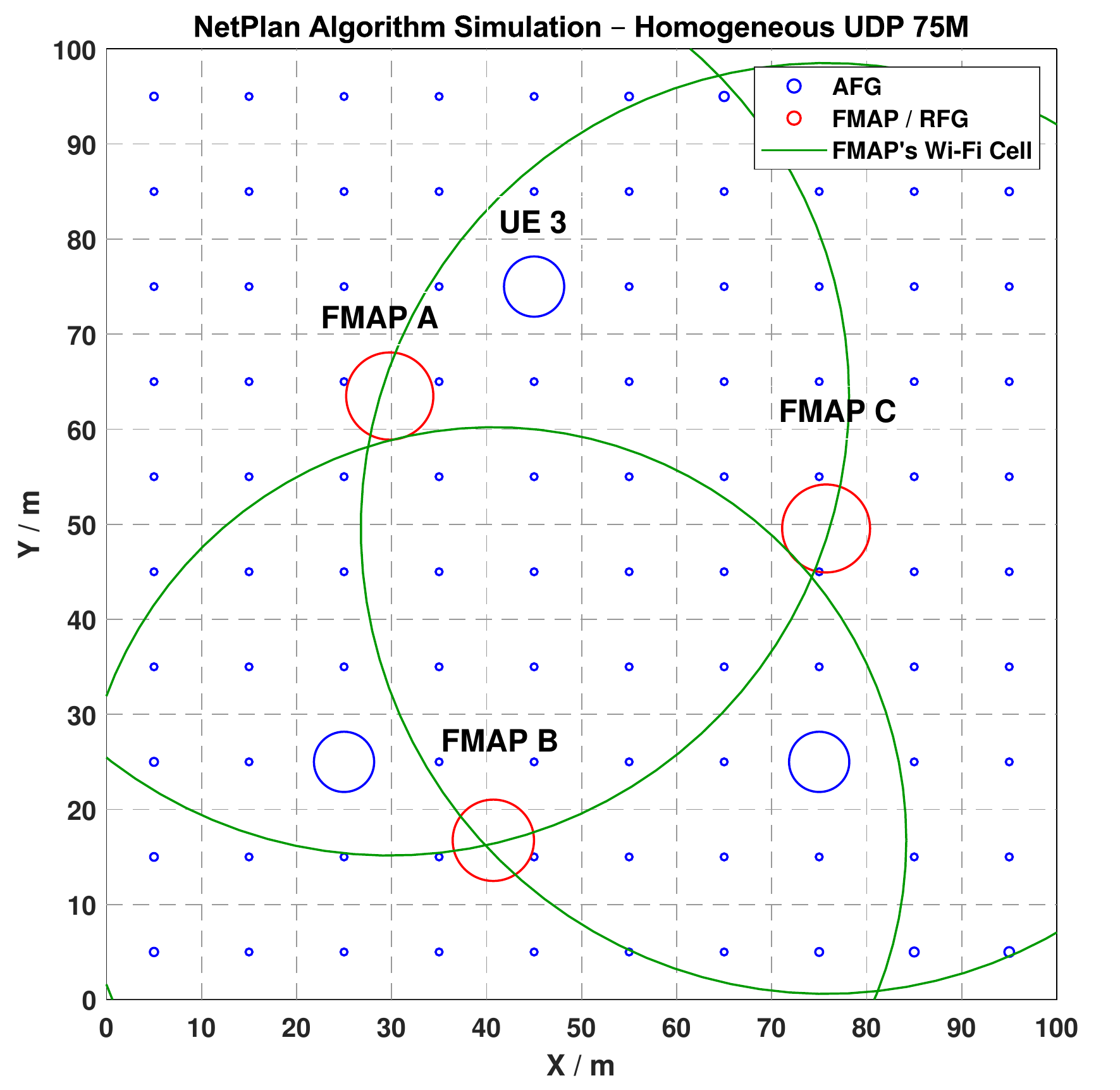}
        \label{NetplanAlgorithmEvaluation-Figure: Homogeneous matlab simulations UDP 75M}
    }

    \caption{Homogeneous traffic demand scenario showing the final FMAP positions determined by the NetPlan algorithm, according to different users' traffic demand.}
    \label{NetplanAlgorithmEvaluation-Figure: Homogeneous matlab simulations}
\end{figure*}
%%%%%%%%%%%%%%%%%%

%%%%% FIGURE %%%%%
\begin{figure*}[t]
    \centering

    \subfloat[Throughput CDF.] {
        \includegraphics[width=0.3\linewidth]{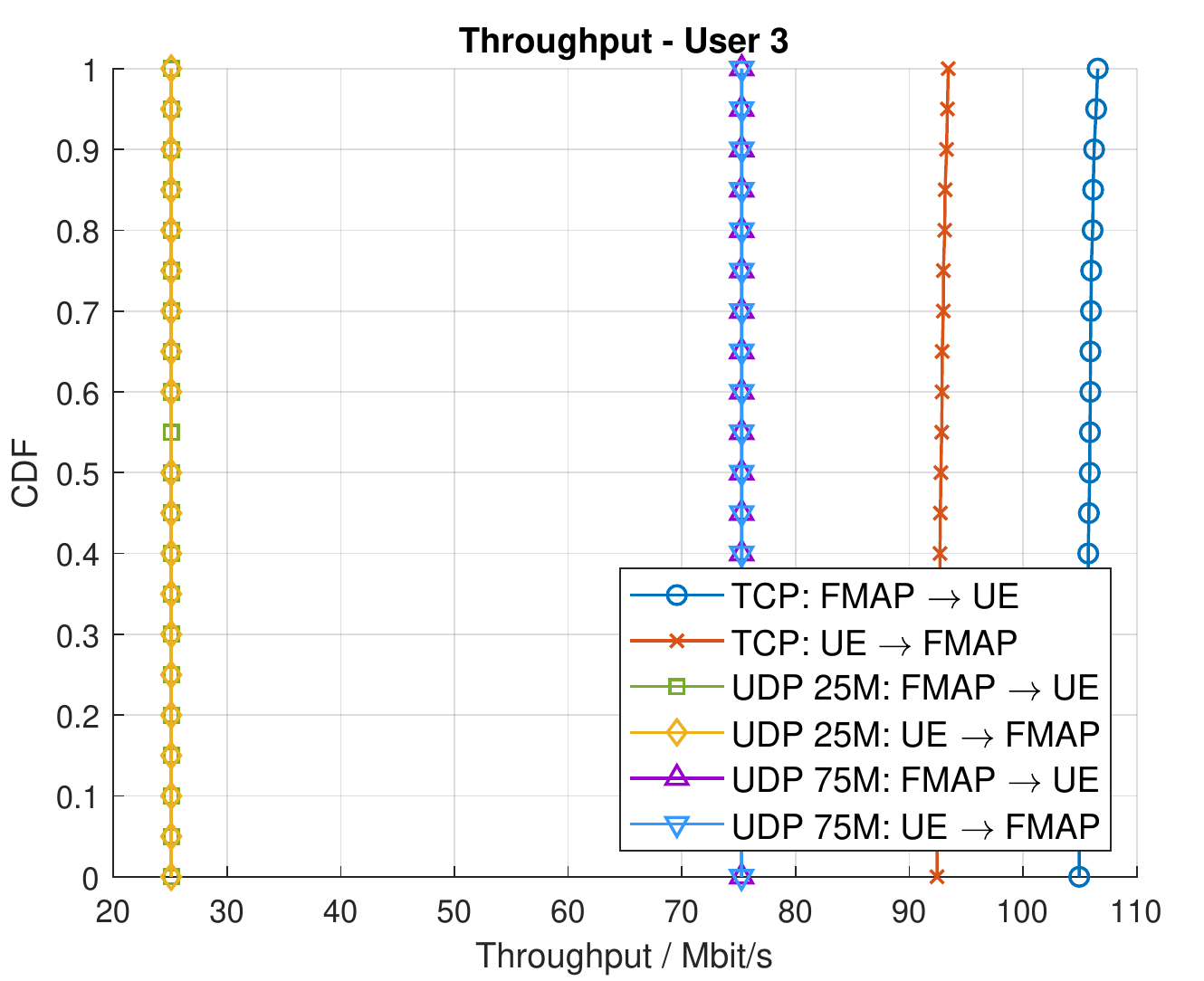}
        \label{NetplanAlgorithmEvaluation-Figure: Homogeneous simulation results throughput}
    }
    \hfill
    \subfloat[Delay CDF.] {
        \includegraphics[width=0.3\linewidth]{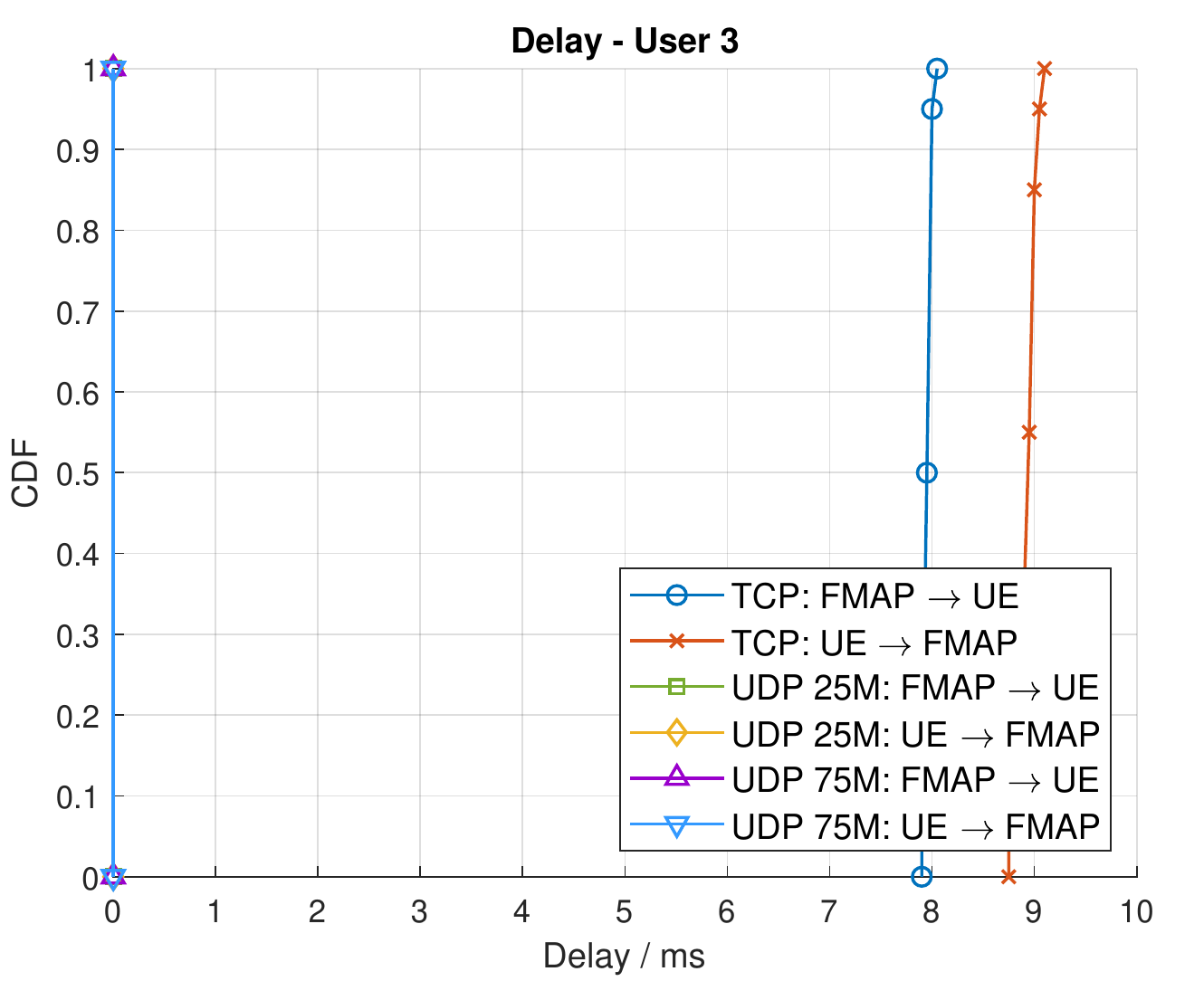}
        \label{NetplanAlgorithmEvaluation-Figure: Homogeneous simulation results delay}
    }
    \hfill
    \subfloat[PLR CDF.] {
        \includegraphics[width=0.3\linewidth]{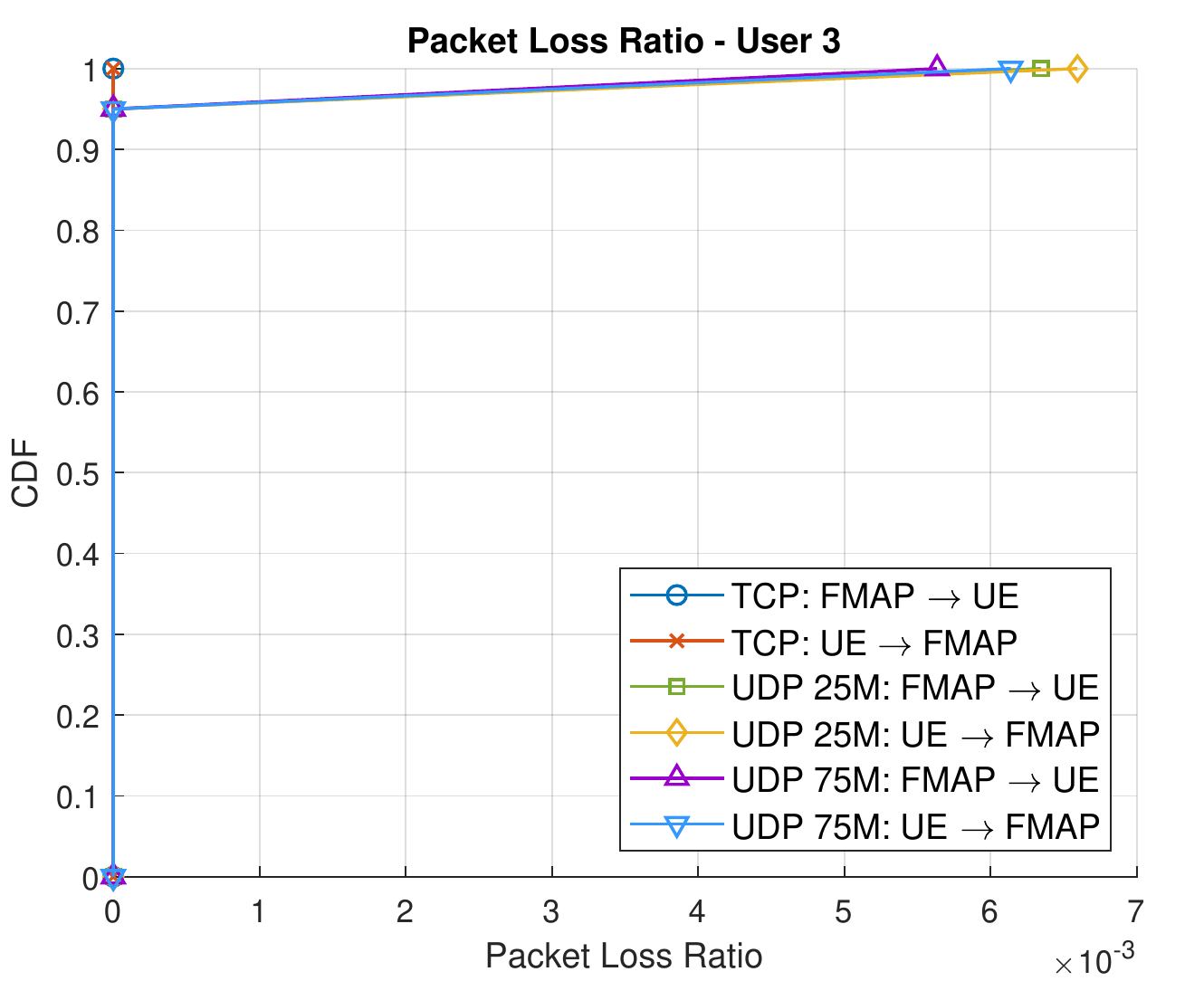}
        \label{NetplanAlgorithmEvaluation-Figure: Homogeneous simulation results PLR}
    }

    \caption{Network performance of UE 3 in the homogeneous traffic demand scenarios, obtained by means of ns-3 simulation.}
    \label{NetplanAlgorithmEvaluation-Figure: Homogeneous simulation results}
\end{figure*}
%%%%%%%%%%%%%%%%%%

\cref{NetplanAlgorithmEvaluation-Figure: Homogeneous matlab simulations} contains the results of Matlab simulations depicting the user positions, their offered traffic and the final FMAP positions determined by the NetPlan algorithm in the homogeneous traffic demand scenario for the three offered traffic types analyzed. Since the users are homogeneously distributed throughout the coverage area generating the same traffic demand, the FMAPs are also homogeneously distributed throughout the area.

The network performance results of UE 3 are displayed in \cref{NetplanAlgorithmEvaluation-Figure: Homogeneous simulation results}. Since each FMAP's Wi-Fi cell operates in a dedicated and orthogonal channel and each user is associated to a different FMAP, all users are provided with the full channel capacity.

For each offered traffic, both the uplink and downlink flows present similar results. Moreover, the TCP flow demonstrates that the maximum achievable throughput in this scenario is $\approx$~\SI{105}{Mbit/s}, with 0\% PLR and a delay of $\approx$~\SI{9}{ms}. The UDP 75M flow generates a high offered traffic and achieves a throughput of $\approx$~\SI{75}{Mbit/s} with 0\% PLR and a delay of \SI{0}{ms}. These results reveal that the UDP 75M flow does not saturate the channel. Unlike UDP, TCP includes a congestion control mechanism. Due to the congestion control mechanism, TCP dynamically adjusts the traffic offered to the IP layer so that the channel is fully utilized but not saturated. Moreover, TCP guarantees reliable end-to-end communications by means of acknowledgement packets and retransmissions in case of errors and packet losses. As a consequence, the PLR is 0\% but the packet delay increases relative to the UDP 75M traffic flow. The UDP 25M flow demonstrates that the channel has enough capacity to transport all the offered traffic with a throughput of $\approx$~\SI{25}{Mbit/s}, with 0\% PLR and a delay of $\approx$~\SI{0}{ms}.

\subsubsection{Experimental Results}

%%%%% FIGURE %%%%%
\begin{figure*}[t]
    \centering
    \subfloat[Throughput CDF.] {
        \includegraphics[width=0.48\linewidth]{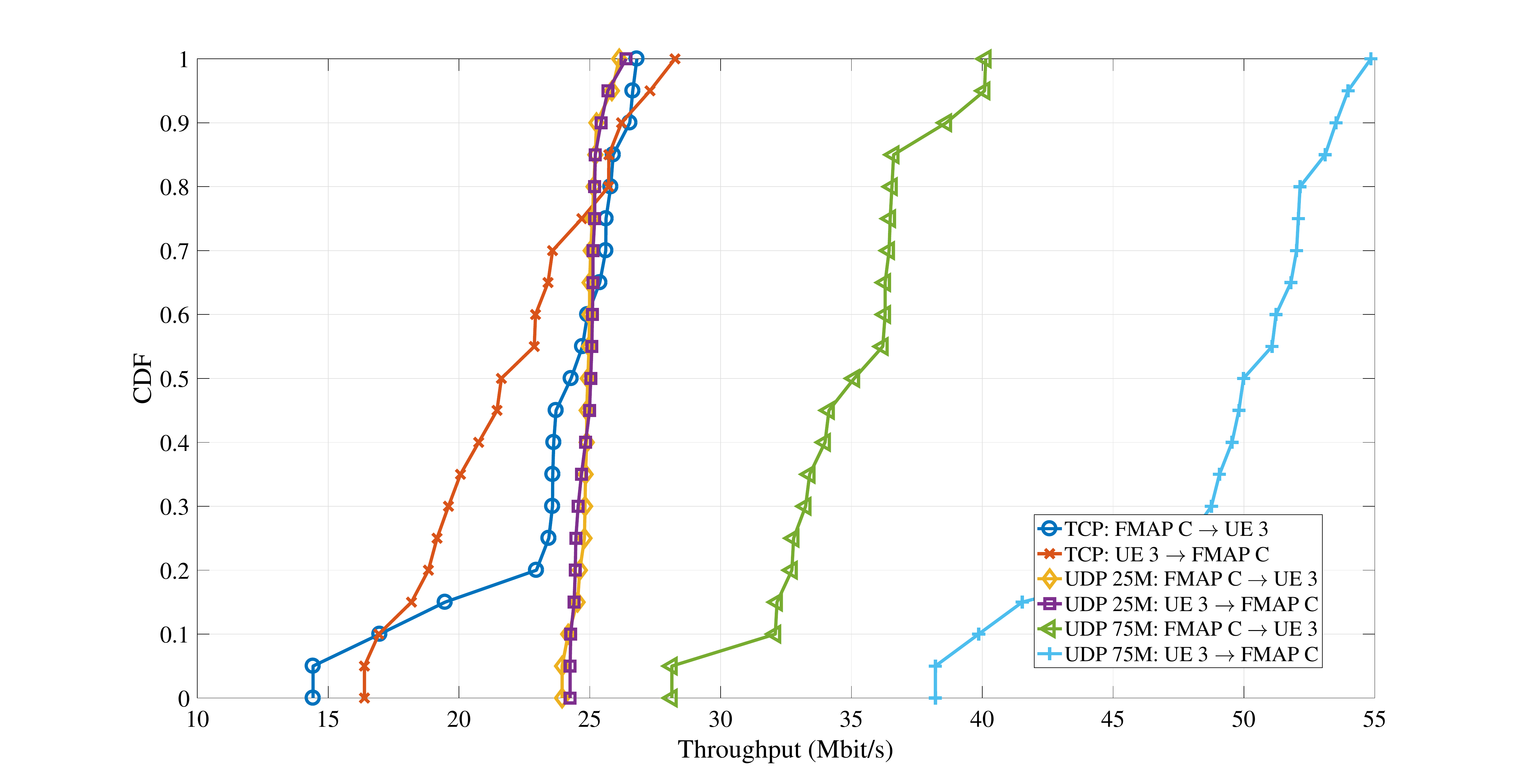}
        \label{NetplanAlgorithmEvaluation-Figure: Traffic results homogeneous throughput}
    }
    \hfill
    \subfloat[Physical data rate histogram.] {
        \includegraphics[width=0.47\linewidth]{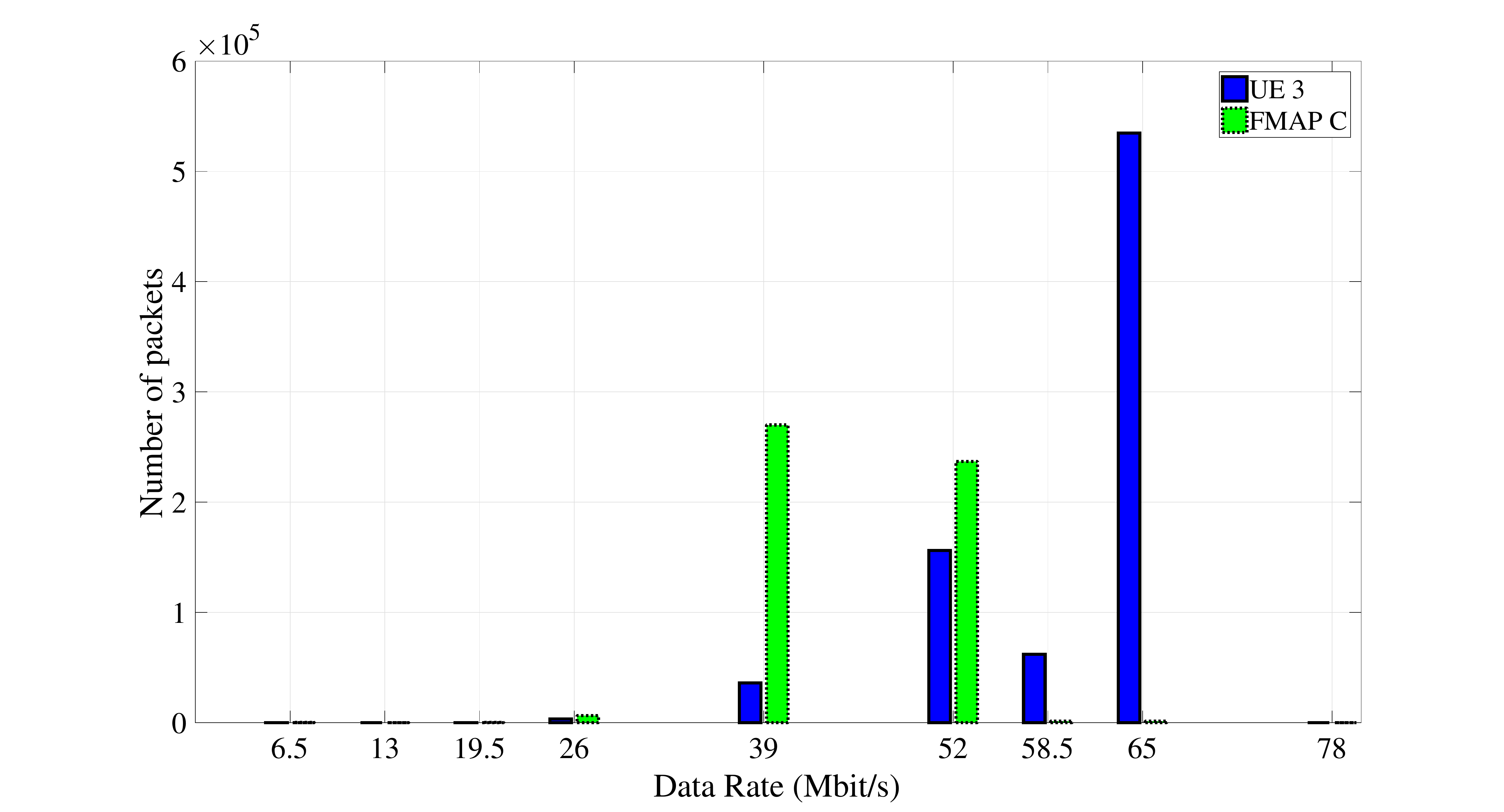}
        \label{NetplanAlgorithmEvaluation-Figure: Traffic results homogeneous data rate}
    }
    \caption{Experimental traffic results for FMAP C, considering the homogeneous traffic demand scenario. The results include the throughput CDFs and the histogram of the physical data rate of the packets generated by both the UE 3 (solid bar) and FMAP C (dotted bar) during the experiment.}
    \label{NetplanAlgorithmEvaluation-Figure: Traffic results homogeneous}
\end{figure*}
%%%%%%%%%%%%%%%%%%

In this scenario, the positions of the FMAPs were defined in order to allow that each UE on the ground was in the coverage area of a single FMAP. Since the FMAPs were configured on orthogonal channels, each UE was able to take advantage of the full channel capacity provided by a single FMAP.

The experimental throughput results are presented in \cref{NetplanAlgorithmEvaluation-Figure: Traffic results homogeneous}, represented by means of histograms of the physical data rates corresponding to the MCS for the IEEE 802.11n standard, \SI{800}{ns} guard interval, and \SI{20}{MHz} channel bandwidth, according to the configurations presented in \cref{NetplanAlgorithmEvaluation-Section: Test scenarios}.

It is possible to observe a significant asymmetry between the downlink direction (FMAP to UE) and the uplink direction (UE to FMAP), especially for the TCP traffic flow and UDP 75M traffic flow. This is denoted by the higher physical data rate values being used by most of the packets sent by the UEs to the FMAPs during the experiment. The physical data rate used to transmit a packet is selected by the MinstrelHt MAC auto-rate mechanism, which evaluates the channel conditions and selects the highest physical data rate that allows the transmission of the packet without errors. Since the conditions of the channel are variable, the selected physical data rate value will also be variable, which results in a variable throughput and PLR.

Overall, the throughput in the uplink is higher than in the downlink. This can be justified by several reasons. First, as concluded in \cref{NetplanAlgorithmEvaluation-Section: Experimental channel model} the communications link is asymmetric, since the Rician K-factor in the UE is higher than the K-factor in the FMAP. Moreover, different transmission power were used in the UE and the FMAP. Since in our testbed two commercial smartphones and a laptop were used as UEs, they were employing the default transmission power of \SI{20}{dBm}, based on the premise that our solution does not rely on modifications performed in the UEs. Conversely, in the FMAPs the transmission power was set to \SI{0}{dBm}, in order to enable short range Wi-Fi cells, required to validate the NetPlan algorithm. Link asymmetry can also be justified by the usage of different antennas in the receiver and in the transmitter. While the FMAPs were using two \SI{5}{dBi} omni-directional external antennas, the smartphones performing the role of UE were using their internal antennas, which typically have a lower reception gain. Besides the expected differences in the antennas' gain and radiation pattern, the communications performance is also affected by the firmware used to control antenna diversity, which is part of MIMO. Finally, the noise floor should not be neglected as well. It is determined by the receiver's sensitivity and by the performance of the low noise amplifier, which is in charge of amplifying weak received signals into stronger signals.

\subsection{Concentrated Traffic Demand}

This scenario, illustrated in \cref{NetplanAlgorithmEvaluation-Figure: Concentrated traffic demand scenario without NetPlan,NetplanAlgorithmEvaluation-Figure: Concentrated traffic demand scenario with NetPlan}, aimed at assessing the performance of the NetPlan algorithm and characterize the network performance when the traffic demand in a concentration area increased compared to the overall traffic demand of the coverage area. To test this scenario, the UE initially located in area A was moved to area C.

The UE--FMAP link was characterized in two cases: i) when the decisions of the NetPlan algorithm were not considered; and ii) when they were considered. In order to accommodate the traffic demand, the NetPlan algorithm positioned FMAP A near area C, which was experiencing a higher traffic demand than previously while traffic demand in area A was reduced to zero. FMAP B maintained its position, in order to provide coverage to the UE in area B. In area C, FMAP A and FMAP C enabled two cells in orthogonal channels, which were used by each UE. The UE--FMAP link was able to provide higher throughput, lower PLR, and lower Round Trip Time (RTT), measured at the application layer, when the decisions of the NetPlan algorithm were employed.

\subsubsection{Simulation Results}

%%%%% FIGURE %%%%%
\begin{figure*}
    \centering
    \subfloat[TCP offered traffic.] {
        \includegraphics[width=0.3\linewidth]{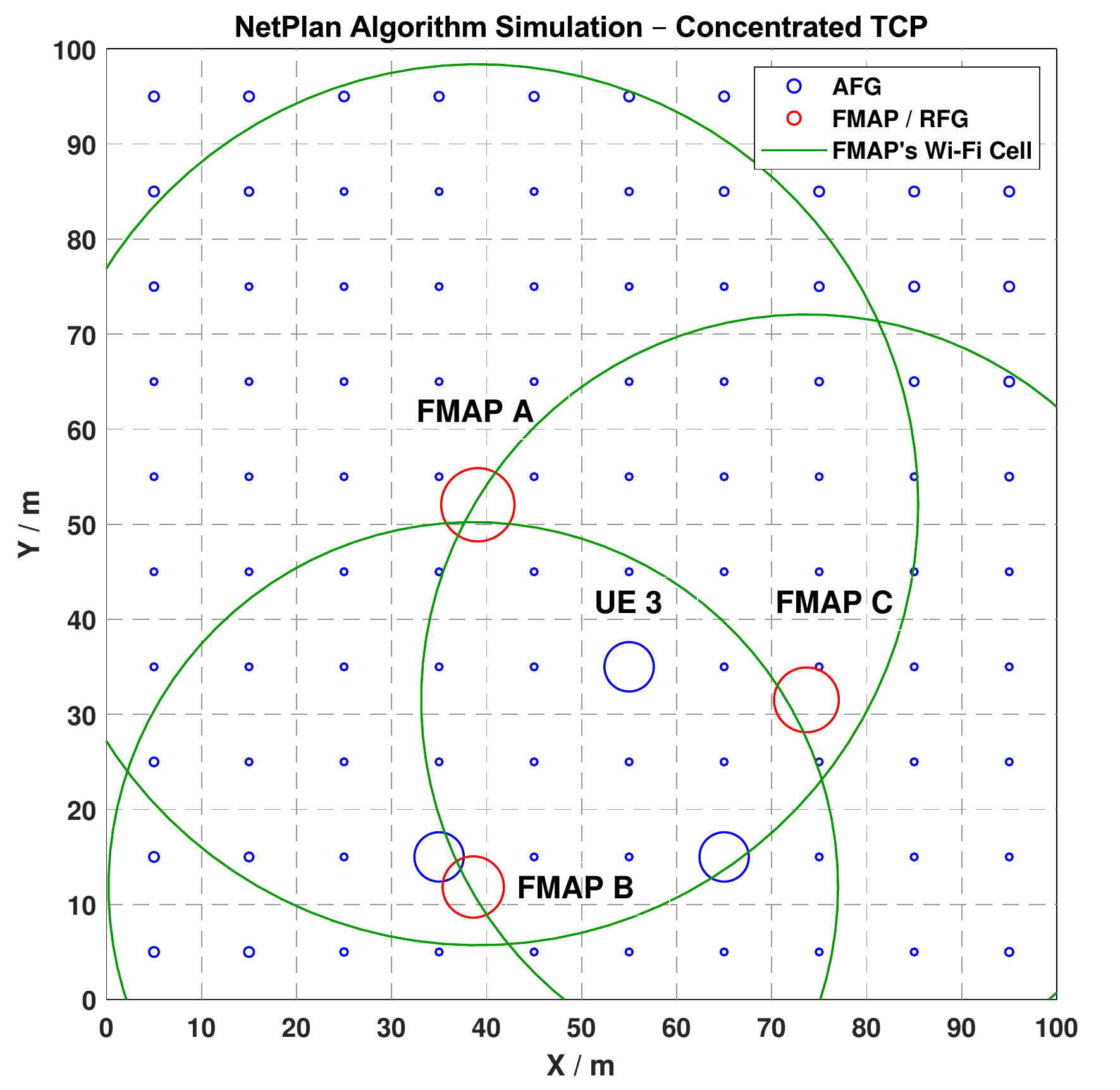}
        \label{NetplanAlgorithmEvaluation-Figure: Matlab simulations concentrated TCP}
    }
    \hfill
    \subfloat[UDP \SI{25}{Mbit/s} constant bitrate offered traffic.] {
        \includegraphics[width=0.3\linewidth]{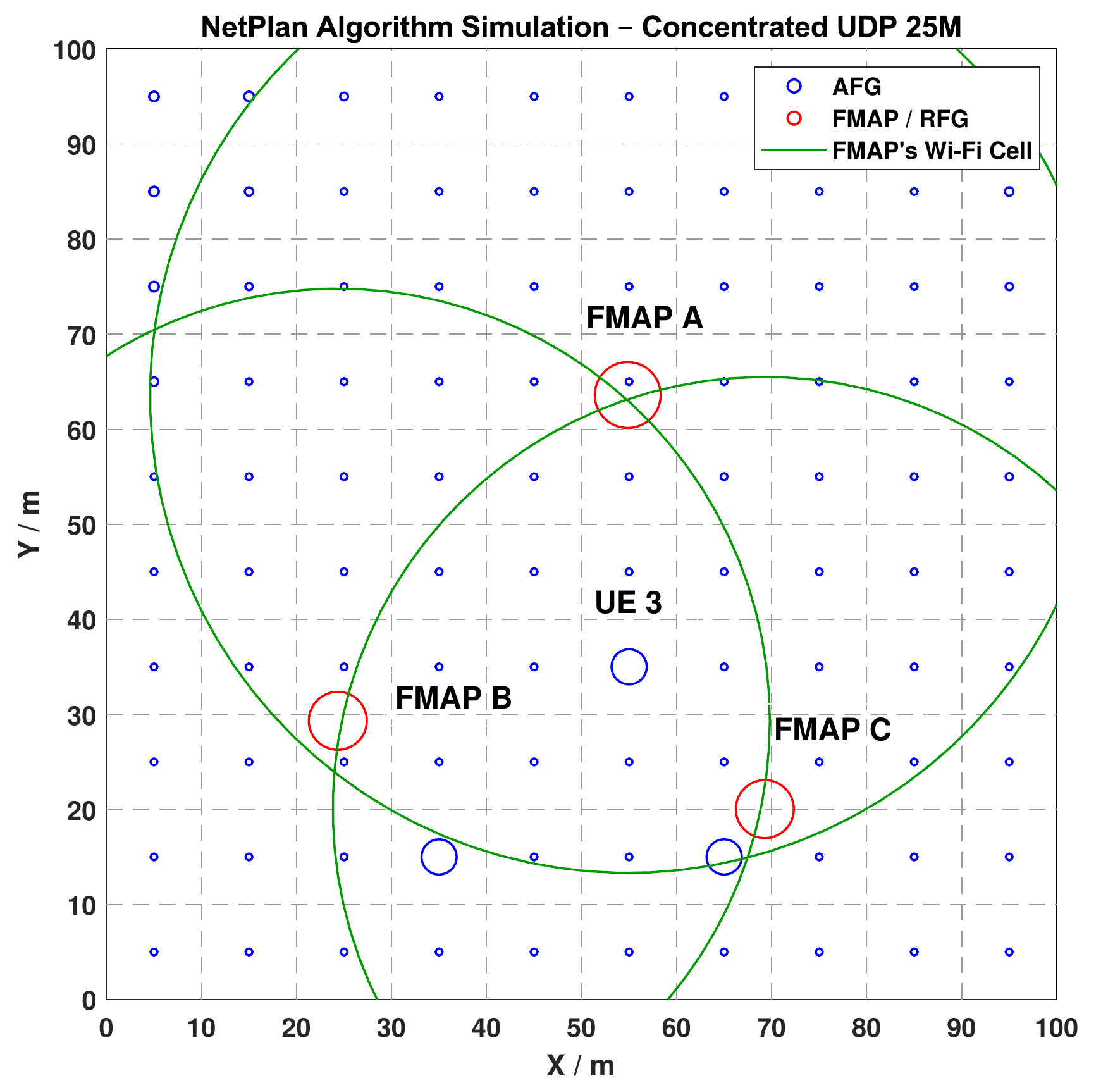}
        \label{NetplanAlgorithmEvaluation-Figure: Matlab simulations concentrated UDP 25M}
    }
    \hfill
    \subfloat[UDP \SI{75}{Mbit/s} bitrate offered traffic.] {
        \includegraphics[width=0.3\linewidth]{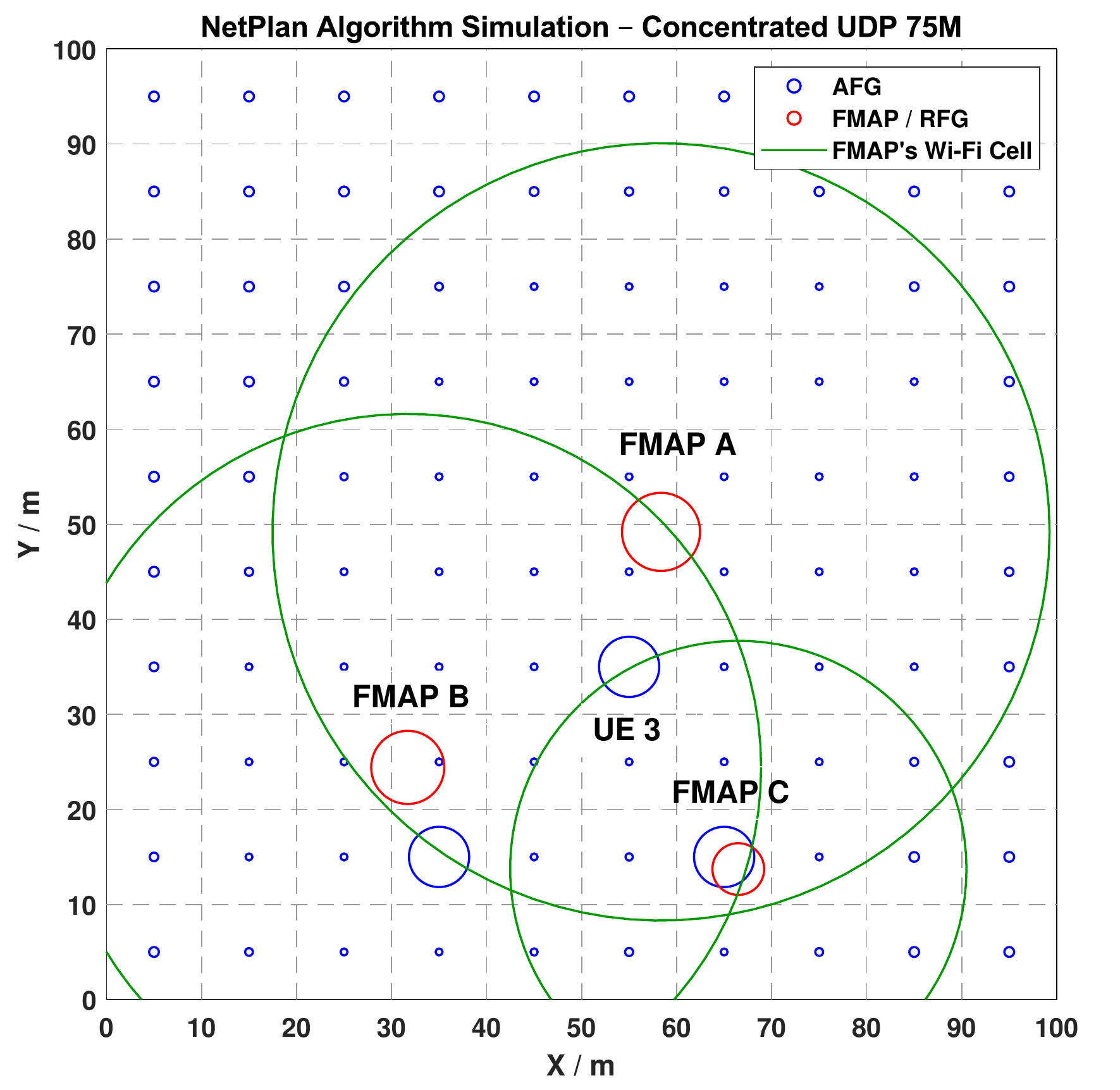}
        \label{NetplanAlgorithmEvaluation-Figure: Matlab simulations concentrated UDP 75M}
    }
    \caption{Concentrated traffic scenario showing the final FMAP positions determined by the NetPlan algorithm, according to different users' traffic demand.}
    \label{NetplanAlgorithmEvaluation-Figure: Matlab simulations concentrated}
\end{figure*}
%%%%%%%%%%%%%%%%%%

%%%%% FIGURE %%%%%
\begin{figure*}
    \centering
    \subfloat[Throughput CDF (without NetPlan).] {
        \includegraphics[width=0.3\linewidth]{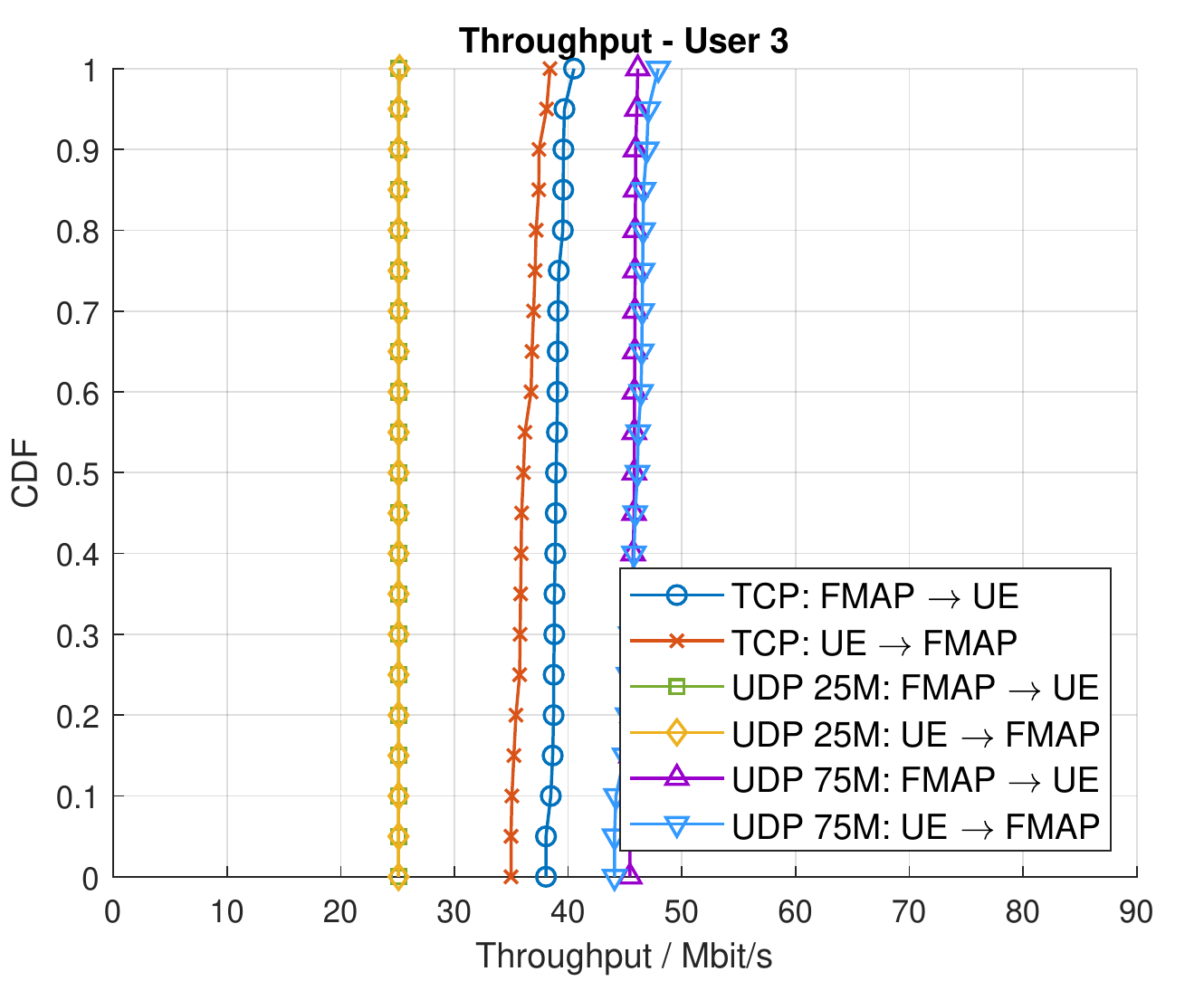}
        \label{NetplanAlgorithmEvaluation-Figure: Concentrated simulation results throughput without NetPlan}
    }
    \hfill
    \subfloat[Delay CDF (without NetPlan).] {
        \includegraphics[width=0.3\linewidth]{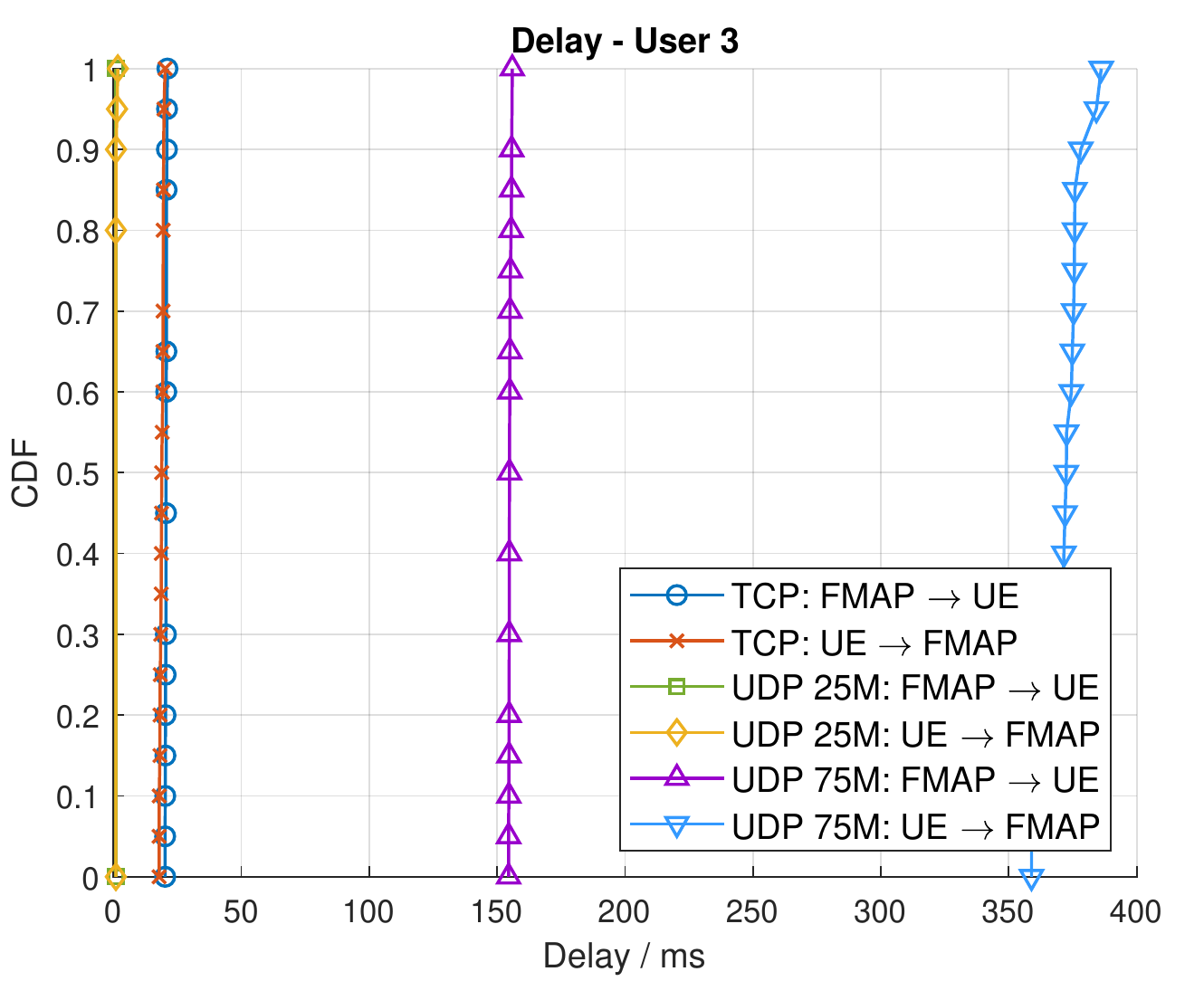}
        \label{NetplanAlgorithmEvaluation-Figure: Concentrated simulation results delay without NetPlan}
    }
    \hfill
    \subfloat[PLR CDF (without NetPlan).] {
        \includegraphics[width=0.3\linewidth]{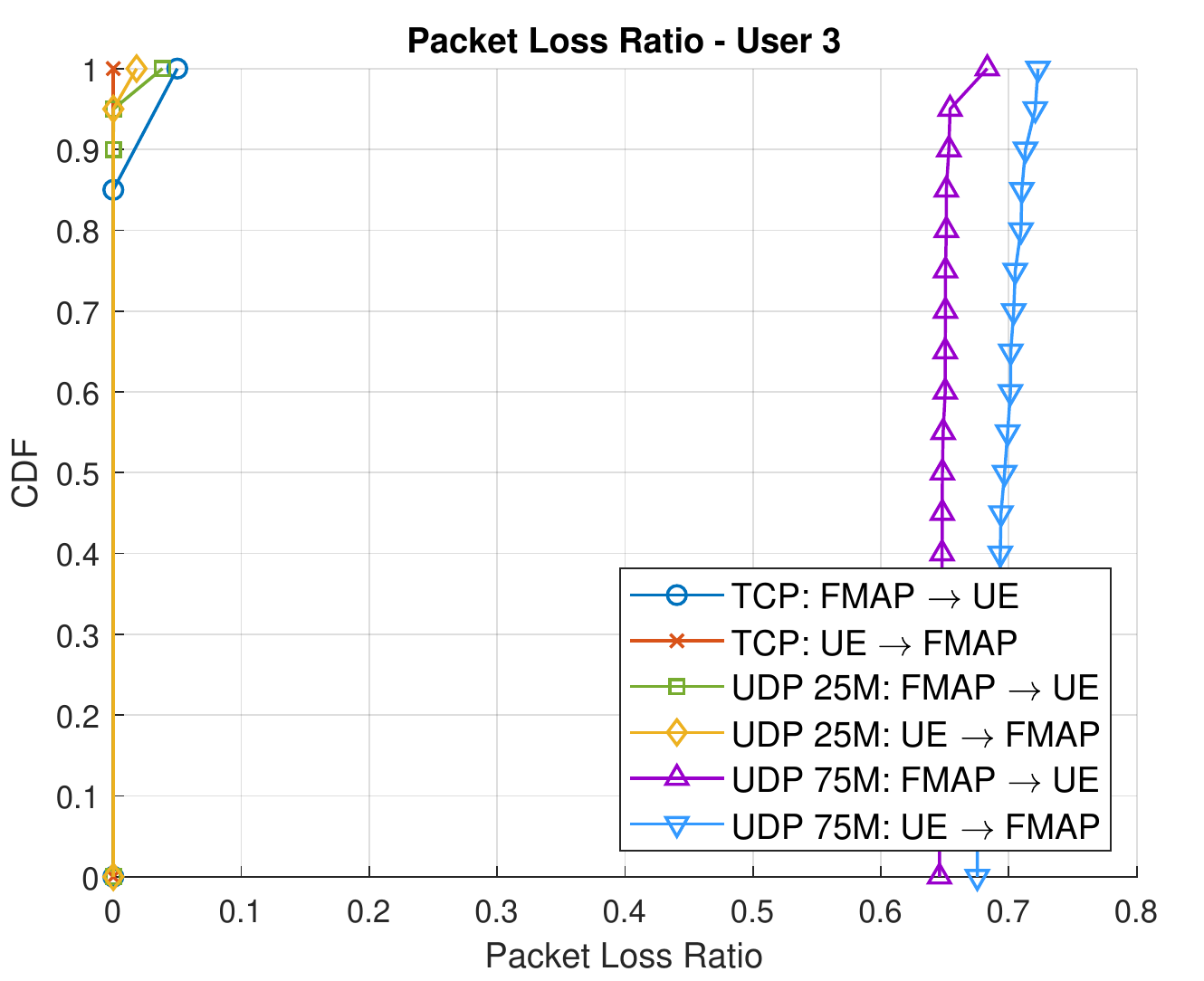}
        \label{NetplanAlgorithmEvaluation-Figure: Concentrated simulation results PLR without NetPlan}
    }
    %%%
    \vfil
    %%%
    \subfloat[Throughput CDF (with NetPlan).] {
        \includegraphics[width=0.3\linewidth]{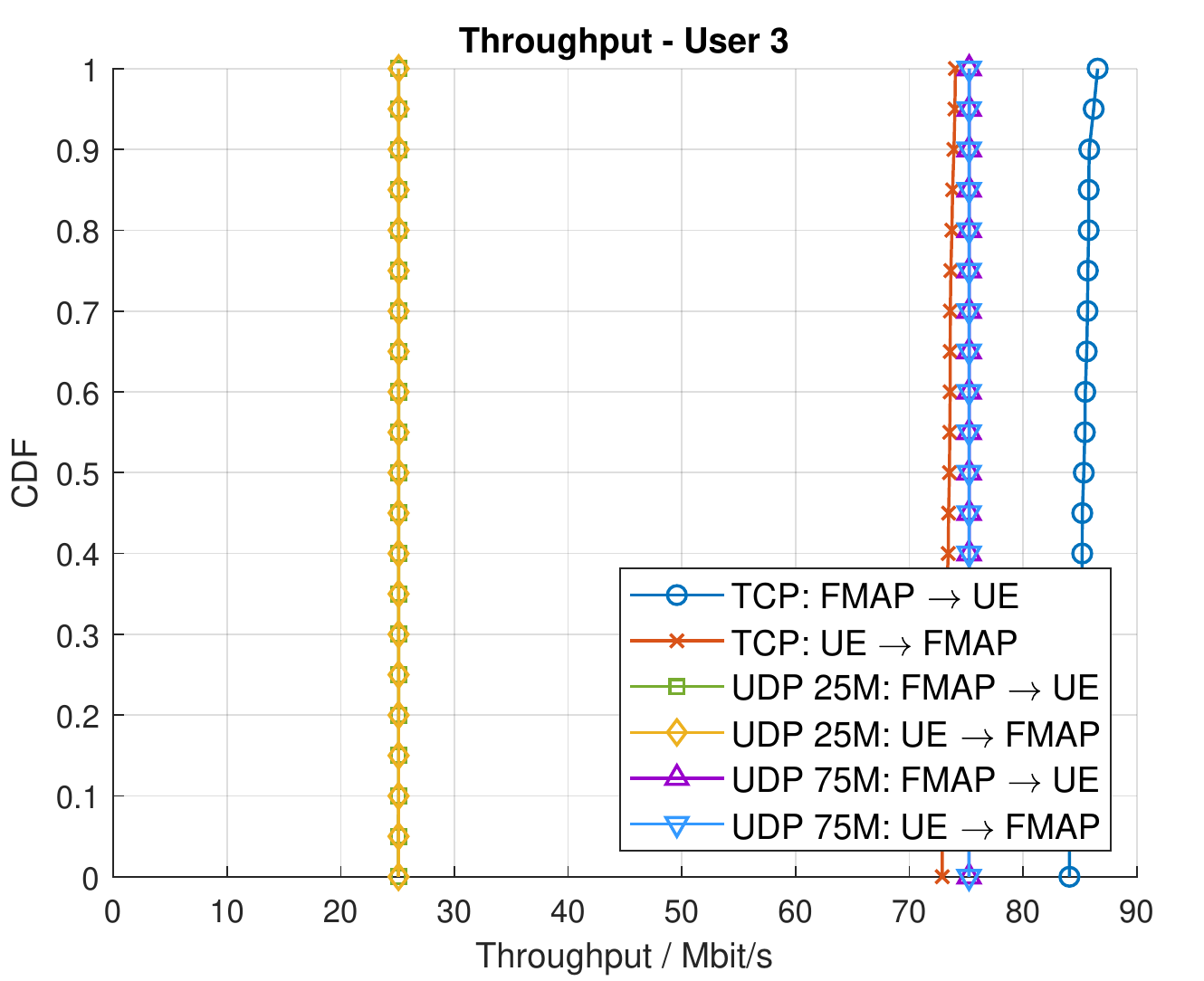}
        \label{NetplanAlgorithmEvaluation-Figure: Concentrated simulation results throughput with NetPlan}
    }
    \hfill
    \subfloat[Delay CDF (with NetPlan).] {
        \includegraphics[width=0.3\linewidth]{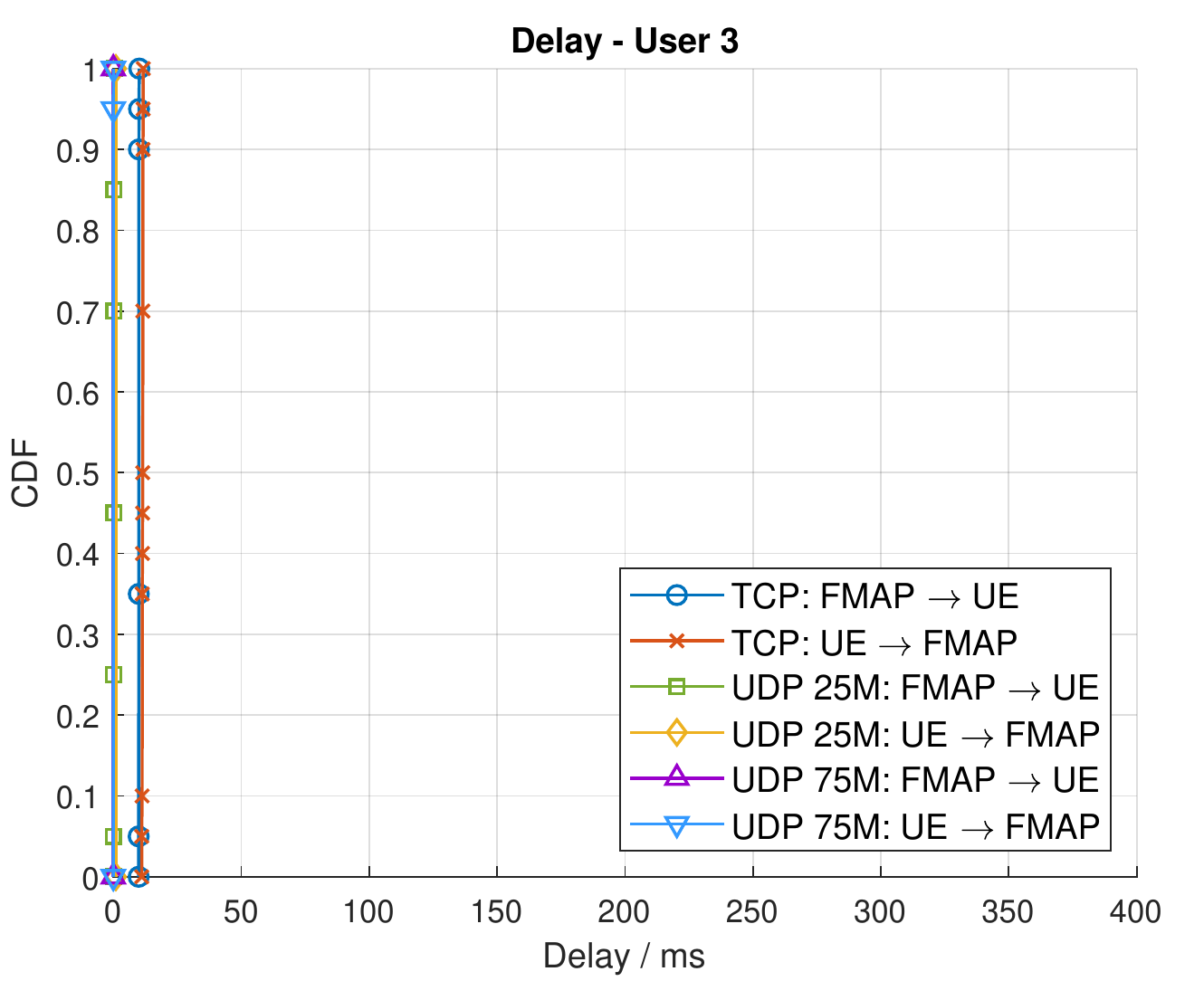}
        \label{NetplanAlgorithmEvaluation-Figure: Concentrated simulation results delay with NetPlan}
    }
    \hfill
    \subfloat[PLR CDF (with NetPlan).] {
        \includegraphics[width=0.3\linewidth]{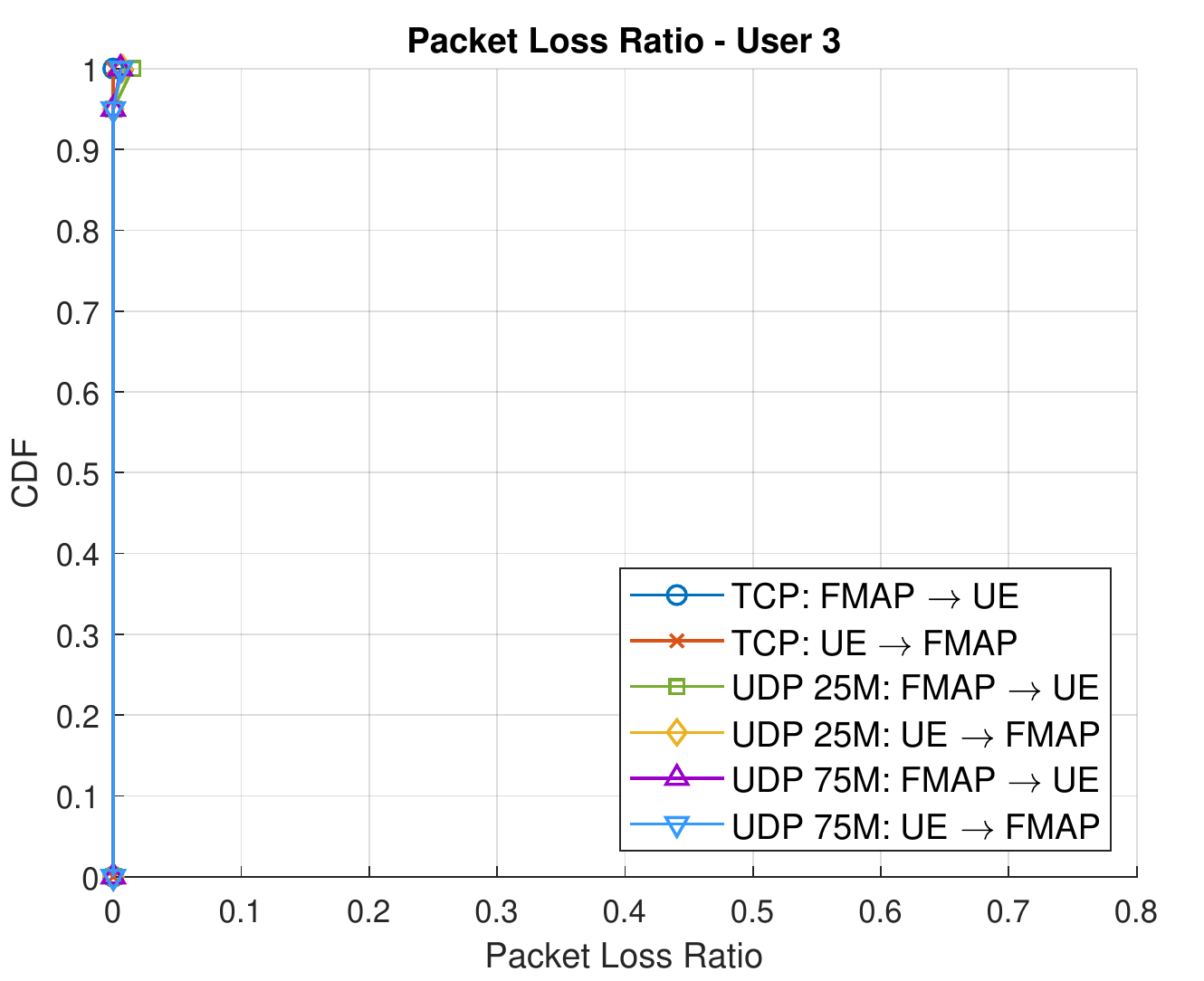}
        \label{NetplanAlgorithmEvaluation-Figure: Concentrated simulation results PLR with NetPlan}
    }

    %%% CAPTION FOR THE WHOLE FIGURE
    \caption{Network performance of UE 3 in the concentrated traffic demand scenario, obtained by means of ns-3 simulation. The first row represents the results without the NetPlan algorithm, whereas the second row represents the results with the NetPlan algorithm.}
    \label{NetplanAlgorithmEvaluation-Figure: Concentrated simulation results}
\end{figure*}
%%%%%%%%%%%%%%%%%%

The Matlab simulations for the concentrated traffic demand scenario are illustrated in \cref{NetplanAlgorithmEvaluation-Figure: Matlab simulations concentrated}. The images show the UE positions, their offered traffic and the final FMAP positions determined by the NetPlan algorithm for the three offered traffic types analyzed. The baseline for this scenario, in which the NetPlan algorithm is not used, consists in evenly distributing the FMAPs throughout the coverage area adopting the same positions as in \cref{NetplanAlgorithmEvaluation-Figure: Concentrated traffic demand scenario without NetPlan}. Since the UEs are concentrated in a small area, the FMAPs are also concentrated around that area, without compromising the overall coverage of the area.

In order to analyze the network performance gains in the concentrated traffic demand scenario due to the NetPlan algorithm, UE 3 is analyzed since this is the UE that benefits the most with the FMAPs' repositioning. \cref{NetplanAlgorithmEvaluation-Figure: Concentrated simulation results} presents the network performance results of UE 3 with and without the NetPlan algorithm. In order to analyze the network performance gains, the TCP flow is analyzed. As previously explained, when the NetPlan algorithm is not used, UE 2 and UE 3 are both associated to FMAP C. Hence, the capacity of FMAP C has to be shared among both UEs. However, when the NetPlan algorithm is used, FMAP A is repositioned closer to UE 3, so that this UE can associate to this FMAP. As a result, since both FMAPs now only have one UE, each UE can take advantage of the full FMAP channel capacity.

The analysis and comparison of the median values of the QoS metrics obtained with and without the NetPlan algorithm demonstrate that the theoretical expectations are valid. In fact, for the TCP flow the median throughput of UE 3 increased to $\approx$~2x when the TMFN was positioned according to the NetPlan algorithm, compared to the baseline of not using the NetPlan algorithm. Furthermore, the median delay reduced to $\approx$~0.4x when not using the NetPlan algorithm. The PLR was 0\% for both cases, since TCP guarantees reliable packet delivery.

Similar to the homogeneous traffic demand scenario, the UDP 25M flow does not demonstrate performance gains, since the channel still has capacity to transport the traffic being offered by the three UEs. In both cases, UE 3 achieves a throughput of $\approx$~\SI{25}{Mbit/s}, with a delay of $\approx$~\SI{0}{ms} and a PLR of $\approx$~0\%. Moreover, the downlink and uplink flows have similar results.

The UDP 75M traffic flow saturates the channel when the NetPlan algorithm was not used, but does not saturate when the NetPlan algorithm is used, since each UE was associated to a different FMAP. For the downlink flow (FMAP to UE), the median throughput increased to $\approx$~1.7x compared to the baseline of not using the NetPlan algorithm, the delay reduced from $\approx$~\SI{150}{ms} to $\approx$~\SI{0}{ms} and the PLR reduced from $\approx$~65\% to $\approx$~0\%. For the uplink flow (UE to FMAP), the median throughput increased to $\approx$~1.7x compared to the baseline of not using the NetPlan algorithm, the delay reduced from $\approx$~\SI{350}{ms} to $\approx$~\SI{0}{ms} and the PLR reduced from $\approx$~70\% to $\approx$~0\%. In order to control the delay of the UDP traffic flows, active queue management applied to the MAC queues could be used at the expense of an increased PLR.

Similar to the TCP flow, the performance gains are explained by the repositioning of FMAP A. Moreover, as analyzed in \cref{NetplanAlgorithmEvaluation-Section: Experimental channel model}, the channel is modeled by the Friis propagation loss model with a Rician fast-fading characterized by the Rician K-Factor, which is the ratio between the Rx power of the LoS component and the sum of the Rx power of the Non Line-of-Sight (NLoS) components. Since the downlink Rician K-factor (\SI{40}{dB}) is much higher than the uplink Rician K-factor (\SI{13}{dB}), the downlink channel benefits from better propagation conditions. In this sense, when the FMAP is transmitting to the UE, there is a higher probability of using higher physical data rates, thus improving the network performance.

\subsubsection{Experimental Results}

%%%%% FIGURE %%%%%
\begin{figure*}
    \centering

    \subfloat[Throughput CDF (without NetPlan).] {
        \includegraphics[width=0.48\linewidth]{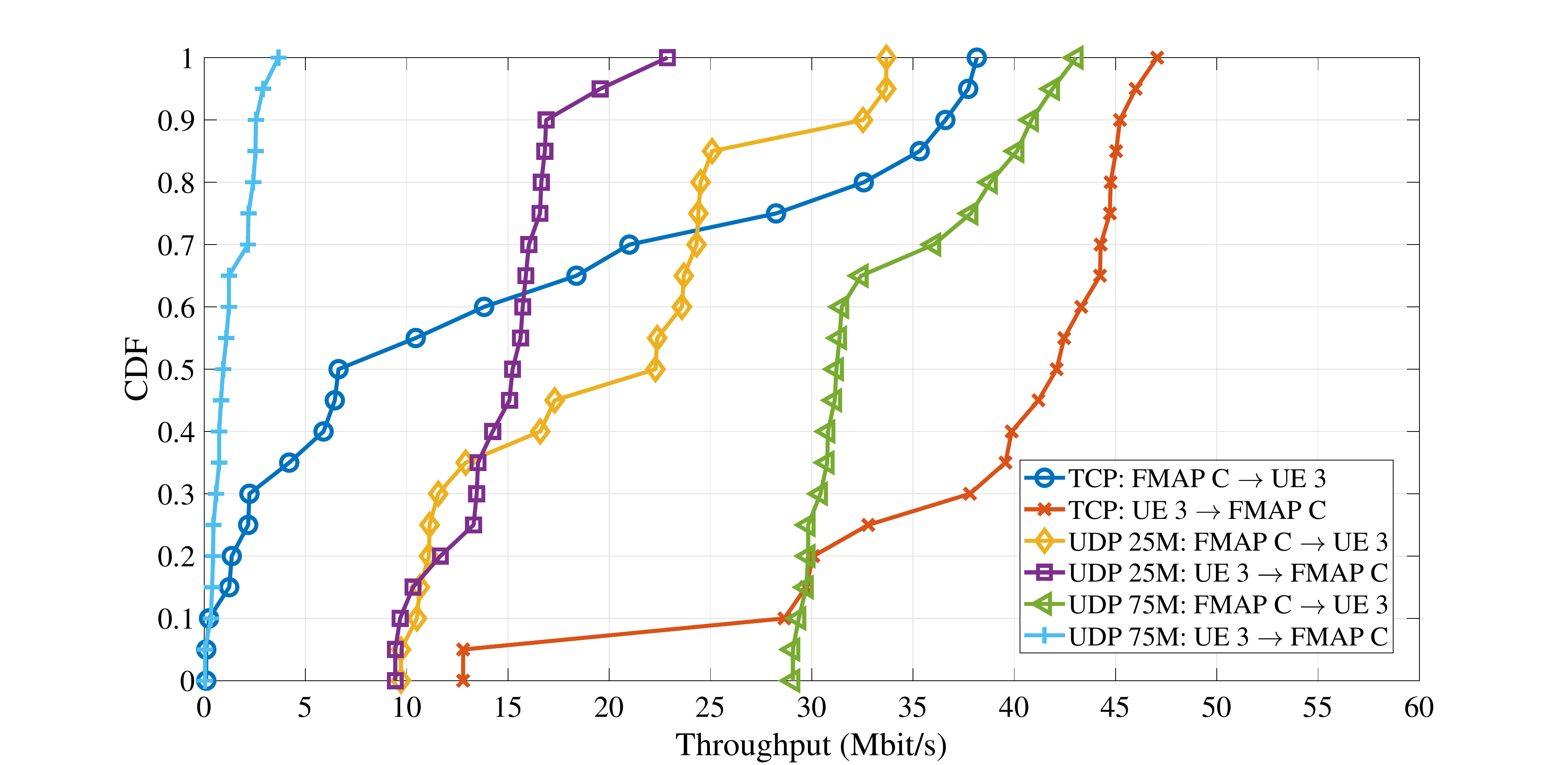}
        \label{NetplanAlgorithmEvaluation-Figure: Traffic results concentrated throughput without NetPlan}
    }
    \hfill
    \subfloat[Physical data rate histogram (without NetPlan).] {
        \includegraphics[width=0.47\linewidth]{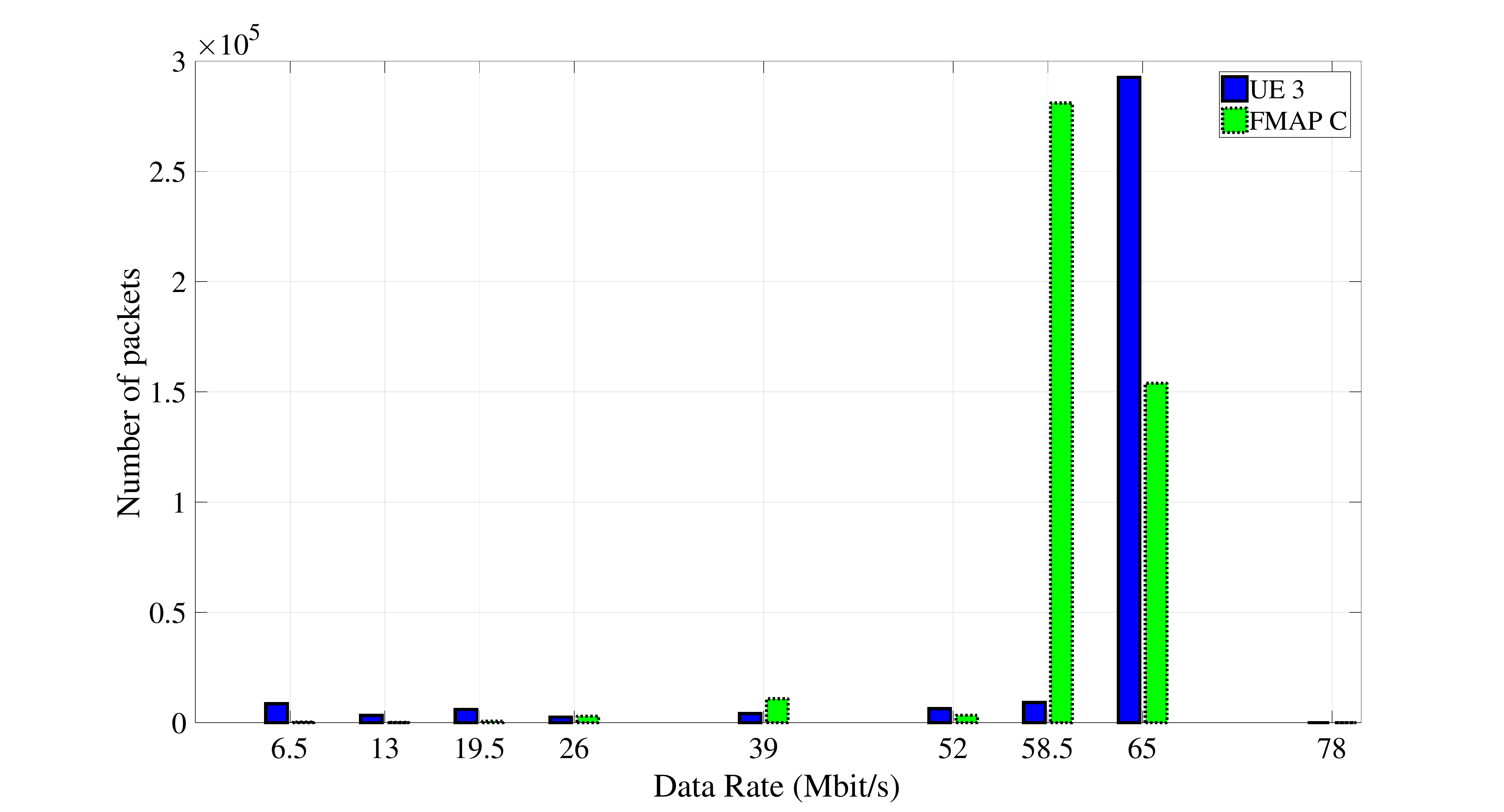}
        \label{NetplanAlgorithmEvaluation-Figure: Traffic results concentrated data rates without NetPlan}
    }
    %%%
    \vfil
    %%%
    \subfloat[Throughput CDF (with NetPlan).] {
        \includegraphics[width=0.48\linewidth]{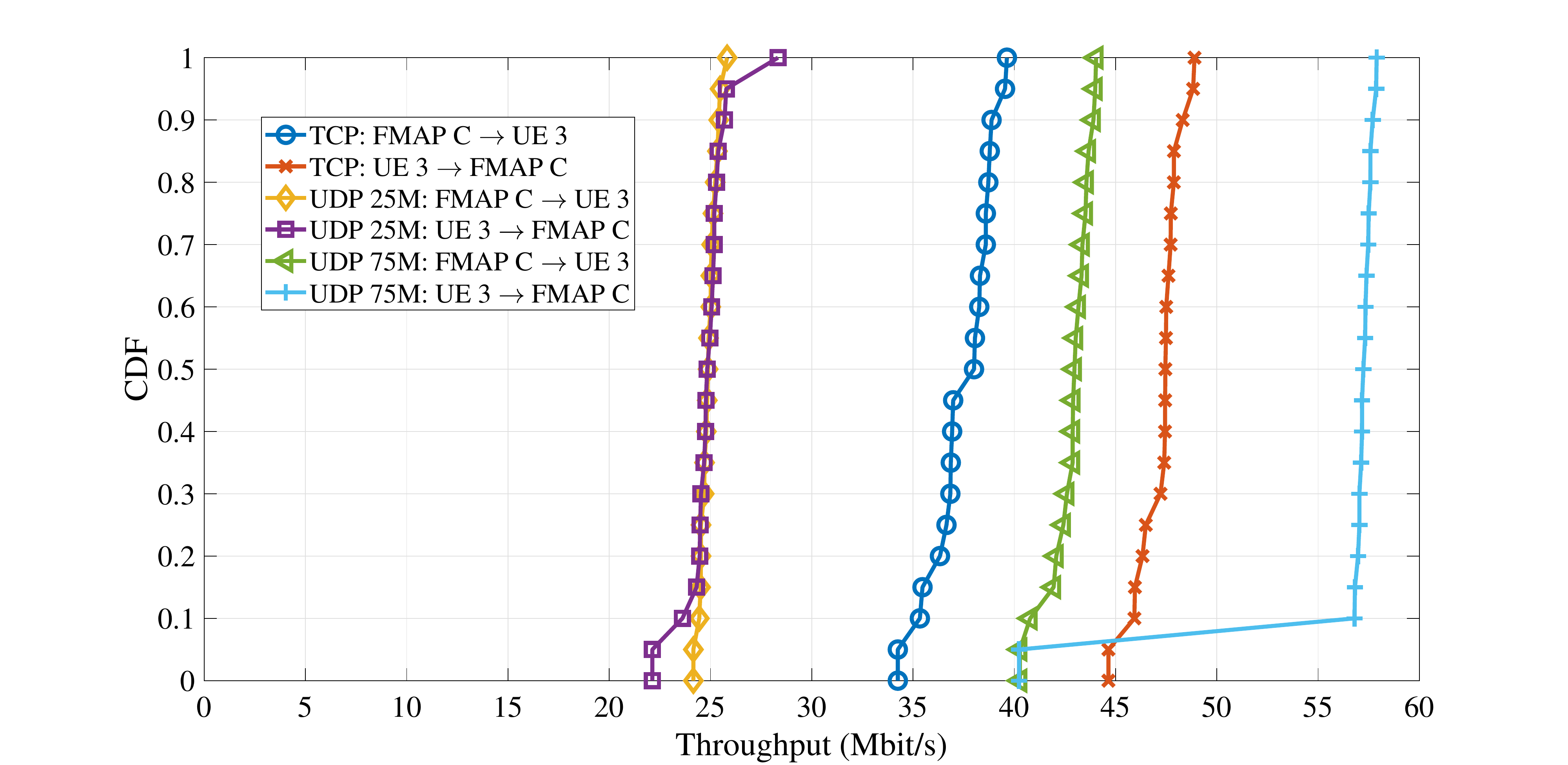}
        \label{NetplanAlgorithmEvaluation-Figure: Traffic results concentrated throughput with NetPlan}
    }
    \hfill
    \subfloat[Physical data rate histogram (with NetPlan).] {
        \includegraphics[width=0.47\linewidth]{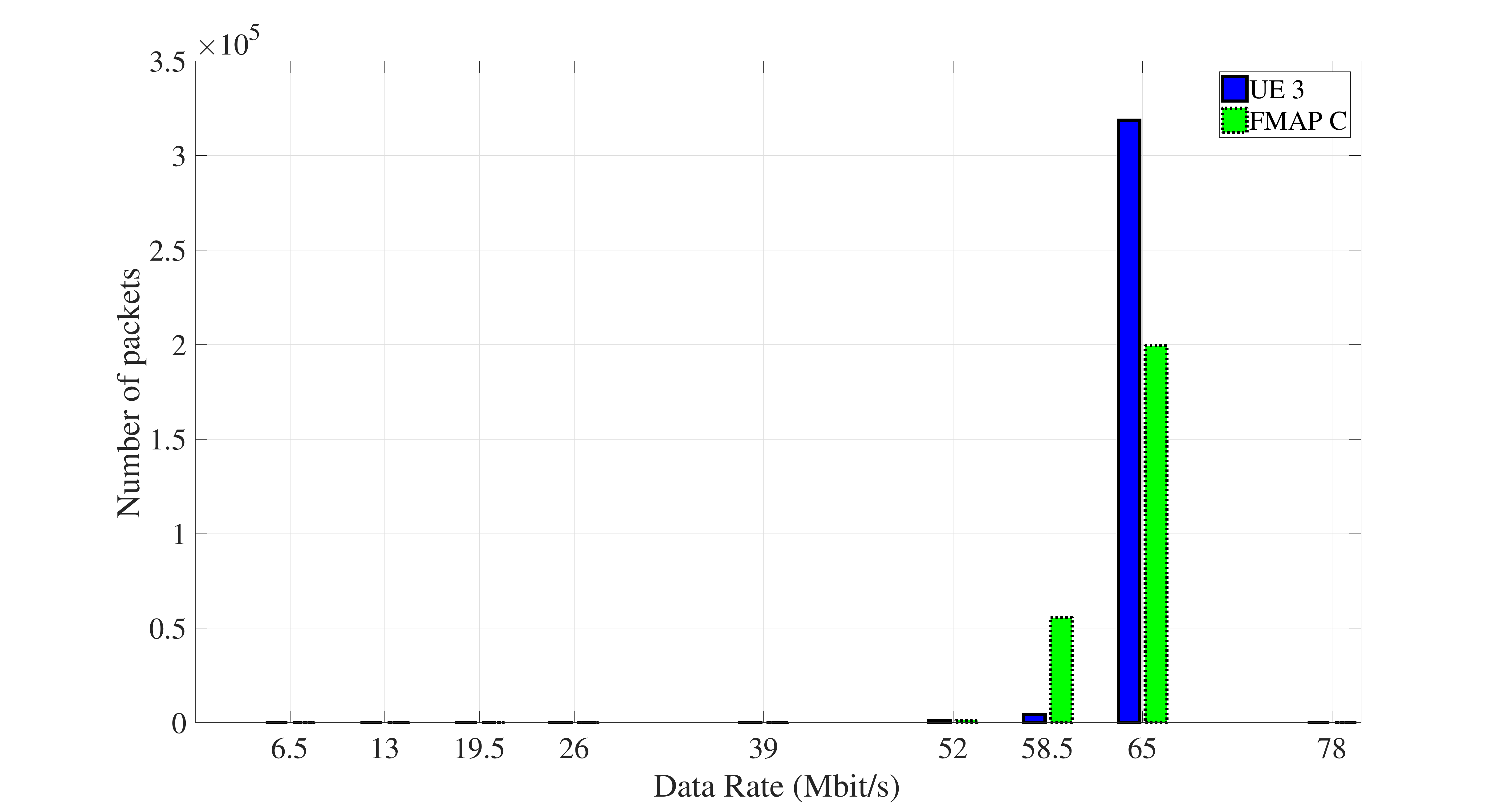}
        \label{NetplanAlgorithmEvaluation-Figure: Traffic results concentrated data rates with NetPlan}
    }

    %%% CAPTION FOR THE WHOLE FIGURE
    \caption{Experimental traffic results of FMAP C for the concentrated traffic demand scenario with and without the NetPlan algorithm. The results include the throughput CDFs and the histogram of the physical data rates of the packets generated by both the UE 3 (solid bar) and FMAP C (dotted bar) during the experiments.}
    \label{NetplanAlgorithmEvaluation-Figure: Traffic results concentrated}

\end{figure*}
%%%%%%%%%%%%%%%%%%

%%%%% FIGURE %%%%%
\begin{figure*}
    \centering

    \subfloat[UDP \SI{25}{Mbit/s} constant bitrate flow.] {
        \includegraphics[width=0.47\linewidth]{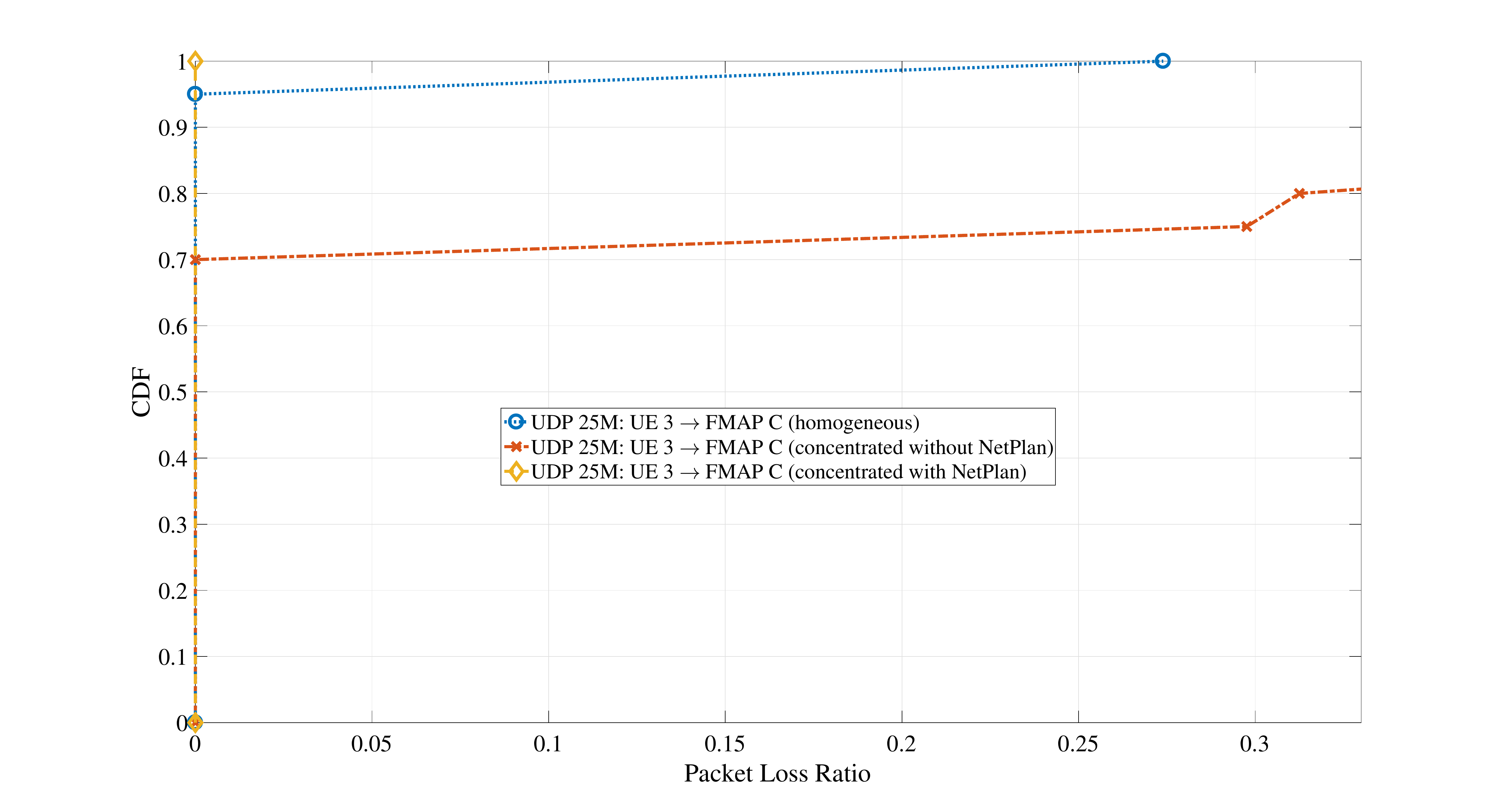}
        %\label{}
    }
    \hfill
    \subfloat[UDP \SI{75}{Mbit/s} constant bitrate flow.] {
        \includegraphics[width=0.47\linewidth]{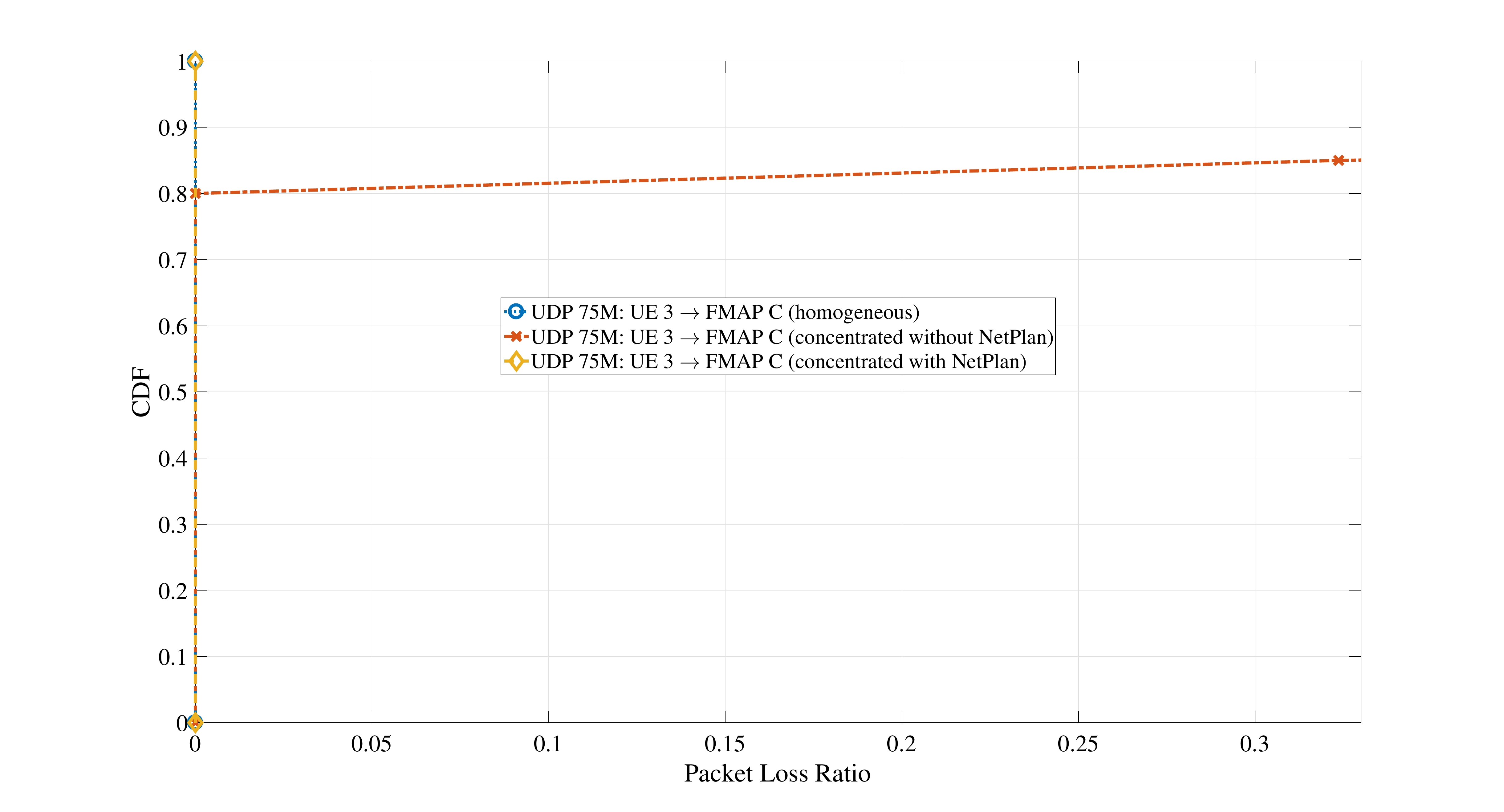}
        %\label{}
    }

    \caption{Packet Loss Ratio (PLR) CDF in FMAP C for different traffic demands, considering the homogeneous traffic demand and the concentrated traffic demand (with and without the NetPlan algorithm).}
    \label{NetplanAlgorithmEvaluation-Figure: Experimental PLR}
\end{figure*}
%%%%%%%%%%%%%%%%%%

%%%%% FIGURE %%%%%
\begin{figure}
    \centering
    \includegraphics[width=1\linewidth]{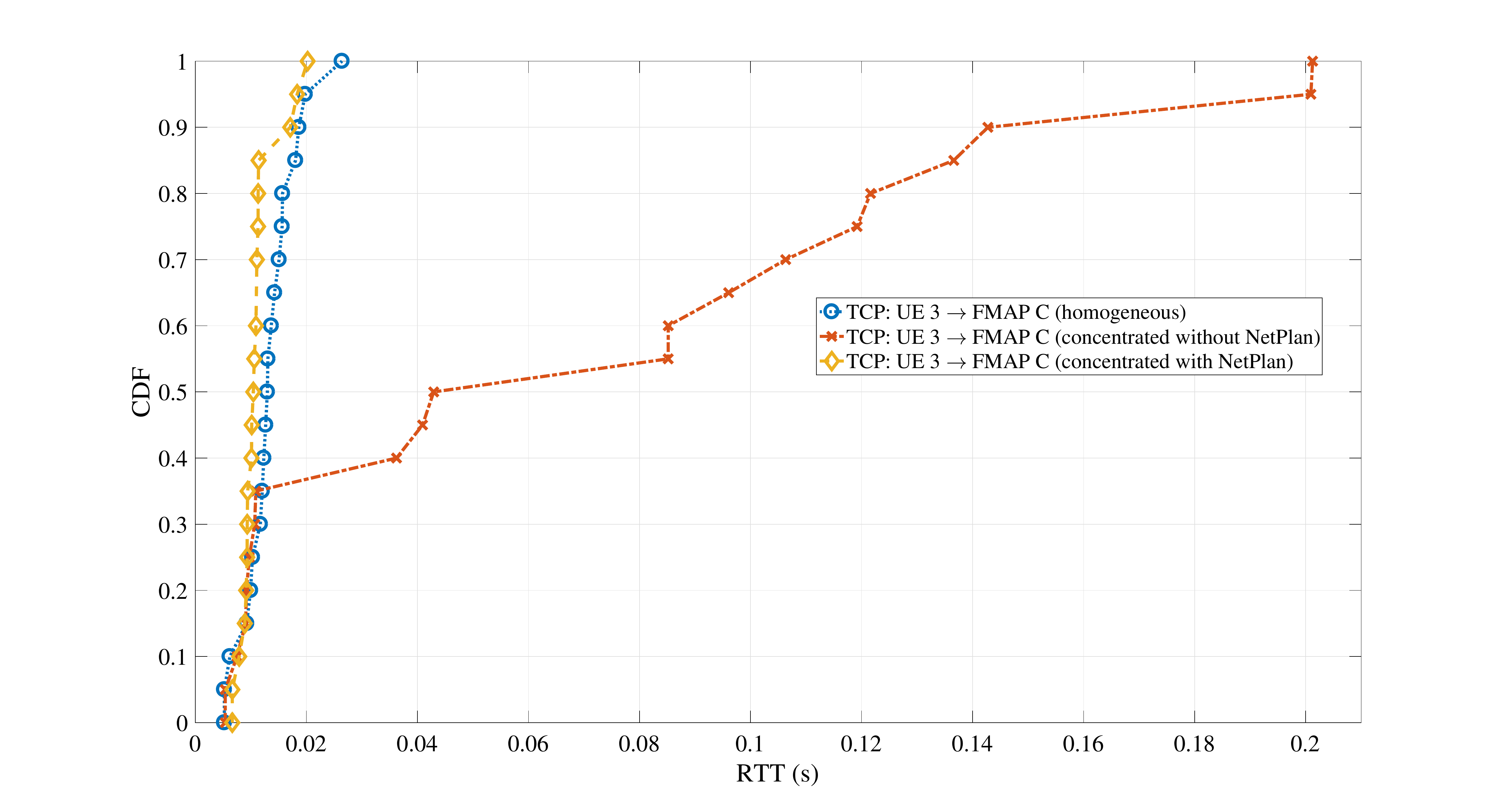}
    \caption{Round Trip Time (RTT) CDF in FMAP C considering the homogeneous traffic demand and the concentrated traffic demand (with and without the NetPlan algorithm).}
    \label{NetplanAlgorithmEvaluation-Figure: Experimental RTT}
\end{figure}
%%%%%%%%%%%%%%%%%%

In this scenario, two UEs were generating traffic in the same area. This scenario motivates the usage of the NetPlan algorithm, in order to position the FMAPs according to the traffic demand, in order to improve the network performance. When the NetPlan algorithm was not employed, the two UEs generating traffic in the same area were sharing the same Wi-Fi channel. This resulted in one of the FMAPs being overloaded with two UEs, whereas the second FMAP did not have any user associated. Hence, the UEs could not take full advantage of the aggregate network capacity provided by the two FMAPs, resulting in a degraded network performance. When the decisions of the NetPlan algorithm were considered, the FMAPs were positioned in order to meet the traffic demand of the UEs. In this case, each UE was able to take advantage of the full channel capacity provided.

In the concentrated traffic demand scenario without the NetPlan algorithm, two UEs inside area C (\cref{NetplanAlgorithmEvaluation-Figure: Concentrated traffic demand scenario without NetPlan}) were associated to FMAP C simultaneously. On the other hand, in the concentrated traffic demand scenario with the NetPlan algorithm (\cref{NetplanAlgorithmEvaluation-Figure: Concentrated traffic demand scenario with NetPlan}), a single UE was associated to FMAP C, since the other UE, previously associated to FMAP C, became client of FMAP A, which has been moved closer to area C, in order to enhance the network capacity and meet the traffic demand of the UEs inside this area. The performance results for both scenarios (with and without NetPlan) are depicted in \cref{NetplanAlgorithmEvaluation-Figure: Traffic results concentrated}.

The results with the NetPlan algorithm significantly outperform those when the NetPlan algorithm is not used. By comparing the 25th, 50th, and 75th percentiles of the throughput CDFs (\cref{NetplanAlgorithmEvaluation-Figure: Traffic results concentrated throughput without NetPlan,NetplanAlgorithmEvaluation-Figure: Traffic results concentrated throughput with NetPlan}), it can be concluded that the NetPlan algorithm allows a throughput improvement up to 1.84x, 1.63x, and 1.52x, respectively. Nevertheless, the link asymmetry is once again observed, which is justified by the same reasons pointed out in the homogeneous scenario. In \cref{NetplanAlgorithmEvaluation-Figure: Traffic results concentrated throughput without NetPlan}, the lower throughput values of some flows, including the UDP 75M flow between the UE and FMAP C, is justified by the simultaneous exchange of traffic in both directions, with both traffic flows competing to access the Wi-Fi medium.

Regarding the RTT, which is depicted in \cref{NetplanAlgorithmEvaluation-Figure: Experimental RTT}, the results for the concentrated traffic demand scenario denote the superior performance of the NetPlan algorithm (yellow diamond) compared to the case where the NetPlan is not employed (orange cross). This can be concluded by observing the 50th and 75th percentiles of the RTT CDFs, for which the RTT is 0.24x and 0.1x lower, respectively. The higher RTT when the NetPlan algorithm is not employed denotes packets being held longer in the transmission queue and packet retransmissions due to link congestion. The same behavior is observed for PLR, which is represented in \cref{NetplanAlgorithmEvaluation-Figure: Experimental PLR}, considering UDP flows with different bitrates, for which there is no packet loss. Overall, the network performance when the NetPlan algorithm is used in a concentrated area is similar to the one obtained for the baseline scenario, which considers homogeneous traffic demand in the different network areas; this is the main contribution of the proposed solution.

%%%%%%%%%%%%%%%%%%%%%%%%%%%%%%%%%%%%%%%%%%%%%%%%%%%%%%%%%%%%%%%%%
% REDEFINE ROUTING SOLUTION
%%%%%%%%%%%%%%%%%%%%%%%%%%%%%%%%%%%%%%%%%%%%%%%%%%%%%%%%%%%%%%%%%
\section{RedeFINE Routing Solution} \label{RedeFINE-Section}

RedeFINE, which was initially proposed in~\cite{coelho2018redefine, coelho2019routing}, is presented in this section.

\subsection{Problem Formulation} \label{RedeFINE-Section: Problem Formulation}

In the following, we formulate the problem addressed by RedeFINE. At a time instant $t_k = k \cdot \Delta t, k \in N_0$ and $\Delta t \in \mathbb{R}$, which is defined according to the TMFN update period imposed by the NetPlan algorithm, the TMFN is represented by a graph $G(t_k)=(V, E(t_k), w(t_k))$, where $V=\{0, ..., N-1\}$ represents the set of UAVs forming the TMFN, $E(t_k) \subseteq V \times V$ represents the communications links, and $w(t_k)$ represents the cost assigned to the communications links of $G$. $(i,j)_{t_k} \in E(t_k)$ represents the directional communications link from\ $i$ to $j$ available at $t_k$, where $i,j \in V$, and $w_{i,j}(t_k)$ represents the cost of link $(i,j)_{t_k}$. $X_i(t_k)$ represents the position of UAV $i$ at time $t_k$ and depends on its initial position at instant $t_0$ and its associated velocity and acceleration vectors defined by the CS that controls the UAV positions.
The availability of the wireless links connecting the UAVs changes along the time. Hence, the directional wireless communications link $(i,j)_{t_k}$ exists if and only if the power $P_{R_i}(t_k)$ received by UAV $i$ at time $t_k$ divided by the noise power $N_i$ satisfies \cref{RedeFINE-Equation: SNR Threshold}, that is, if the Signal-to-Noise Ratio (SNR) is higher than a threshold $S$. The received power at UAV $i$, $P_{R_i}(t_k)$, results from the Free-space path loss model defined in \cref{RedeFINE-Equation: Friis propagation model}, where $P_{T_j}(t_k)$ describes the power transmitted by UAV $j$ at time $t_k$, $c$ represents the speed of light in vacuum, $f_{i,j}$ denotes the carrier frequency, and $d_{i,j}(t_k)$ expresses the Euclidean distance between UAV $i$ and UAV $j$ at time $t_k$.

\begin{equation}
    \frac{P_{R_i}(t_k)}{N_i} > S
    \label{RedeFINE-Equation: SNR Threshold}
\end{equation}

\begin{equation}
    \frac{P_{R_i}(t_k)}{P_{T_j}(t_k)}= \left[ \frac{c}{4\pi \times d_{i,j}(t_k) \times f_{i,j}} \right]^2
    \label{RedeFINE-Equation: Friis propagation model}
\end{equation}

We define a path as a set of adjacent links connecting UAV $i$ to the GW. Multiple paths may be available for UAV $i$ at time $t_k$, but only one of them is used. We also define $C_{i,j}(t_k)$ as the maximum capacity, in bit/s, of the communications link available between UAV $i$ and UAV $j$ at time $t_k$, considering a constant value for the link bandwidth $B_{i,j}$ in \SI{}{\hertz}; Shannon-Hartley theorem is used for this purpose, as given by~\cref{RedeFINE-Equation: Shannon-Hartley theorem}.

\begin{equation}
    C_{i,j}(t_k) = B_{i,j} \times \log_{2} \left[ 1+\frac{P_{R_i}(t_k)}{N_i} \right]
    \label{RedeFINE-Equation: Shannon-Hartley theorem}
\end{equation}

Considering the throughput $R_i(t_k)$, in bit/s, as the bitrate of the flow from UAV $i$ received at the GW at time $t_k$, and $N$ UAVs generating traffic towards the GW, we aim at maximizing at any time instant $t_k$ the amount of bits received by the GW during time interval $\Delta t$. As such, our objective function can be defined as:

\begin{equation}
    \text{maximize} \qquad R(t_k) = \sum_{i=0}^{N-1}R_i(t_k)
    \label{RedeFINE-Equation: Objective function}
\end{equation}

The factors influencing $R_i(t_k)$ include the capacity of the path used by UAV $i$, which should be limited by the link in the path having the smallest capacity, the number of flows traversing the links, medium access protocol behaviour, and interference between the communications nodes. This is a complex optimization problem since the last two factors are not easily characterized.

In order to solve this problem we will attempt to find a path for each UAV $i$ for each time instant $t_k$, so that we meet \cref{RedeFINE-Equation: Objective function}.

\subsection{Concept} \label{RedeFINE-Section: Concept}

RedeFINE was designed to take advantage of the centralized view of the TMFN available at the CS. By considering a) the future positions of the FMAPs composing the TMFN, which are defined by the NetPlan algorithm to fulfill the traffic demand of the UEs on the ground, and b) the velocity of the FMAPs following a straight line between source and destination, RedeFINE selects periodically the best path for each FMAP. We will define the best path as the path that has the smallest cost, according to metrics that will be discussed later on.

RedeFINE assumes a strong LoS component between the nodes, which is characteristic of the links between UAVs flying dozens of meters above the ground. To find the shortest path between UAVs, RedeFINE employs the Dijkstra's algorithm~\cite{walker1992implementing}.

\subsection{Analysis of a Reference Case} \label{RedeFINE-Section: Analysis of reference case}

\begin{figure}[t]
	\centering
	\includegraphics[width=1\linewidth]{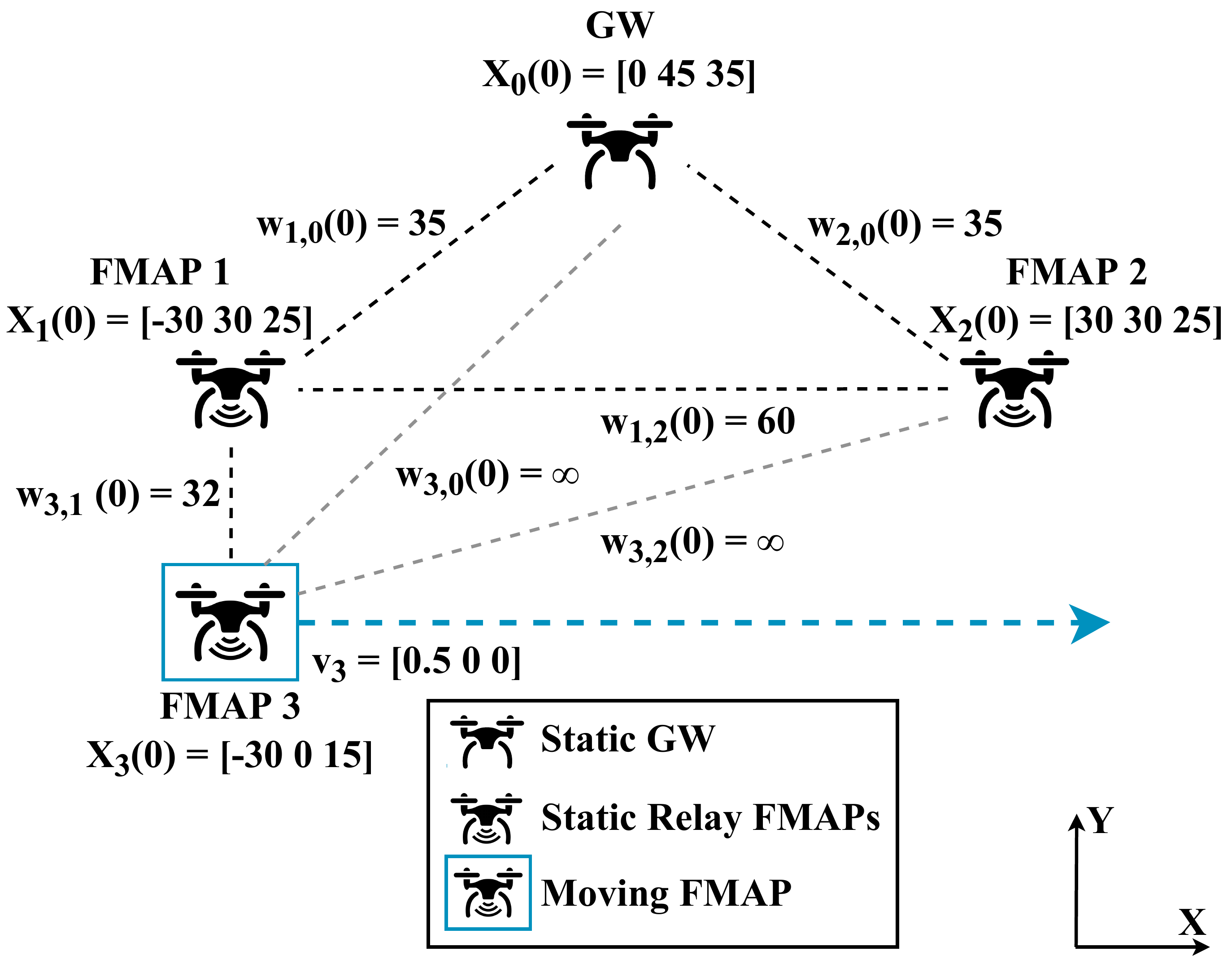}
	\caption{Scenario used for the theoretical validation of RedeFINE. For the initial position of FMAP 3, the wireless links between FMAP 3 and both FMAP 2 and GW are not available, due to the SNR threshold constraint.}
	\label{RedeFINE-Figure: Conceptual Scenario}
\end{figure}

\begin{figure}[t]
	\centering
	\includegraphics[width=1\linewidth]{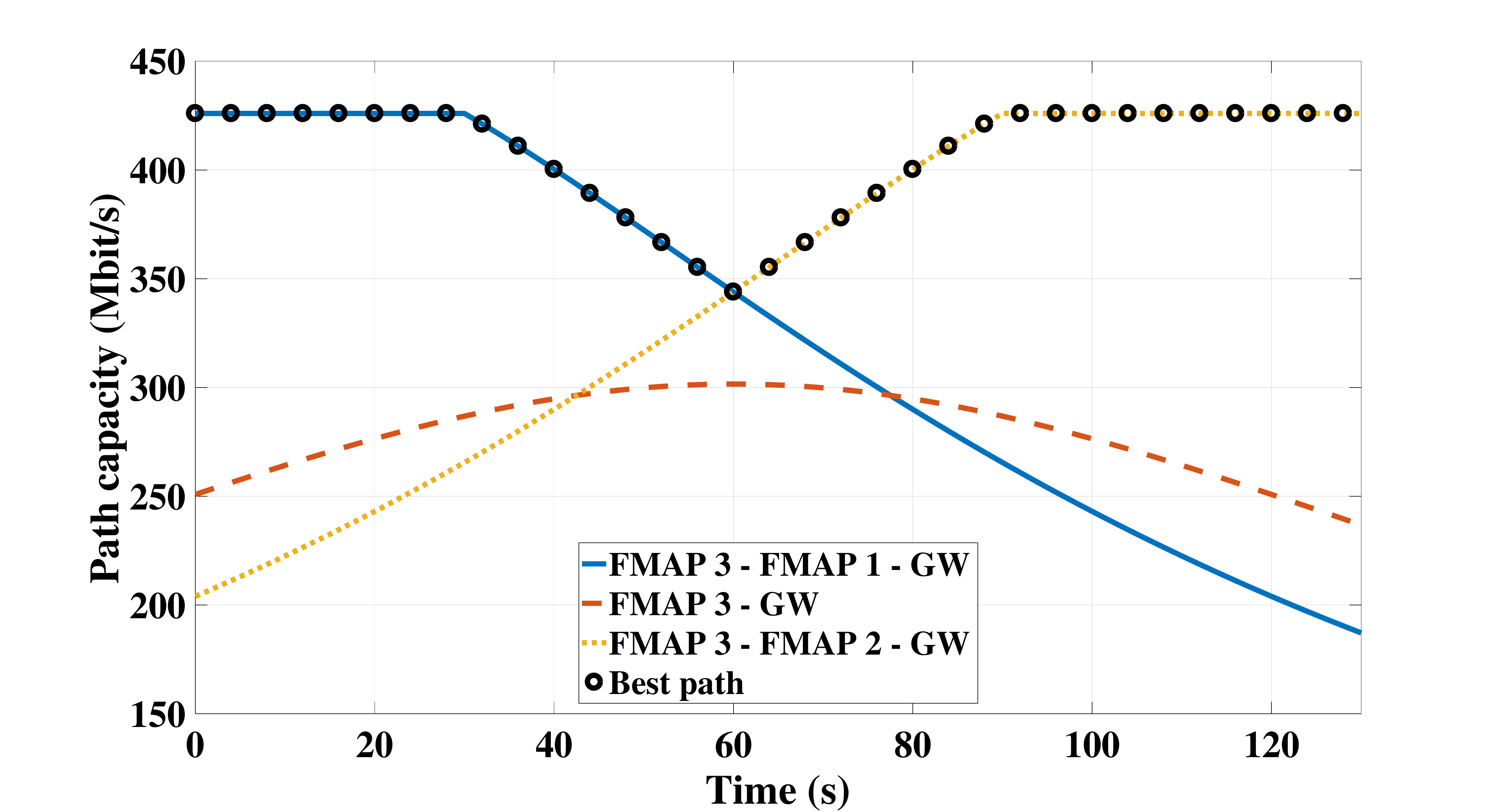}
	\caption{Theoretical capacity values for the paths between FMAP 3 and the GW. In order to reach the GW, FMAP 3 uses as relay nodes FMAP 1 until $t=\SI{60}{\second}$ and FMAP 2 from $t=\SI{60}{\second}$ to $t=\SI{130}{\second}$. The path with the highest capacity at each instant is highlighted by the circle symbol.}
	\label{RedeFINE-Figure: RedeFINE concept}
\end{figure}

In order to demonstrate the RedeFINE concept and perform a preliminary evaluation, a theoretical validation is made herein. For the sake of simplicity, we consider the scenario depicted in~\cref{RedeFINE-Figure: Conceptual Scenario}. It is formed by: 1) a static GW; 2) two static relay FMAPs (FMAP 1 and FMAP 2); and 3) a moving FMAP (FMAP 3). This scenario aims at illustrating a TMFN reconfiguration that causes link disruptions. In particular, FMAP 3 follows a straight line at \SI{0.5}{m/s} in the direction from FMAP 1 to FMAP 2, in order to reach the new location defined by the NetPlan algorithm. FMAP 1 and FMAP 2 remain in the same positions and they are able to forward the traffic received from FMAP 3 towards the GW, while simultaneously providing connectivity to users on the ground.

The initial position of FMAP $i$ is represented by $X_i(0)$ and the constant velocity is denoted by $v_i$. The initial cost of the wireless link between FMAP $i$ and FMAP $j$ is represented by $w_{i,j}(0)$, which for demonstration purposes is the linear Euclidean distance between their initial positions. The wireless link availability is restricted by a \SI{5}{dB} SNR, that is $S \approx 3.16$ threshold (cf.~\cref{RedeFINE-Equation: SNR Threshold}). The SNR is derived from the Free-space path loss model (cf.~\cref{RedeFINE-Equation: Friis propagation model}). We assume that the maximum capacity for the wireless links is given by the Shannon-Hartley theorem (cf.~\cref{RedeFINE-Equation: Shannon-Hartley theorem}). The maximum capacity of a wireless link is computed considering an average noise power equal to \SI{-85}{dBm}.
In turn, the capacity of a path is restricted by the link with lower capacity among the set of links forming the path. Additionally, we consider the transmission power equal to \SI{0}{dBm}, the carrier frequency equal to \SI{5250}{MHz}, and the channel bandwidth equal to \SI{160}{MHz}, which are values compatible with the IEEE 802.11ac standard.

Based on these assumptions, the theoretical maximum capacity values for the paths between FMAP 3 and the GW are plotted in \cref{RedeFINE-Figure: RedeFINE concept}.
With RedeFINE, in order to reach the GW, FMAP 3 uses as relay nodes FMAP 1 until $t=\SI{60}{\second}$ and FMAP 2 from $t=\SI{60}{\second}$ to $t=\SI{130}{\second}$. These are the highest-capacity paths, which are highlighted by the circle symbol in \cref{RedeFINE-Figure: RedeFINE concept}.

Considering as baseline the static routing configuration that uses FMAP 1 as relay node for the communications between FMAP 3 and the GW during \SI{130}{s}, RedeFINE allows a gain of $\approx$~20\% regarding the total amount of bits received in the GW. The areas under the curves in \cref{RedeFINE-Figure: RedeFINE concept} give the total amount of bits carried by the respective path. We compute the amount of information received on each case (static and RedeFINE) as $\sum_{k=0}^{130}R_0(t_k)\times\Delta t$, considering $\Delta t=\SI{1}{\second}$. In this analysis, FMAP $1$ and FMAP $2$ do not generate traffic. They only forward traffic received from FMAP $0$.

\subsection{Interference Model} \label{RedeFINE-Section: Interference model}

Taking into account the IEEE 802.11 MAC protocol, for a packet transmission to be successful neither the transmitter nor the receiver should be interfered by other nodes. Hence, the transmissions on links $(i, j)_{t_k}$ and $(k, l)_{t_k}$ are both successful at $t_k$ if and only if both $i$ and $j$ are outside the interference range of $k$ and $l$ at $t_k$. This is expressed by the Transmitter-Receiver Conflict Avoidance (TRCA) interference model~\cite{houaidia2017inter}.

\begin{figure}
    \centering
    \includegraphics[width=0.70\linewidth]{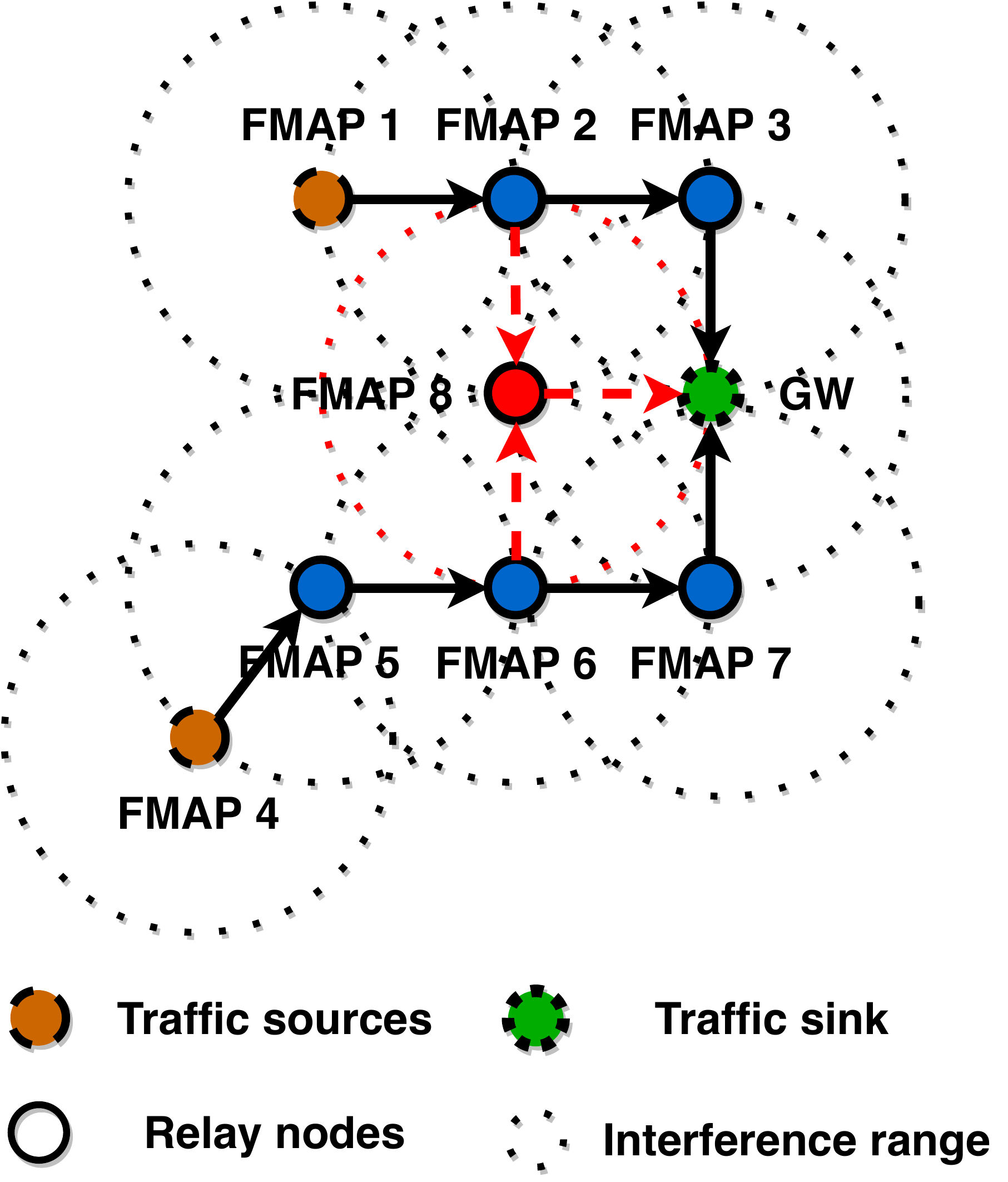}
    \caption{Network graph illustrating the TRCA interference model. If FMAP 8 is used as relay node, the network performance will be reduced up to 50\%, since FMAP 8 is in the interference range of FMAP 2 and FMAP 6.}
    \label{RedeFINE-Figure: Interference model}
\end{figure}

In order to demonstrate how the selection of relay nodes affects the network performance, let us analyze a reference case. For the sake of simplicity, we consider the scenario depicted in~\cref{RedeFINE-Figure: Interference model}. It is formed by: 1) two FMAPs generating traffic -- FMAP 1 and FMAP 4; 2) a GW; and 3) six FMAPs able to forward traffic. The interference range of each node is represented by a dashed circumference around that node. %We consider the interference range equal to the transmission range of each node.
Firstly, we consider two paths for the flows between the FMAPs generating traffic and the GW: $p_1$:$<$FMAP 1, FMAP 2, FMAP 3, GW$>$ and $p_2$:$<$FMAP 4, FMAP 5, FMAP 6, FMAP 8, GW$>$. For these paths, there is no inter-flow interference, excluding the nodes competing for the access to the GW, which is addressed by the MAC protocol. Hence, the throughput achieved by each flow is only limited by the link with the lowest capacity among the ones forming the path. Conversely, if FMAP 8 is chosen to be part of a path $p_1'$:$<$FMAP 1, FMAP 2, FMAP 8, GW$>$, then inter-flow interference will be introduced in the network. For instance, since FMAP 8 is in the interference range of FMAP 6, the links $<$FMAP 2, FMAP 8$>$ and $<$FMAP 5, FMAP 6$>$ become mutually interfered. Therefore, the network performance is reduced up to 50\%, when compared with the previous routing configuration.

This reference case motivates the definition of an inter-flow interference-aware routing approach to improve the performance of RedeFINE. In fact, by using the Euclidean distance as routing metric in the reference scenario depicted in~\cref{RedeFINE-Figure: Interference model}, it becomes indifferent selecting FMAP 7, or FMAP 2 and FMAP 6, respectively, to forward the traffic from FMAP 1 and FMAP 5, since the minimal Euclidean distance is equal for both in this reference case.

\subsection{I2R Routing Metric} \label{RedeFINE-Section: I2R concept}

Motivated by the problem presented in~\cref{RedeFINE-Section: Interference model}, an inter-flow interference-aware routing metric tailored for centralized routing in TMFNs with controllable topology, named I2R, was incorporated into RedeFINE.

I2R consists of two factors: the distance-aware factor and the inter-flow interference-aware factor. Both factors are fed by the centralized view of the TMFN provided by the CS running the NetPlan algorithm, which defines the future locations of the FMAPs that will serve the mobile UEs on the ground. Since a strong LoS component characterizes the wireless links between the UAVs flying dozens of meters above the ground, we use Free-space path loss to model the links between the FMAPs, and estimate the SNR and number of neighboring FMAPs. This holistic knowledge avoids the usage of control packets for neighbor discovery and interference estimation. The distance-aware factor is based on the Euclidean distance of the links between each pair of FMAPs, at time instant $t_k$, which is denoted by $d_{i,j}(t_k)$. As such, this factor includes the sum of the Euclidean distances of the set of links forming a path $p$, considering in advance the future trajectories that FMAPs will follow, which were calculated and pre-defined by the CS to fulfill the traffic demand of the ground users. Using the Euclidean distance as part of the routing metric is compliant with the objective of selecting high-capacity paths, since the link capacity increases as the Euclidean distance decreases, according to the Shannon-Hartley theorem. $d_{i,j} (t_k)$ is normalized to the maximum Euclidean distance among all the usable links of the TMFN, at $t_k$. In turn, the inter-flow interference aware factor is a value $\gamma(t_k)$ that is added to the Euclidean distance of the link between FMAP $i$ and FMAP $j$, at $t_k$. $\gamma_j(t_k)$ represents the number of neighboring nodes of FMAP $j$, excluding FMAP $i$, at $t_k$. We assume as neighboring nodes the FMAPs in carrier-sense range that are generating or forwarding traffic. $\gamma_j(t_k)$ considers that the level of interference is equal either the neighboring nodes are close or far away, as the TRCA model denotes. $\gamma_j(t_k)$ is normalized to the maximum number of neighbors that any FMAP composing the TMFN has at $t_k$.

The path cost using I2R is defined in \cref{RedeFINE-Equation: Routing metric}, where $0 \leq \alpha \leq 1$ is a tunable parameter that weights the influence of the distance-aware and interference-aware factors. To calculate the path between any FMAP and the GW, the Dijkstra's algorithm~\cite{walker1992implementing} is used. Considering the reference case depicted in~\cref{RedeFINE-Figure: Interference model}, I2R uses the factor $\gamma$ to increase the cost of the links $<$FMAP 2, FMAP 8$>$ and $<$FMAP 6, FMAP 8$>$, since FMAP 8 is in the interference range of FMAP 2 and FMAP 6.

\begin{equation}
    I2R = (1-\alpha)\times\sum_{\forall (i,j) \in p}d_{i,j}(t_k) + \alpha \times \sum_{\forall j \in p} \gamma_{j}(t_k)
    \label{RedeFINE-Equation: Routing metric}
\end{equation}

%%%%%%%%%%%%%%%%%%%%%%%%%%%%%%%%%%%%%%%%%%%%%%%%%%%%%%%%%%%%%%%%%
% REDEFINE EVALUATION
%%%%%%%%%%%%%%%%%%%%%%%%%%%%%%%%%%%%%%%%%%%%%%%%%%%%%%%%%%%%%%%%%
\section{RedeFINE Evaluation} \label{RedeFINEEvaluation-Section}

The performance evaluation of RedeFINE is presented in this section, including the simulation setup, the simulation scenarios, and the performance metrics considered.

\subsection{Simulation Setup} \label{RedeFINEEvaluation-Section: Simulation setup}

The ns-3 simulator was used to evaluate RedeFINE in complex networking scenarios formed by a TMFN composed of 1 GW and 20 FMAPs. In each node, a Network Interface Card (NIC) was configured in Ad Hoc mode, using the IEEE 802.11ac standard in channel 50, which allows \SI{160}{\mega\hertz} channel bandwidth. The data rate was defined by the \emph{IdealWifiManager} mechanism. The wireless links were modeled by the Free-space path loss model; only links with SNR above \SI{5}{\deci\bel} were considered as usable. The transmission power of the NICs was set to \SI{0}{dBm}.

One IEEE 802.11ac spatial stream was used for the wireless links. With one spatial stream, the data corresponding to the maximum Modulation and Coding Scheme (MCS) index is \SI{780}{Mbit/s}, considering \SI{800}{\nano\second} Guard Interval. Taking into account the dimensions of the simulated scenarios, we assume an average number of 2 hops between the FMAPs generating traffic and the GW, in order to calculate the maximum achievable data rate per flow; this results in $(\frac{780}{N_{tx}} / 2) \SI{}{Mbit/s}$, where $N_{tx}$ denotes the number of FMAPs generating traffic. Based on that, the maximum offered load for each scenario was set to 75\% of the maximum achievable data rate per flow, for a total number of FMAPs generating traffic between 5 and 10. The traffic generated was UDP with arrival process modeled as Poisson, for a constant packet size of 1400 bytes; the traffic generation was only triggered after \SI{30}{\second} of simulation, in order to stabilize the OLSR routing tables. In addition, different values for the tunable parameter $\alpha$, between 0.2 and 1, were considered.

A summary of the ns-3.27 simulation parameters used is presented in \cref{RedeFINEEvaluation-Table: ns-3 simulation parameters}.

\begin{table}
    \centering
    \caption{Summary of the ns-3.27 simulation parameters.}
    \label{RedeFINEEvaluation-Table: ns-3 simulation parameters}
    \begin{tabular}{l l}
        \hline % --------------------------------------------------------
        Simulation time        & (\SI{30}{\second} init. +) \SI{130}{\second}   \\
        Wi-Fi standard         & IEEE 802.11ac    \\
        Wi-Fi mode             & Ad Hoc   \\
        Wi-Fi Channel          & 50 [\SI{5250}{\mega\hertz}]  \\
        Channel bandwidth      & \SI{160}{\mega\hertz}    \\
        Guard interval         & \SI{800}{\nano\second}   \\
        Tx power               & \SI{0}{dBm}  \\
        Propagation delay      & Constant speed   \\
        Propagation loss       & Friis \\
        Remote station manager & IdealWifiManager \\
        Mobility model         & Waypoint    \\
        Traffic type           & UDP Poisson  \\
        Packet size            & 1400 Bytes   \\
        Traffic control        & CoDel    \\
        \hline % --------------------------------------------------------
    \end{tabular}
\end{table}

\subsection{Simulation Scenarios} \label{RedeFINEEvaluation-Section: Simulation scenarios}

Five scenarios, in which the UAVs were moving according to the Random Waypoint Mobility (RWM) model, were generated to evaluate the performance of RedeFINE in typical crowded events. Under the RWM model, each UAV chooses a random destination and a speed uniformly distributed between a minimum and a maximum value. Then, the UAV moves to the chosen destination at the selected speed; upon arrival, the UAV stops for a specified period of time and repeats the process for a new destination and speed~\cite{camp2002survey}.
Since RedeFINE relies on knowing in advance the movements of the UAVs, instead of generating the random movements during the ns-3 simulation, we used BonnMotion~\cite{aschenbruck2010bonnmotion}, which is a mobility scenario generation tool. BonnMotion was set to create random waypoint 3D movements for 21 nodes (20 FMAPs and 1 GW) within a box of dimensions \SI{80}{\meter} $\times$ \SI{80}{\meter} $\times$ \SI{25}{\meter}  during \SI{160}{\second}, considering a velocity between \SI{0.5}{m/s} and \SI{3}{m/s} for the UAVs. These scenarios were used to calculate in advance the forwarding tables and the instants they shall be updated. Both the forwarding tables and the generated scenarios were finally imported to ns-3 with a sampling period of \SI{1}{\second}. To apply mobility to the UAVs, based on the generated scenarios, the \emph{WaypointMobilityModel} model of ns-3 was used.

\subsection{Performance Metrics} \label{RedeFINEEvaluation-Section: Performance metrics}

RedeFINE using the Euclidean distance, Airtime, and I2R routing metrics was evaluated against two state of the art distributed routing protocols representative of the reactive and proactive routing paradigms -- AODV~\cite{aodv-etx-repository} and OLSR~\cite{olsr-etx-repository}, respectively, using the ETX routing metric. ETX is a link quality-based routing metric that represents the expected number of transmissions required to send a packet over a link, including retransmissions. Airtime, which is the default routing metric specified in the IEEE 802.11s standard~\cite{hiertz2010ieee}, expresses the amount of channel resources consumed for transmitting a frame over a link. Since the theoretical calculation of the Airtime resulting costs is not straightforward, we exported them from ns-3, by running previous simulations for each one of the generated scenarios. Afterwards, we employed the Dijkstra's algorithm to find the shortest paths between each FMAP and the GW, considering a sampling period of \SI{1}{\second} and the corresponding routing metric. The Airtime routing metric was used in our evaluation to ensure that I2R is able to outperform a metric that uses real measurements to estimate data rate, overhead, and frame error ratio of the communications links.

Our performance evaluation considers two metrics:

\begin{itemize}
    \item Aggregate throughput: The mean number of bits received per second by the GW.
    \item End-to-end delay: The mean time taken by the packets to reach the application layer of the GW since the instant they were generated at a given FMAP, measured at each second, including queuing, transmission, and propagation delays.
\end{itemize}

\subsection{Simulation Results} \label{RedeFINEEvaluation-Section: Simulation results}

\begin{figure}
    \centering
    \subfloat[End-to-end delay Cumulative Distribution Function (CDF).]{
        \includegraphics[width=1\linewidth]{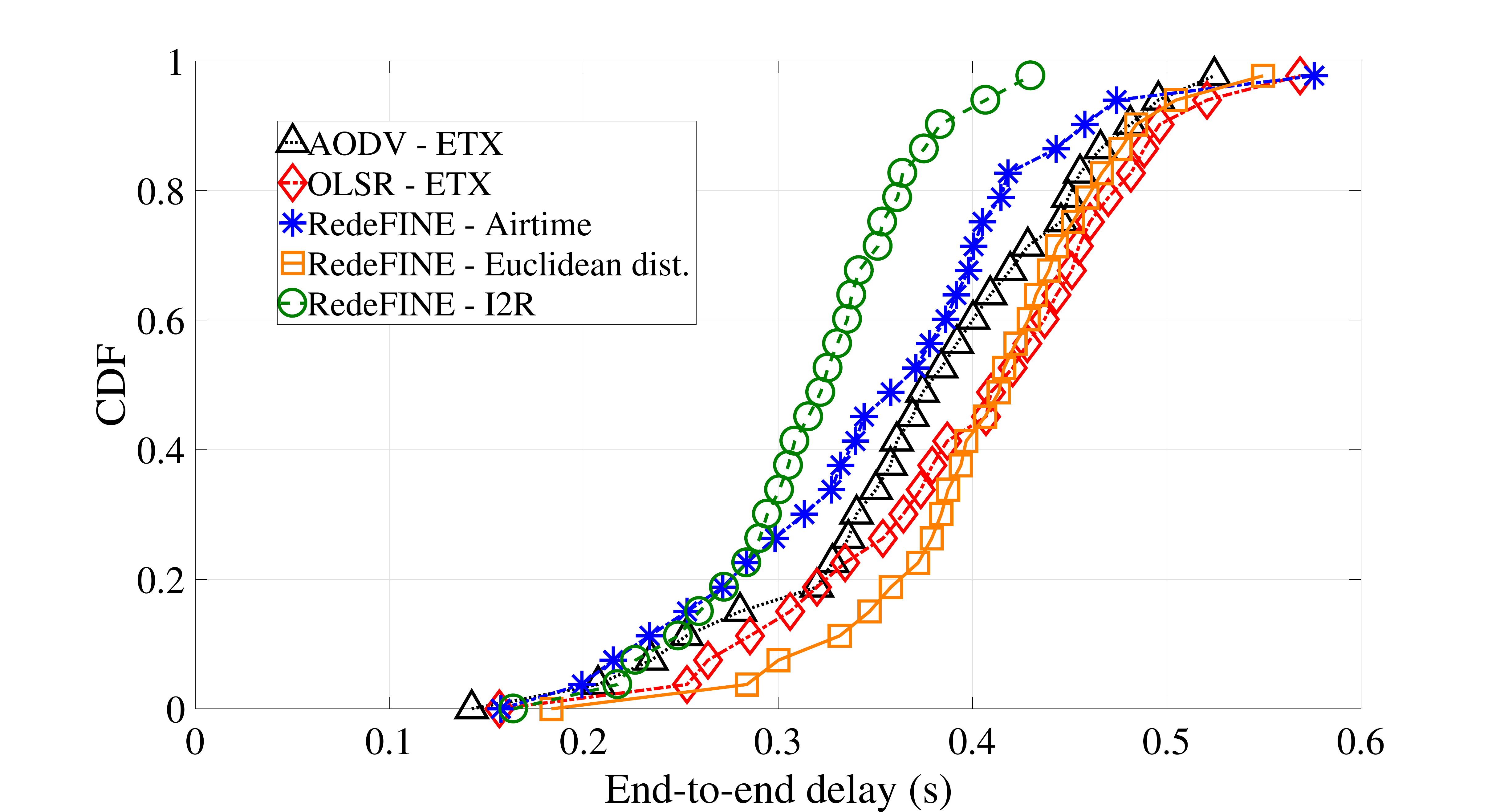}
        \label{RedeFINEEvaluation-Figure: Delay CDF 5 TX nodes}
    }
    \hfill
    \subfloat[Throughput Complementary Cumulative Distribution Function (CCDF).]{
        \includegraphics[width=1\linewidth]{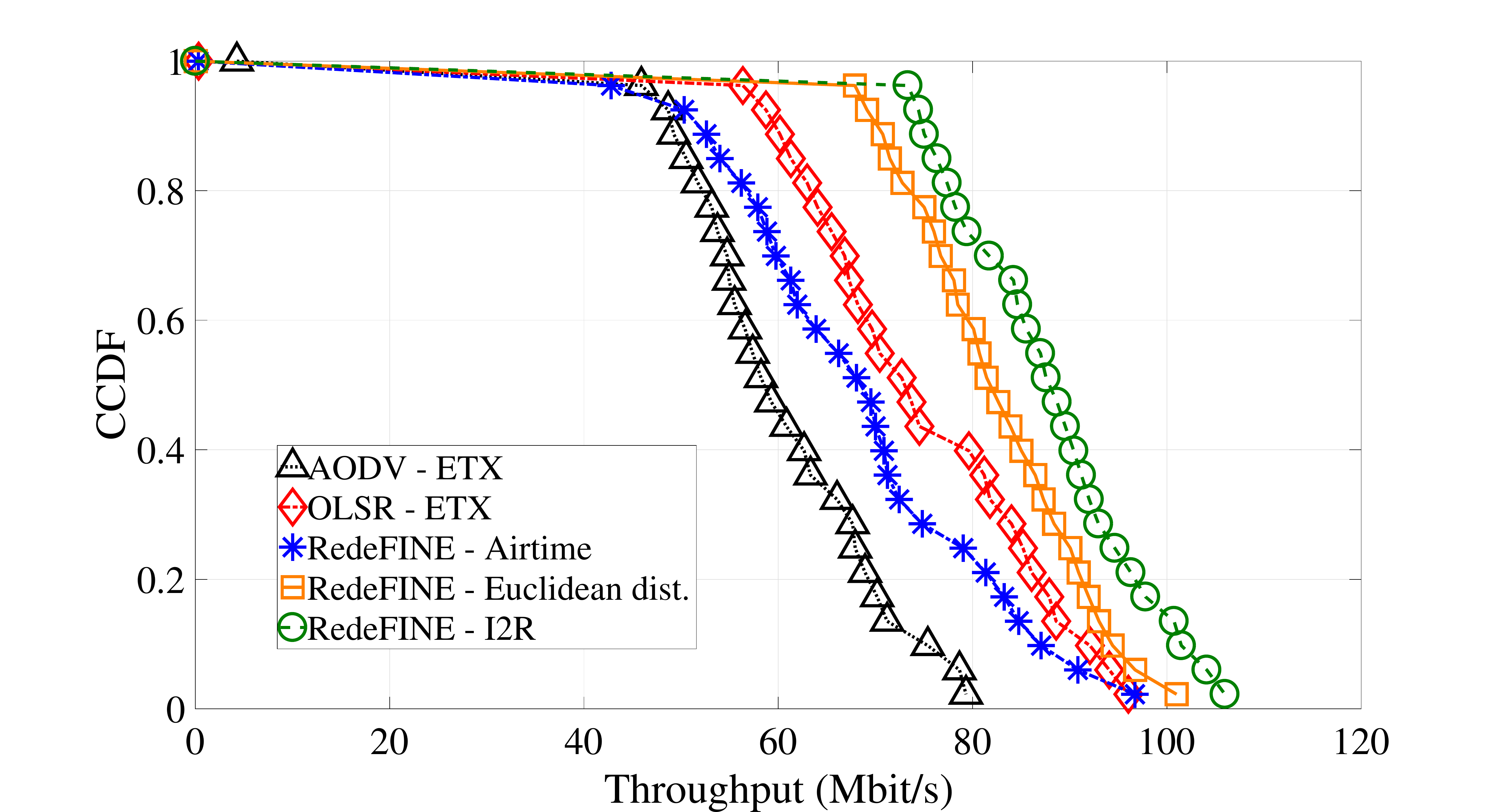}
        \label{RedeFINEEvaluation-Figure: Throughput CCDF 5 TX nodes}
    }
    \hfill
    \subfloat[The 25th, 50th, and 75th percentiles of both the throughput CCDF and end-to-end delay CDF.]{
        \includegraphics[width=1\linewidth]{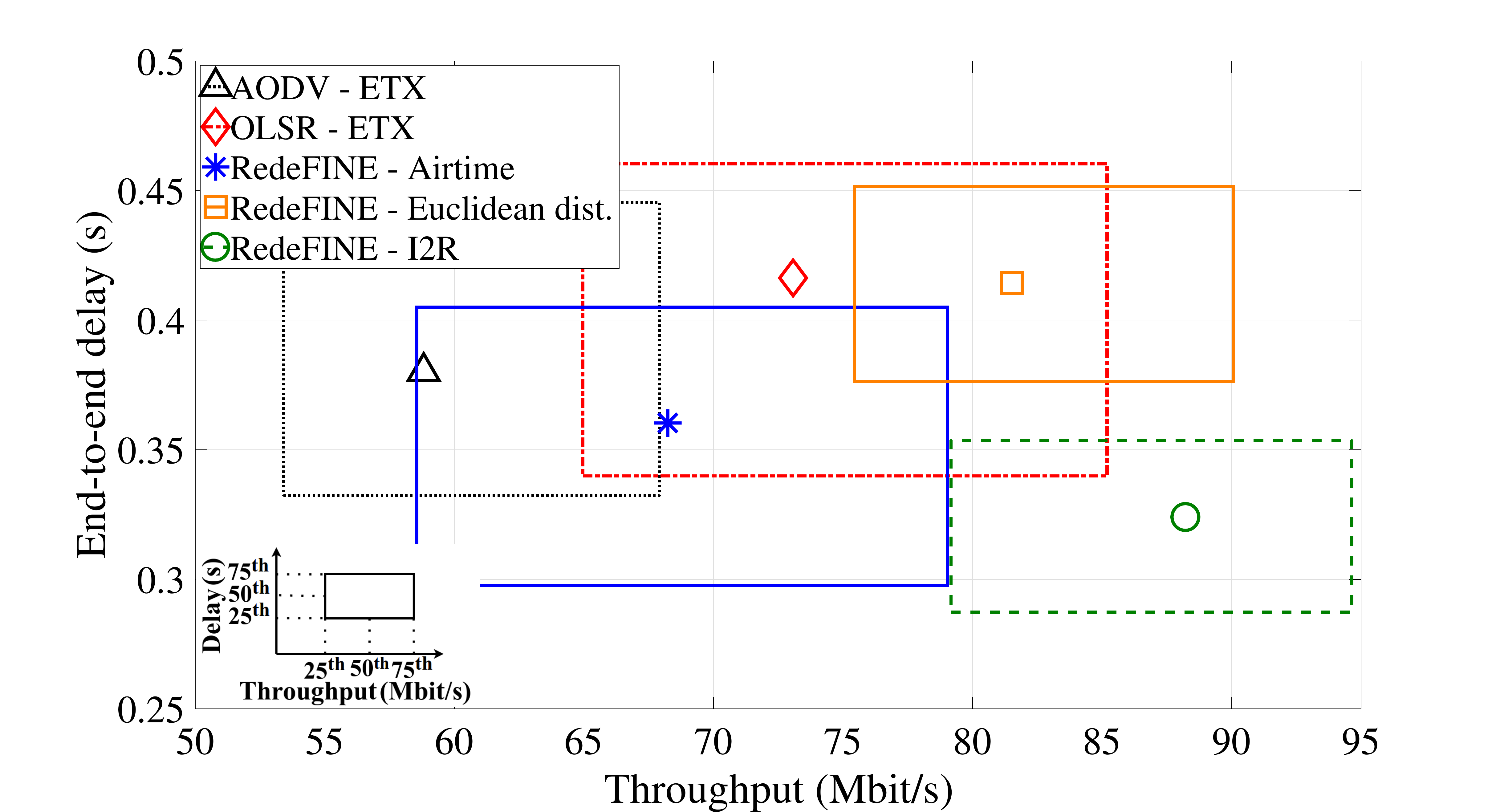}
        \label{RedeFINEEvaluation-Figure: Throughput vs. Delay 5 TX nodes}
    }
    \caption{Results for throughput and end-to-end delay in the GW. The results were obtained considering 5 FMAPs generating traffic, and $\alpha=1$.}
    \label{RedeFINEEvaluation-Figure: Results 5 TX nodes}
\end{figure}

\begin{figure}
    \centering
    \subfloat[End-to-end delay Cumulative Distribution Function (CDF).]{
        \includegraphics[width=1\linewidth]{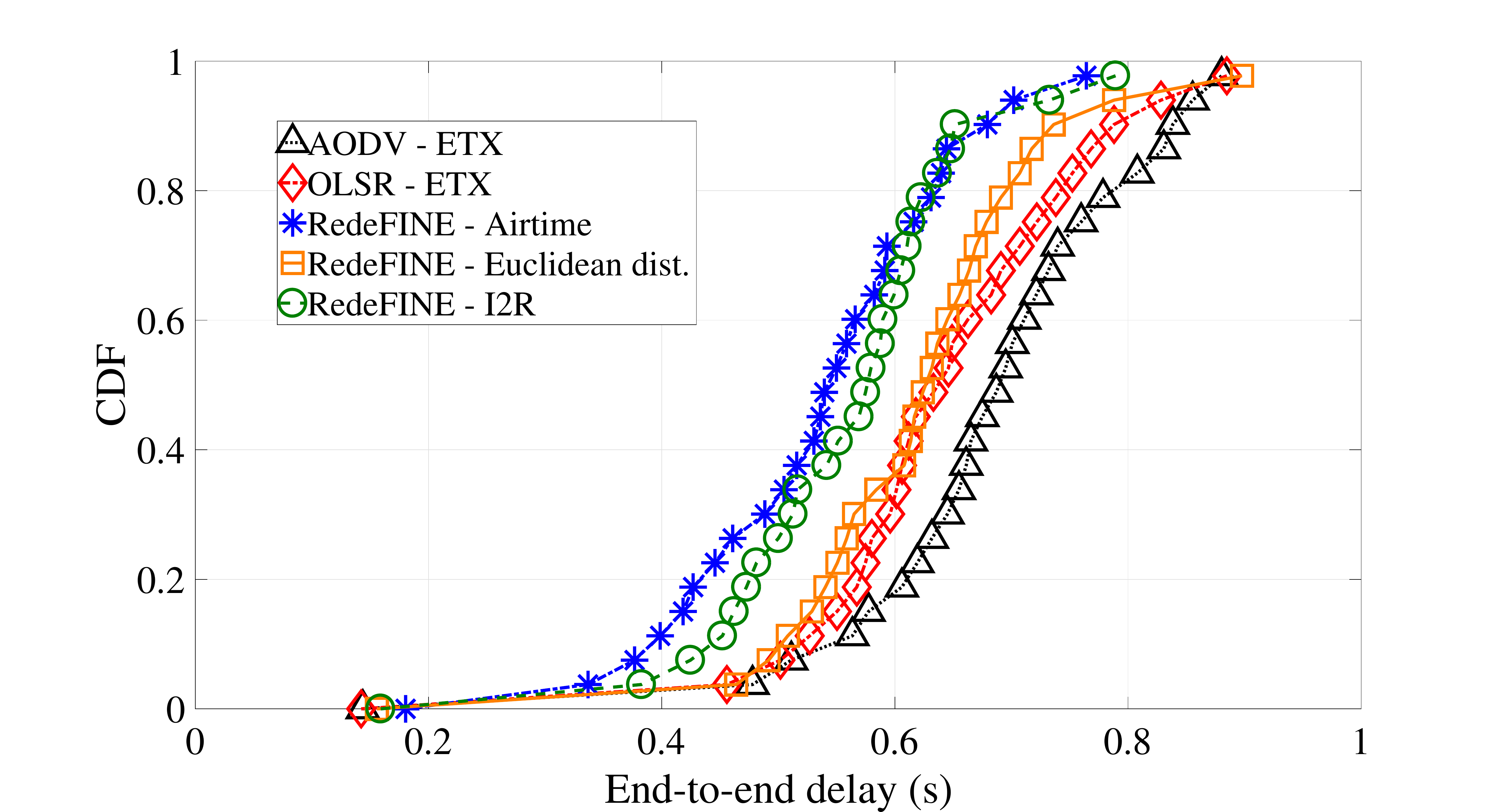}
        \label{RedeFINEEvaluation-Figure: Delay CDF 10 TX nodes}
    }
    \hfill
    \subfloat[Throughput Complementary Cumulative Distribution Function (CCDF).]{
        \includegraphics[width=1\linewidth]{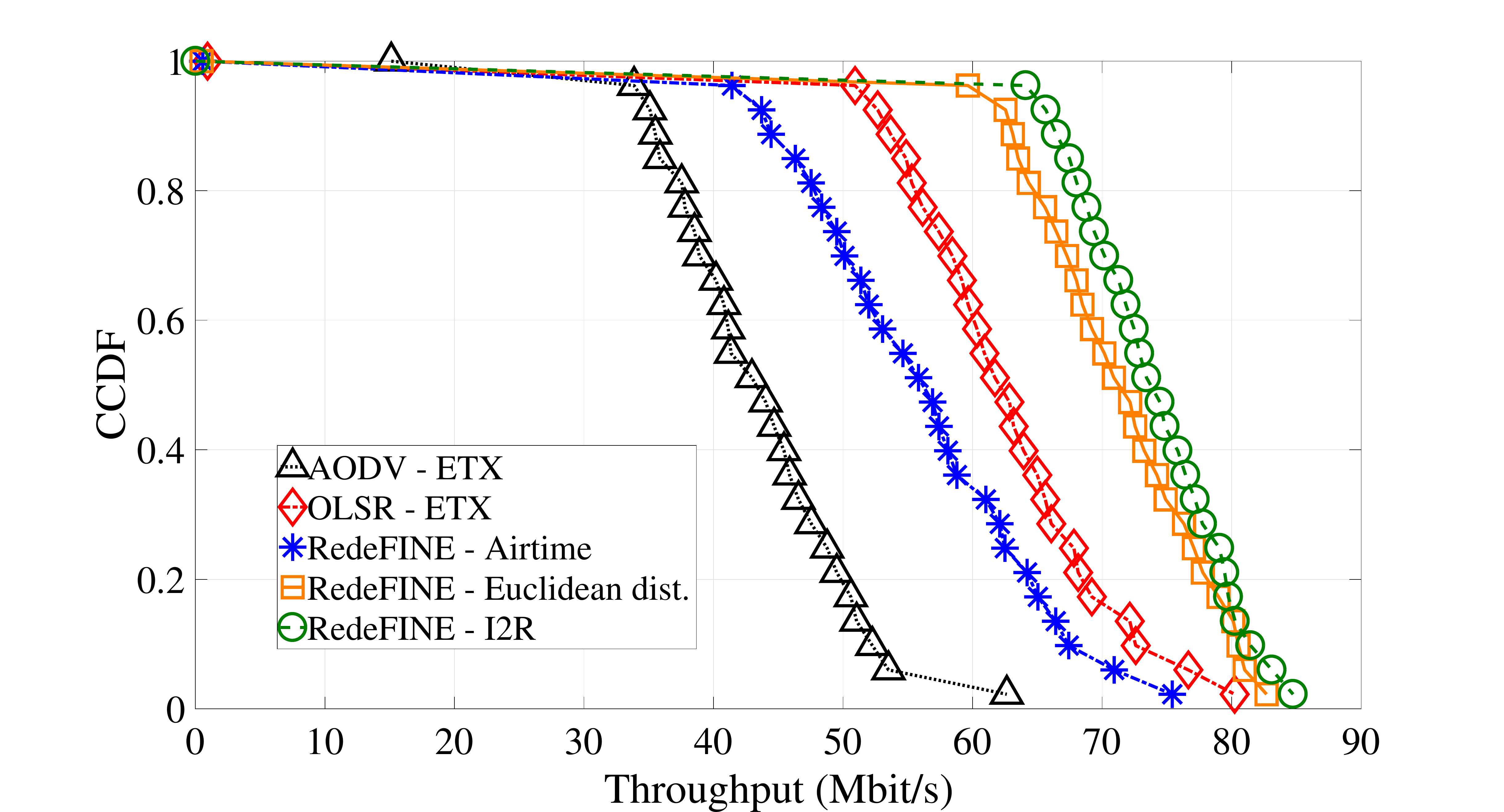}
        \label{RedeFINEEvaluation-Figure: Throughput CCDF 10 TX nodes}
    }
    \hfill
    \subfloat[The 25th, 50th, and 75th percentiles of both the throughput CCDF and end-to-end delay CDF.]{
        \includegraphics[width=1\linewidth]{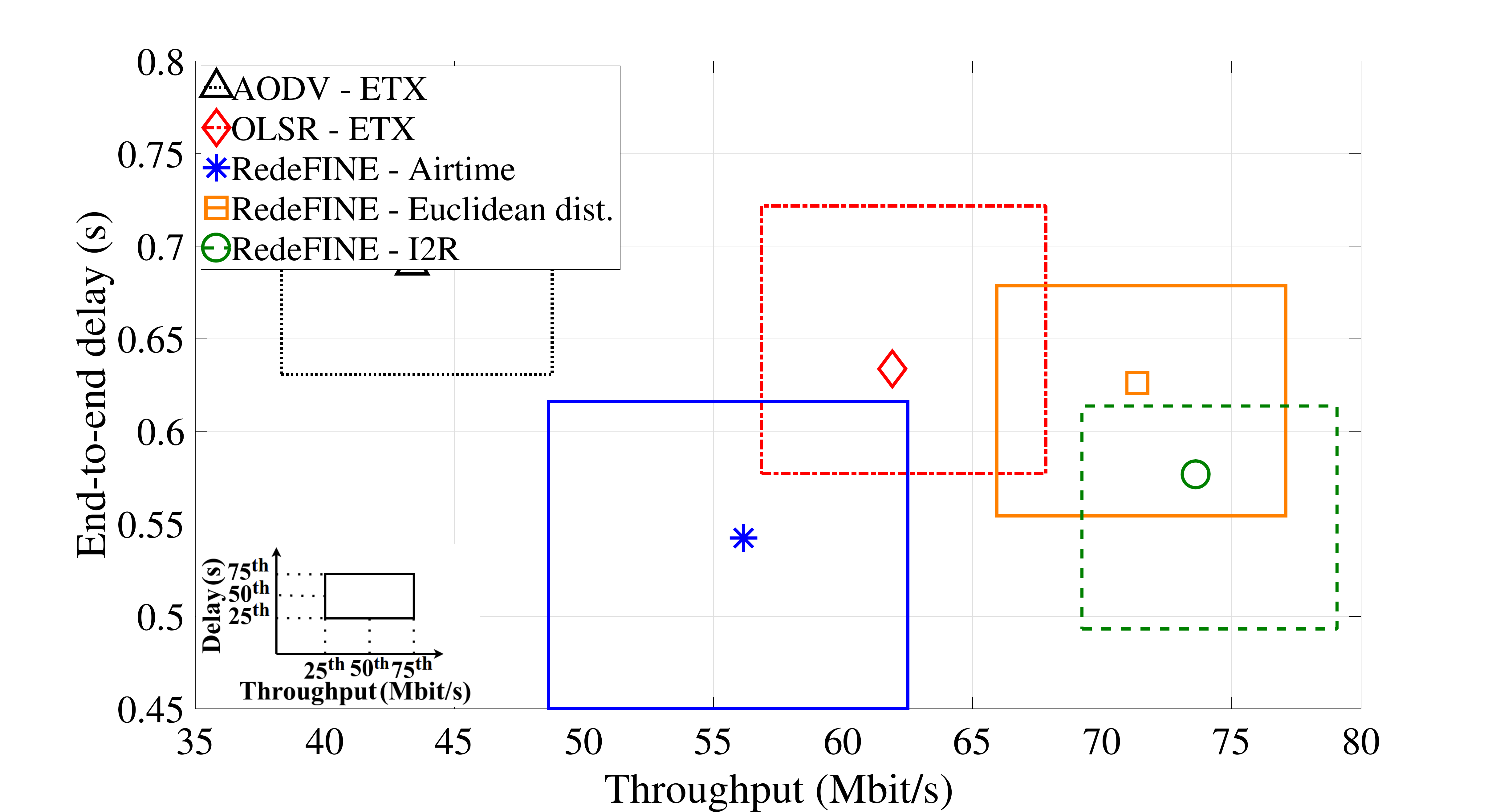}
        \label{RedeFINEEvaluation-Figure: Throughput vs. Delay 10 TX nodes}
    }
    \caption{Results for throughput and end-to-end delay in the GW. The results were obtained considering 10 FMAPs generating traffic, and $\alpha=1$.}
    \label{RedeFINE-Figure: Results 10 TX nodes}
\end{figure}

The results were obtained after 20 simulation runs, using using $RngSeed = 10$ and $RngRun = \{1, ..., 20\}$, for each experimental combination, including different I2R's $\alpha$ values and different number of FMAPs generating traffic. The results are expressed using mean values, considering five random scenarios, as stated in~\cref{RedeFINEEvaluation-Section: Simulation scenarios}. They are represented by means of the CDF for the end-to-end delay and by the complementary CDF (CCDF) for the aggregate throughput, including the values for the 25th, 50th, and 75th percentiles. The CDF $F(x)$ represents the percentage of simulation time for which the mean end-to-end delay was lower than or equal to to $x$, while the CCDF $F'(x)$ represents the percentage of simulation for which the mean aggregate throughput was higher than $x$. %In addition, the gains of I2R in end-to-end delay and throughput, when compared to RedeFINE, are represented by means of bar charts grouped by $\alpha$ value and considering different number of FMAPs generating traffic.
Finally, the influence of the tunable parameter $\alpha$ on the TMFN performance is also evaluated.

When 5 FMAPs are used as traffic sources (cf.~\cref{RedeFINEEvaluation-Figure: Results 5 TX nodes}), the usage of the I2R routing metric improves the end-to-end delay achieved by RedeFINE using the Euclidean distance in approximately 22\%, OLSR and AODV using ETX in 21\% and 15\%, respectively, and RedeFINE using Airtime in 10\%. These values are obtained considering the mean end-to-end delay of the packets received in the GW for the different solutions. The outperforming results of RedeFINE using the I2R routing metric are justified by the selection of paths formed by FMAPs with reduced number of neighbors that are generating or forwarding traffic. Regarding the total amount of bits received in the GW, RedeFINE using the I2R routing metric provides a gain up to 7\% when compared with RedeFINE using the Euclidean distance. In turn, when compared with AODV and OLSR using the ETX routing metric, and with RedeFINE using Airtime, the gains are even more relevant: approximately 45\%, 17\%, and 28\%, respectively.
When 10 FMAPs are generating traffic (cf.~\cref{RedeFINE-Figure: Results 10 TX nodes}), RedeFINE using the I2R routing metric improves end-to-end delay in approximately 10\% with respect to RedeFINE using the Euclidean distance, while the gain over AODV and OLSR using ETX is approximately 18\% and 13\%, respectively. The gain in end-to-end delay of I2R over the Airtime routing metric applied to RedeFINE is negligible.
%Overall, I2R does not achieve the low end-to-end delay as AODV, OLSR, and Airtime, due to the higher amount of data that it allows to forward. As result, the links are subject to a higher level of congestion, resulting in packets held in the transmission queue longer.
Regarding the total amount of bits received in the GW, the gain of RedeFINE using I2R over OLSR using ETX is still approximately 18\%, and over RedeFINE using the Euclidean distance is negligible ($\approx$~4\%). Conversely, the gain over AODV using ETX is increased to approximately 68\%, while with respect to RedeFINE using Airtime it is approximately 31\%.

\begin{figure}
	\centering
	\includegraphics[width=1\linewidth]{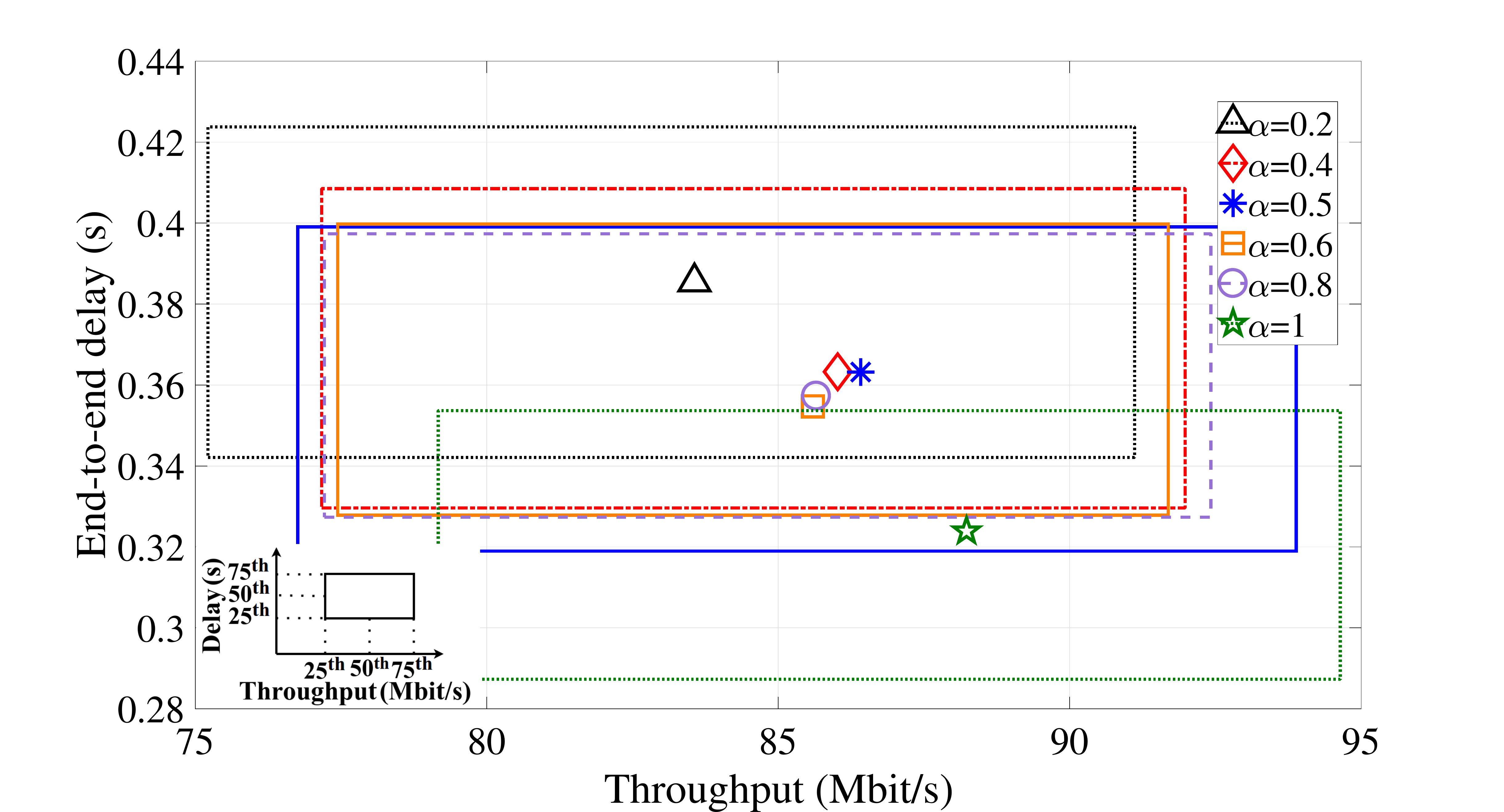}
	\caption{The 25th, 50th, and 75th percentiles of both the throughput CCDF and delay CDF, considering different I2R's $\alpha$ values. The results consider 5 FMAPs generating traffic.}
	\label{RedeFINEEvaluation-Figure: Throughput vs. Delay Multiple Alpha}
\end{figure}

The relation between aggregate throughput and end-to-end delay for the different combinations of protocols and routing metrics is depicted in~\cref{RedeFINEEvaluation-Figure: Throughput vs. Delay 5 TX nodes} and~\cref{RedeFINEEvaluation-Figure: Throughput vs. Delay 10 TX nodes}, where the 25th, 50th, and 75th percentiles of both the throughput CCDF and delay CDF are represented. Overall, the gains in end-to-end delay and throughput of RedeFINE using I2R are reduced when the number of transmission FMAPs increases.
The performance evaluation carried out allowed to conclude that I2R selects preferably as relay nodes the FMAPs that are also sources of traffic; for instance, in a scenario where 5 FMAPs are generating traffic, if any of these FMAPs need a relay to reach the GW, then I2R will give preference to any of the remaining 4 FMAPs that are generating traffic, thus avoiding that a sixth FMAP introduces interference in the TMFN. This effect is faded when the number of FMAPs in the TMFN increases.

Regarding the tunable parameter $\alpha$ of I2R, it must be set to a value close to 1 for higher throughput and lower end-to-end delay values. As $\alpha$ decreases, the performance worsens, as exacerbated by $\alpha=0.2$, in~\cref{RedeFINEEvaluation-Figure: Throughput vs. Delay Multiple Alpha}. This demonstrates how the selection of paths formed by the minimum number of neighboring FMAPs in carrier-sense range contributes to improve the performance of a TMFN, rather than the selection based only on the Euclidean distance.

%%%%%%%%%%%%%%%%%%%%%%%%%%%%%%%%%%%%%%%%%%%%%%%%%%%%%%%%%%%%%%%%%
% TRAFFIC-AWARE GWP ALGORITHM
%%%%%%%%%%%%%%%%%%%%%%%%%%%%%%%%%%%%%%%%%%%%%%%%%%%%%%%%%%%%%%%%%
\section{Traffic-Aware GW Placement Algorithm} \label{GWPAlgorithm-Section}

Even though users are directly affected by the QoS and QoE provided by the access network, the backhaul network, including the GW placement, needs to be carefully designed in order to meet the variable traffic demand of the access network. In this section, a centralized traffic-aware GW Placement (GWP) algorithm for the TMFN, which takes advantage of the knowledge of the placement of the FMAPs and offered traffic to enable communications paths with high enough capacity is presented.

\subsection{Problem Formulation} \label{GWPAlgorithm-Section: Problem formulation}

In the following, the problem addressed in this section is formulated. At time $t_k = k \cdot \Delta t, k \in N_0$ and $\Delta t \in \mathbb{R}$, which is defined according to the TMFN update period imposed by the NetPlan algorithm, the TMFN is represented by a directed graph $G(t_k)=(V, E(t_k))$, where $V=\{0, ..., N-1\}$ is the set of UAVs $i$ positioned at $P_i=(x_i, y_i, z_i)$ inside a cuboid $ X \times Y \times Z$, $E(t_k) \subseteq  V \times V $ is the set of directional links between UAVs $i$ and $j$ at $t_k$, $i, j \in V$, and $(i,j)\in E(t_k)$. The wireless channel between two UAVs is modeled by the Free-space path loss model, since a strong LoS component dominates the links between UAVs flying dozens of meters above the ground.
%We define $C_{i,j}(t_k)$ as the capacity, in bit/s, of the wireless channel available from $\textrm{UAV}_j$ to $\textrm{UAV}_i$ at time $t_k$, considering a constant channel bandwidth $B$ in \si{\hertz}. The Shannon-Hartley theorem is used for this purpose, as given by~\cref{RedeFINE-Equation: Shannon-Hartley theorem}, where $P_{R_{i,j}}(t_k)$ is the average power received at $\textrm{UAV}_i$ transmitted from $\textrm{UAV}_j$ at $t_k$ and $N_i$ is the noise floor power at $\textrm{UAV}_i$, which is assumed to be constant.

%\begin{equation}
%	C_{i,j}(t_k) = B \times \log_{2}\begin{pmatrix}1+\frac{P_{R_i}(t_k)}{N_i}\end{pmatrix}
%	\label{GWPAlgorithm-Equation: Shannon-Hartley}
%\end{equation}

Let us assume that $\textrm{UAV}_i$, $i\in \{1, ..., N-1\}$, performs the role of FMAP and transmits or forwards from other FMAP a traffic flow of bitrate $T_i(t_k)$~\si{bit/s} during time slot $t_k$ towards $\textrm{UAV}_0$ that performs the role of GW. In this case, we have a tree $T(V, E_T)$ that is a subgraph of $G$, where $E_T \subset E$ is the set of direct links between $\textrm{UAV}_i$ and $\textrm{UAV}_0$. This sub-tree defines the TMFN active topology. The flow $F_{0,i}$ is received at $\textrm{UAV}_0$ from $\textrm{UAV}_i$ with bitrate $R_i(t_k)$ bit/s. The wireless medium is shared and we assume that every $\textrm{UAV}_i$ is in the same collision domain, including $\textrm{UAV}_0$. The Carrier Sense Multiple Access with Collision Avoidance (CSMA/CA) mechanism is employed for Medium Access Control (MAC), which enables transmissions only when the channel is sensed to be idle, in order to avoid collisions of network packets.

Considering the throughput $R_i(t_k)$ as the bitrate of the flow $F_{0,i}$ at time $t_k$, and $N-1$ FMAPs generating or forwarding traffic towards $\textrm{UAV}_0$, we aim at determining at any time instant $t_k$ the position of $\textrm{UAV}_0$, $P_0 = (x_0, y_0, z_0)$, and the transmission power $P_T$ of the UAVs, such that the aggregate throughput, $R(t_k) = \sum_{i=1}^{N-1} R_i(t_k)$ is maximized. Our objective function is defined in \cref{GWPAlgorithm-Equation: Objective funtion}.

\begin{equation} \label{GWPAlgorithm-Equation: Objective funtion}
\begin{aligned}
    & \underset{P_T, (x_0, y_0, z_0)}{\textrm{maximize}} && R(t_k)=\sum_{i=1}^{N-1}R_i(t_k) \\
    %%%
    & \textrm{subject to:} \\
    & && \hspace{-4em} (0, i), (i, 0) \in E(t_k), & \hspace{-2em} i \in \{1, ..., N-1\} \\
    & && \hspace{-4em} T_i(t_k) > 0, & \hspace{-2em} i \in \{1,...,N-1\} \\
    & && \hspace{-4em} R_i(t_k) \leq T_i(t_k), & \hspace{-2em} i \in \{1,...,N-1\} \\
    & && \hspace{-4em} 0 \leq x_i \leq X, & \hspace{-2em} i \in \{0,...,N-1\} \\
    & && \hspace{-4em} 0 \leq y_i \leq Y, & \hspace{-2em} i \in \{0,...,N-1\} \\
    & && \hspace{-4em} 0 \leq z_i \leq Z, & \hspace{-2em} i \in \{0,...,N-1\} \\
    & && \hspace{-4em} (x_0, y_0, z_0) \neq (x_i, y_i, z_i), & \hspace{-2em} i \in\{1,...,N-1\}
\end{aligned}
\end{equation}

\subsection{Rationale} \label{GWPAlgorithm-Section: Rationale}

The GWP algorithm takes advantage of the centralized view of the TMFN provided by the NetPlan algorithm. For the sake of simplicity we omit $t_k$ thereafter. Considering the future positions of $\textrm{UAV}_i$ and the bitrate of the traffic flow $F_{0,i}$, $T_i$, we aim at guaranteeing that the wireless link between $\textrm{UAV}_i$ and $\textrm{UAV}_0$ (GW) has a minimum SNR, $SNR_i$, which enables the usage of a MCS index, $\textrm{MCS}_i$, capable of transmitting $T_i$~\SI{}{bit/s}. Conceptually, if $\textrm{MCS}_i$ is ensured by the network, then $R_i \approx T_i$ and $R_i$ is maximized; this is according to the objective function defined in~\cref{GWPAlgorithm-Equation: Objective funtion}.

The minimum $SNR_i$ required for using $\textrm{MCS}_i$ imposes a minimum received power $P_{R_0,i}$. Then, if the transmission power $P_{T_i}$ is known, we can calculate the maximum distance $d_{\max_i}$ between $\textrm{UAV}_i$ and $\textrm{UAV}_0$, using the Free-space path loss model defined in~\cref{GWPAlgorithm-Equation: Free-space Path Loss Model}.

\begin{equation}
    \frac{P_{R_{0,i}}}{P_{T_i}}=\left( \frac{c}{4\pi \times d_{\max_i} \times f_i} \right) ^2
    \label{GWPAlgorithm-Equation: Free-space Path Loss Model}
\end{equation}

In the three-Dimensional (3D) space, $d_{\max_i}$ corresponds to the radius of a sphere, centered at $\textrm{UAV}_i$, inside which $\textrm{UAV}_0$ should be placed. Considering $N-1$ UAVs, the placement subspace for positioning $\textrm{UAV}_0$ is defined by the intersection of the corresponding spheres $i \in \{1, ..., N-1\}$; we refer to this subspace as the Gateway Placement Subspace, $S_G$. In order to simplify the process of calculating $S_G$, we follow Algorithm~A, which iteratively allows obtaining the point $P_0 = (x_0, y_0, z_0)$ for positioning $\textrm{UAV}_0$ and the transmission power $P_T$ which we assume to be the same for all UAVs.

\begin{algorithm}[t]\label{GWPAlgorithm-Algorithm: GWP Algorithm}
\renewcommand{\thealgorithm}{A}
\caption{-- GWP Algorithm}
\begin{algorithmic}[1]
%\Procedure{$S_G$}{$SNR_i, i\in \{1, ..., N-1\}$}\Comment{The g.c.d. of a and b}
\State $P_T = 0$ \Comment{\SI{0}{dBm} Tx power}
\While {true}
    %\State \emph{loop:}
    \State $P_{T_i} = P_T, i\in \{1, ..., N-1\}$ \Comment{Same UAVs' Tx power}
    \State Calculate $(x_0, y_0, z_0)$ \Comment{System of equations~\cref{GWPAlgorithm-SystemOfEquations: Gateway Placement Subspace}}
    \If{$(x_0, y_0, z_0) \neq \oslash$} \Comment{i.e., $(x_0, y_0, z_0) \in S_G$}
        \State \textbf{return} $P_T, (x_0, y_0, z_0)$ \Comment{Tx power, GW pos.}
    \Else
        \State $P_T = P_T + 1$ \Comment{Increase Tx power by \SI{1}{dBm}}
        %\State \textbf{go to} \emph{loop} \Comment{Try another iteration}
    \EndIf
\EndWhile
\end{algorithmic}
\end{algorithm}

\begin{figure}
	\centering
	\includegraphics[width=1\linewidth]{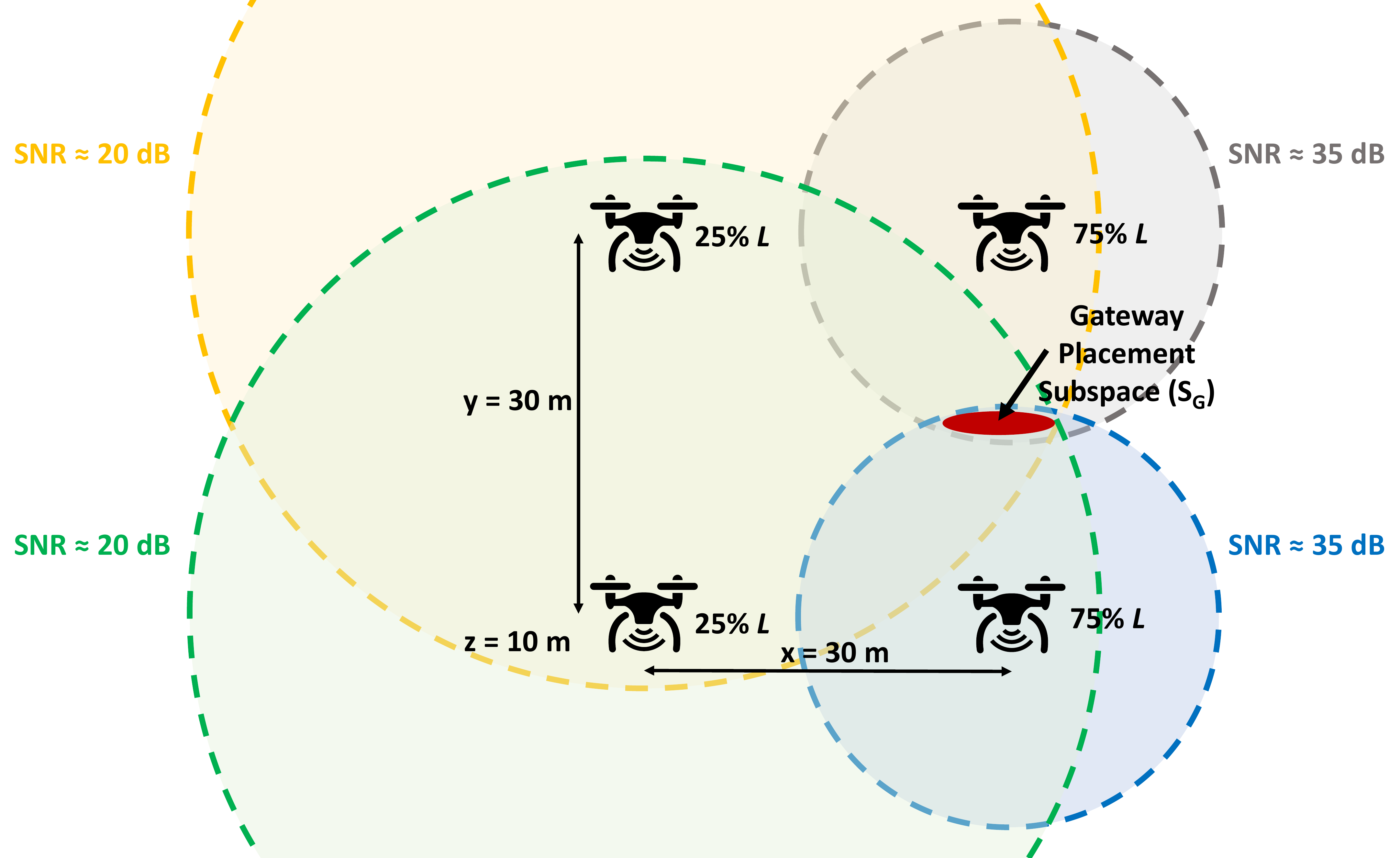}
	\caption{Gateway Placement Subspace ($S_G$) in a two-Dimensional (2D) space, which results from the intersection of the circumferences, centered at each FMAP, with radius equal to the maximum distance compliant with a minimum wireless link's SNR.}
	\label{GWPAlgorithm-Figure: Gateway Placement Subspace}
\end{figure}

The GWP algorithm provides the same output whether downlink or uplink traffic is considered, since all the nodes are configured with the same transmission power and the wireless channel is assumed to be symmetric.

\subsection{Numerical Analysis} \label{GWPAlgorithm-Section: Numerical Analysis}

Without loss of generality, we now exemplify the execution of Algorithm A for the simple scenario shown in \cref{GWPAlgorithm-Figure: Gateway Placement Subspace}; the algorithm is generic and may be applied to any traffic demand and number of FMAPs. The scenario of \cref{GWPAlgorithm-Figure: Gateway Placement Subspace} is composed of 4 FMAPs that are placed within a square of \SI{30}{\meter} sideways, hovering at \SI{10}{\meter} altitude. The capacity of the shared wireless medium is assumed to be equal to the maximum MCS index of the IEEE 802.11ac technology, which is \SI{780}{Mbit/s}, considering one spatial stream, \SI{800}{\nano \second} Guard Interval (GI), and \SI{160}{\mega\hertz} channel bandwidth (channel 50 at \SI{5250}{\mega\hertz}). Since the wireless medium is shared by four FMAPs generating traffic, and assuming a single hop between the FMAPs and the GW ($\textrm{UAV}_0$), this results in a fair share $L = \frac{780}{4} = \SI{195}{Mbit/s}$ for the capacity of the wireless channel between each $\textrm{FMAP}_i, i \in \{1, 2, 3, 4\}$, and the GW. In the scenario of \cref{GWPAlgorithm-Figure: Gateway Placement Subspace}, the FMAPs on the left-side have traffic demand equal to 25\% of the FMAPs' fair share of the wireless channel capacity, and the righ-side FMAPs have traffic demand equal to 75\% of the FMAPs' fair share of the wireless channel capacity. Accordingly, the FMAPs on the left-side transmit at bitrate $T_1 = T_2 = 0.25 \times 195 \approx \SI{49}{Mbit/s}$, and the right-side FMAPs transmit at bitrate $T_3 = T_4 = 0.75 \times 195 \approx \SI{146}{Mbit/s}$.

Taking into account the mapping between SNR, theoretical data rate of the IEEE 802.11ac MCS indexes, and the link capacity for 4 FMAPs sharing the transmission time, from \cref{GWPAlgorithm-Table: Mapping Between SNR Data Rate and Link Capacity} we conclude that the target SNR values in \si{\decibel}, considering a \SI{-85}{dBm} noise floor power, are respectively \SI{20}{\deci\bel} for the left-side FMAPs and \SI{35}{\deci\bel} for the right-side FMAPs. Note that the rationale to calculate the capacity of each individual link, which is presented in the third column of \cref{GWPAlgorithm-Table: Mapping Between SNR Data Rate and Link Capacity}, results from the fact that the average transmission time assigned to each FMAP, as a result of the MAC protocol, is quarter of the transmission time available in the shared wireless channel. For two FMAPs with traffic demand equal to $0.25 \times 195 \approx \SI{49}{Mbit/s}$ and two FMAPs with traffic demand equal to $0.25 \times 195 \approx \SI{146}{Mbit/s}$, this results in $\SI{390}{Mbit/s}$ as the aggregate throughput. Since the shared channel provides $2 \times \SI{78}{Mbit/s}$ $+$ $2 \times \SI{176}{Mbit/s} = \SI{468}{Mbit/s}$ as the maximum capacity, the channel will be occupied during $\frac{390}{468} \approx 0.83$ of the available transmission time.

\begin{table}
	\centering
	\caption{Extract of the mapping between SNR, data rate of the IEEE 802.11ac MCS indexes, and the link capacity values for 4 FMAPs sharing the transmission time~\cite{MCSIndex14:online}.}
	\label{GWPAlgorithm-Table: Mapping Between SNR Data Rate and Link Capacity}
	\begin{tabular}{l l l}
		\hline % --------------------------------------------------------
		SNR         &   MCS data rate   &   Link capacity   \\
		(\si{dB})   &   (\si{Mbit/s})   &   (\si{Mbit/s})   \\
		\hline % --------------------------------------------------------
		12              &   58.5    & $58.5 / 4 \approx 15 $ \\
		\circled{20}    &   234     & 58.5    \\
		\mymk{35}       &   702     & 175.5   \\
		37              &   780     & 195     \\
		\hline % --------------------------------------------------------
	\end{tabular}
\end{table}

\begin{equation} \label{GWPAlgorithm-SystemOfEquations: Gateway Placement Subspace}
	\begin{cases}
		\begin{aligned}
			& (x_0-30)^2+y_0^2+(z_0-10)^2 \leq 10^{\left(\dfrac{K+P_T-\mymk{35}}{20} \right)^2} \\
			& (x_0-30)^2+(y_0-30)^2 \\
			& \qquad + (z_0-10)^2 \leq 10^{\left(\dfrac{K+P_T-\mymk{35}}{20} \right)^2} \\
			& x_0^2+(y_0-30)^2 + (z_0-10)^2 \leq 10^{\left(\dfrac{K+P_T-\circled{20}}{20} \right)^2} \\
			& x_0^2+y_0^2+(z_0-10)^2 \leq 10^{\left(\dfrac{K+P_T-\circled{20}}{20}\right)^2} \\
			& K = -20 \times \log_{10}\left(\frac{4\pi}{3\times10^8}\right)\\
			& \qquad -20 \times \log_{10}(5250\times10^6) - (-85)\\
		\end{aligned}
	\end{cases}
\end{equation}

Solving the system of equations \cref{GWPAlgorithm-SystemOfEquations: Gateway Placement Subspace}, which is derived from \cref{GWPAlgorithm-Equation: Free-space Path Loss Model} in logarithmic scale, we conclude that an optimal placement for the GW is $(x_0,y_0,z_0) \approx (23.3, 15.4, 3.3)$ for a transmission power $P_{T} = $ \SI{22}{dBm}. Note that $P_T$ is the fine tuning parameter in the system of equations \cref{GWPAlgorithm-SystemOfEquations: Gateway Placement Subspace}, so that we can find at least a point $(x_0, y_0, z_0) \in S_G$; otherwise, we may have a system of equations without solution. $P_T$ is initially set to \SI{0}{dBm}; then, it is iteratively increased by \SI{1}{dBm} until a valid solution for the GW position is found. In \cref{GWPAlgorithm-SystemOfEquations: Gateway Placement Subspace}, the carrier frequency $f$ being used is considered: \SI{5250}{\mega\hertz} in this article; however, our conclusions are independent of $f$.

In the GWP algorithm we assume that the overhead introduced by the UDP, IP, and MAC packet headers is negligible; this is compliant with emerging wireless communications technologies, such as IEEE 802.11ax, where Orthogonal Frequency-Division Multiple Access (OFDMA) and frame aggregation mechanisms improve the MAC efficiency~\cite{afaqui2016}.

%%%%%%%%%%%%%%%%%%%%%%%%%%%%%%%%%%%%%%%%%%%%%%%%%%%%%%%%%%%%%%%%%
% GWP ALGORITHM EVALUATION
%%%%%%%%%%%%%%%%%%%%%%%%%%%%%%%%%%%%%%%%%%%%%%%%%%%%%%%%%%%%%%%%%
\section{Evaluation of the GWP Algorithm} \label{GWPAlgorithmEvaluation-Section}

The TMFN performance achieved using the GWP algorithm is presented in this section, including the simulation setup, the simulation scenarios, and the performance metrics considered.

\subsection{Simulation Setup} \label{GWPAlgorithmEvaluation-Section: Simulation setup}

In order to evaluate the TMFN performance achieved with the GWP algorithm, the ns-3 simulator was used. A Network Interface Card (NIC) was configured on each node in Ad Hoc mode, using the IEEE 802.11ac standard in channel 50, with \SI{160}{\mega\hertz} channel bandwidth, and \SI{800}{\nano\second} guard interval. One spatial stream was used for all wireless links. The traffic generated was UDP Poisson for a constant packet size of \SI{1400}{bytes}, during \SI{100}{\second} simulation time. The data rate was automatically defined by the \emph{IdealWifiManager} mechanism. The Controlled Delay (CoDeL) algorithm~\cite{nichols2012}, which is a Linux-based queuing discipline that considers the time that packets are held in the transmission queue to discard packets, was used; it allows mitigating the bufferbloat problem. The default parameters of CoDeL in ns-3 were employed, including 1000 packets as queue size and \SI{5}{\milli\second} as target queue delay \cite{CoDel:online}.

\subsection{Simulation Scenarios} \label{GWPAlgorithmEvaluation-Section: Simulation Scenarios}

\begin{figure}
	\centering
	\includegraphics[width=0.9\linewidth]{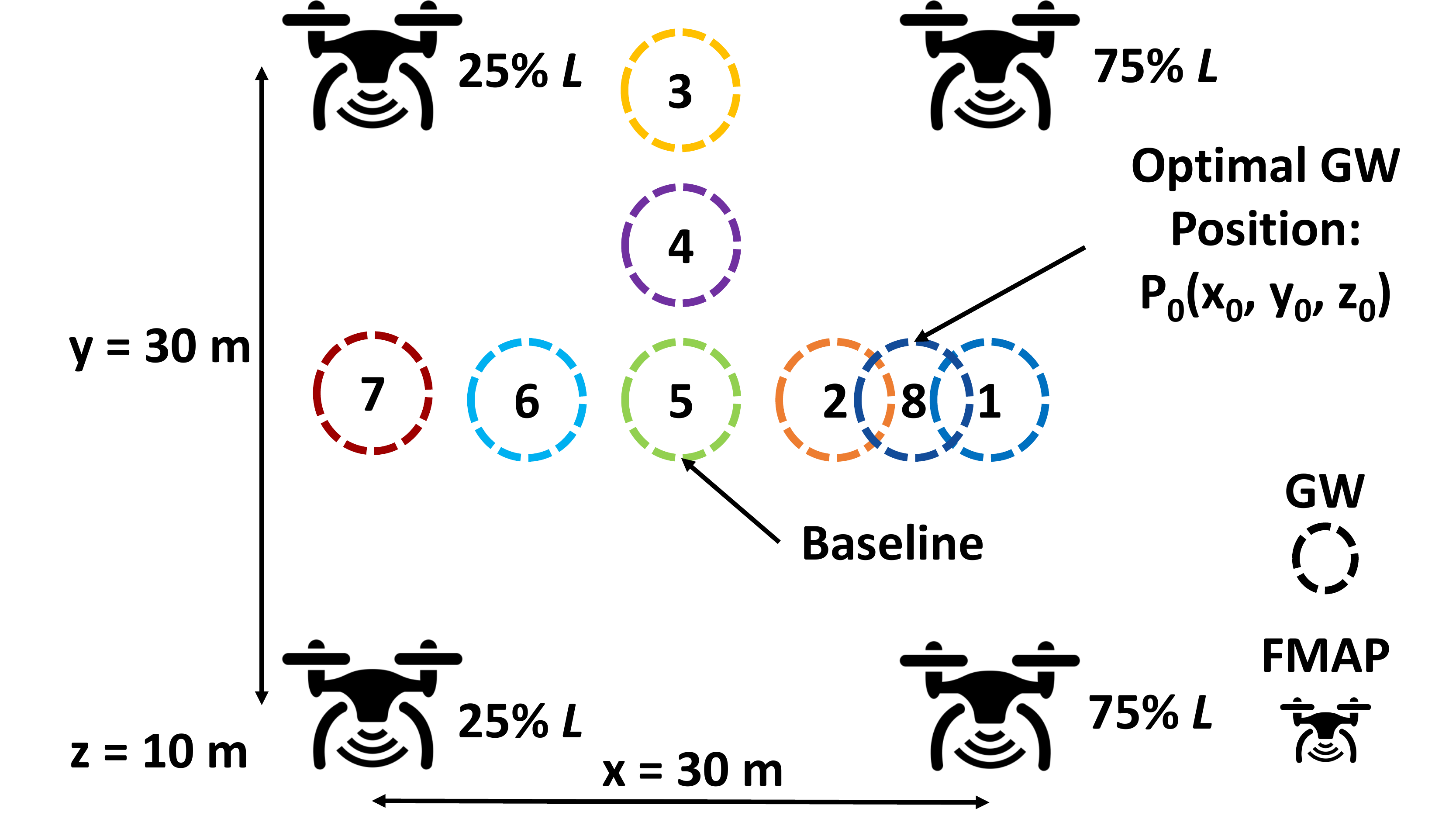}
	\caption{Scenario A, in which different positions for the GW were evaluated. Position 8 corresponds to the optimal GW position, while position 5 corresponds to the baseline -- GW placed in the FMAPs center.}
	\label{GWPAlgorithmEvaluation-Figure: Baseline Scenario}
\end{figure}

\begin{figure}
	\centering
	\includegraphics[width=0.9\linewidth]{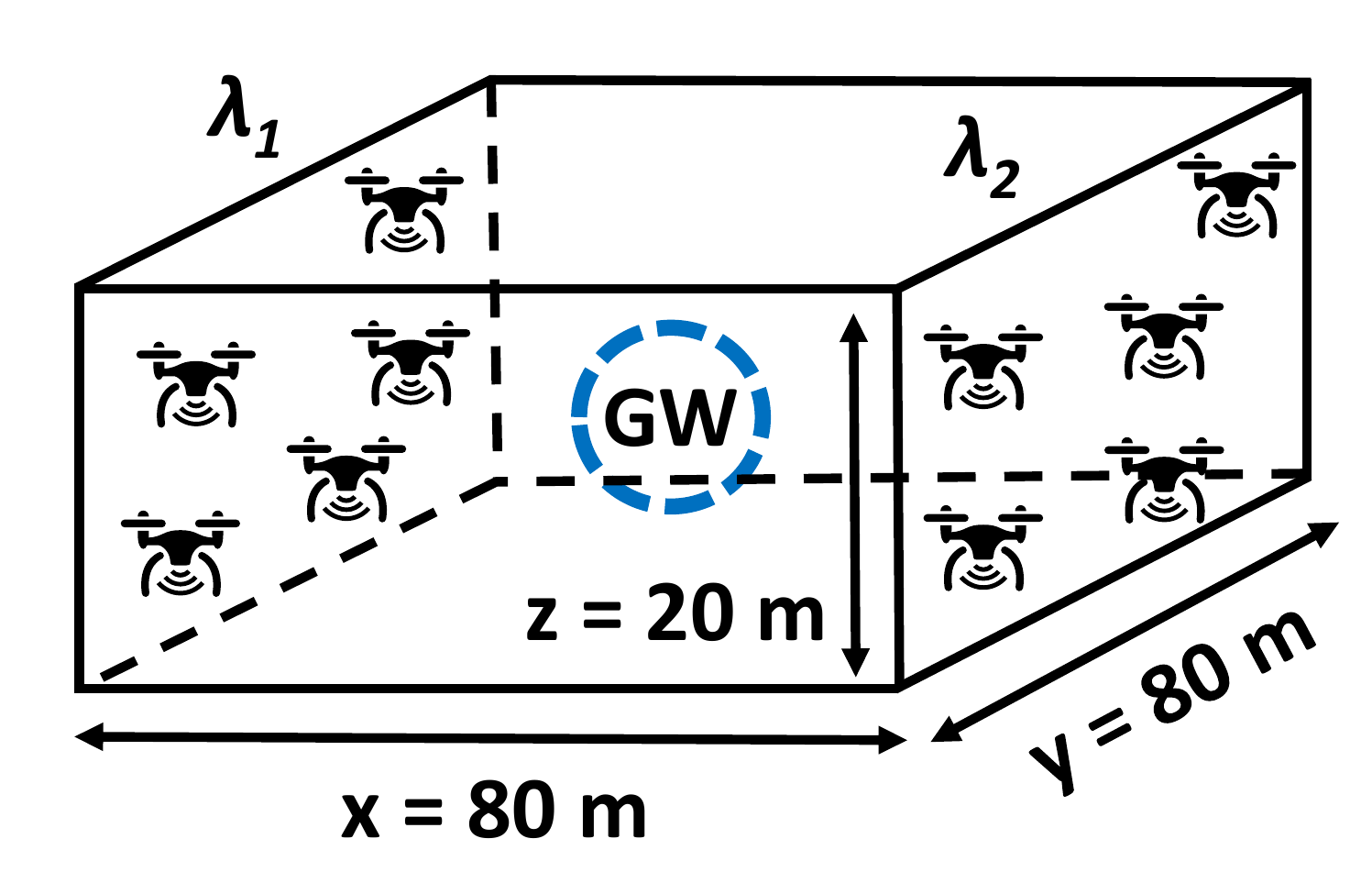}
	\caption{Scenario B, in which 10 FMAPs were randomly positioned in order to form two zones with different traffic demand: $\lambda_1$ and $\lambda_2$. The baseline corresponds to the GW placed in the FMAPs center, which is represented by a dashed circumference.}
	\label{GWPAlgorithmEvaluation-Figure: Complex Scenario}
\end{figure}

In addition to the optimal GW position, which was obtained using the GWP algorithm, other positions for the GW in the venue depicted in ~\cref{GWPAlgorithm-Figure: Gateway Placement Subspace} were evaluated, in order to show the performance gains obtained when using the GWP algorithm; the seven additional positions considered are depicted in ~\cref{GWPAlgorithmEvaluation-Figure: Baseline Scenario} and hereafter referred to as Scenario A. Position 1 to position 7 were defined to allow an inter-position distance of \SI{7.5}{\meter}; they aimed at exploring the vertical and horizontal corridors of the venue. We define as baseline the GW placed in the FMAPs center (i.e.,
three-coordinates average considering all FMAPs). Position 8 represents the optimal GW placement, which was derived from~\cref{GWPAlgorithm-SystemOfEquations: Gateway Placement Subspace}.

\begin{figure*}[t]
	\centering
	\subfloat[Throughput (R) CCDF.]{
		\includegraphics[width=0.47\linewidth]{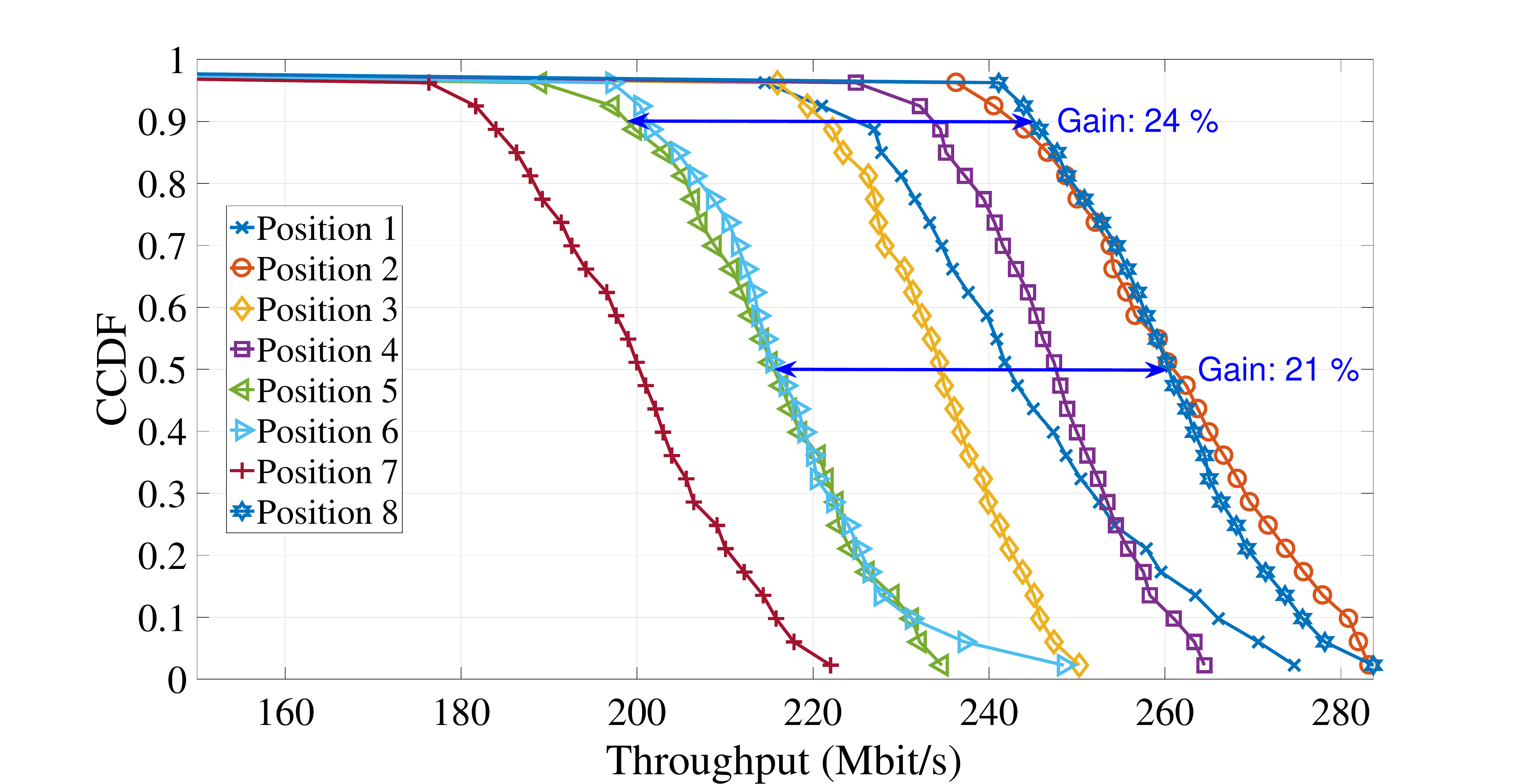}
		\label{GWPAlgorithmEvaluation-Figure: Baseline Scenario Throughput}
	}
	\hfill
	\subfloat[End-to-end delay CDF.]{
		\includegraphics[width=0.47\linewidth]{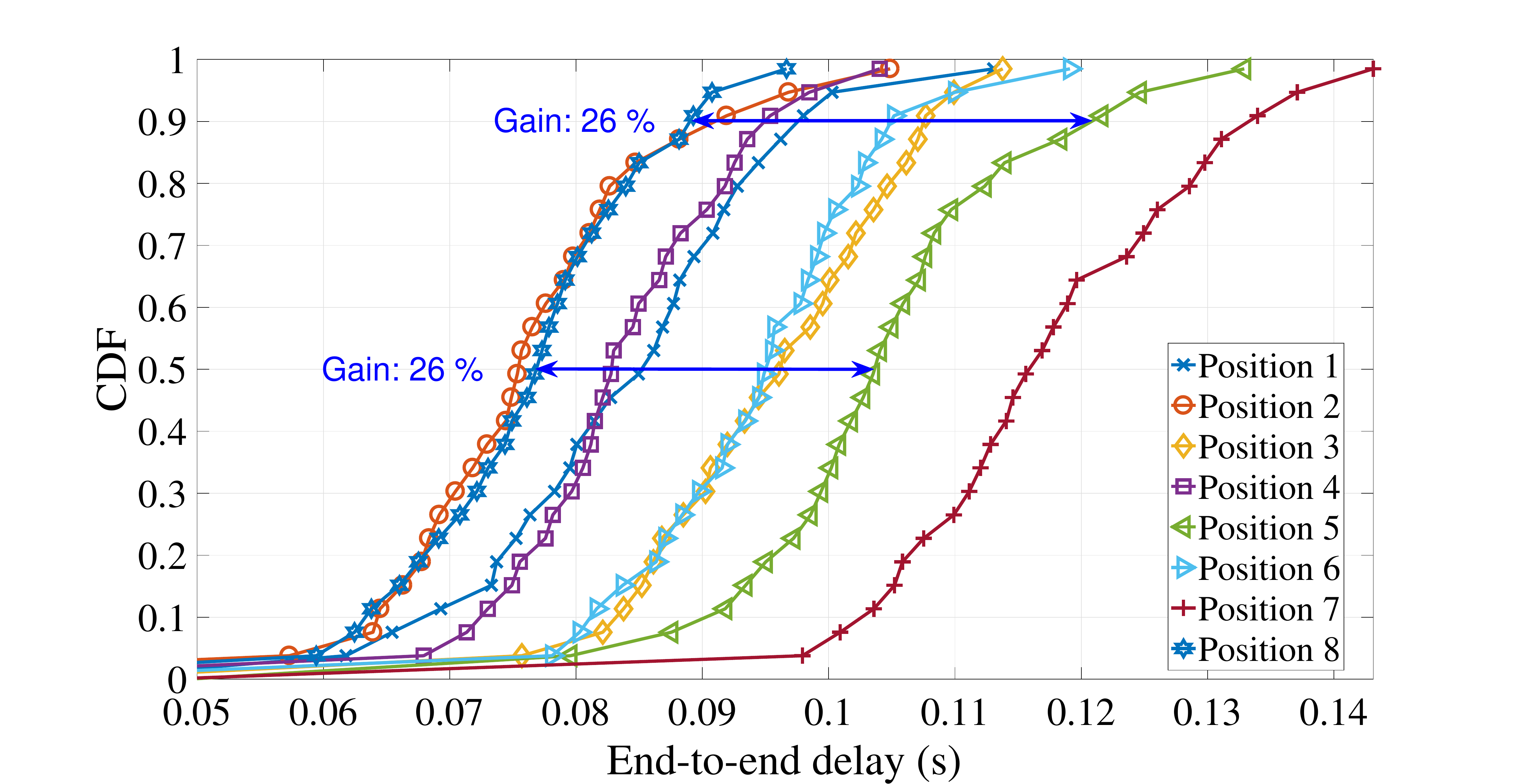}
		\label{GWPAlgorithmEvaluation-Figure: Baseline Scenario Delay}
	}
	\caption{Scenario A - Aggregate Throughput (R) and End-to-end delay results measured in the GW. Position 8 corresponds to our proposal for the GW position.}
	\label{GWPAlgorithmEvaluation-Figure: Baseline Scenario Results}
\end{figure*}

In order to evaluate the performance achieved when using the GWP algorithm in a typical crowded event, a more complex scenario, depicted in \cref{GWPAlgorithmEvaluation-Figure: Complex Scenario} and hereafter named as Scenario B, was also considered. It represents a TMFN composed of 10 FMAPs and 1 GW, inside a cuboid of dimensions \SI{80}{\meter} $\times$ \SI{80}{\meter} $\times$ \SI{20}{\meter}. The FMAPs were randomly positioned in order to form two zones with different traffic demand: $\lambda_1$ and $\lambda_2$ bit/s.
Since the GWP algorithm relies on knowing in advance the positions of the FMAPs, provided by the NetPlan algorithm in a real-world deployment, instead of generating the random waypoints during the ns-3 simulation we used BonnMotion~\cite{aschenbruck2010bonnmotion}, which is a mobility scenario generation tool. These waypoints were considered to calculate in advance the forwarding tables and the optimal GW position. We considered as baseline the GW placed in the FMAPs center. Finally, the forwarding tables and the GW position along the time, as well as the generated scenarios were imported to ns-3, with a sampling period of \SI{1}{\second}. The \emph{WaypointMobilityModel} model of ns-3, which places the UAVs in the positions generated by BonnMotion, was used. Two different traffic demand combinations were considered: a) $\lambda_1$ $= 0.1 \times L$ and $\lambda_2 = 0.9 \times L$; and b) $\lambda_1$ $= 0.25 \times L$ and $\lambda_2 = 0.75 \times L$, where  $L$ is the capacity of the wireless medium divided by the number of FMAPs.

\subsection{Performance Metrics} \label{GWPAlgorithmEvaluation-Section: Performance Metrics}

\begin{figure*}[t]
	\centering
	\subfloat[Throughput (R) CCDF.]{
		\includegraphics[width=0.47\linewidth]{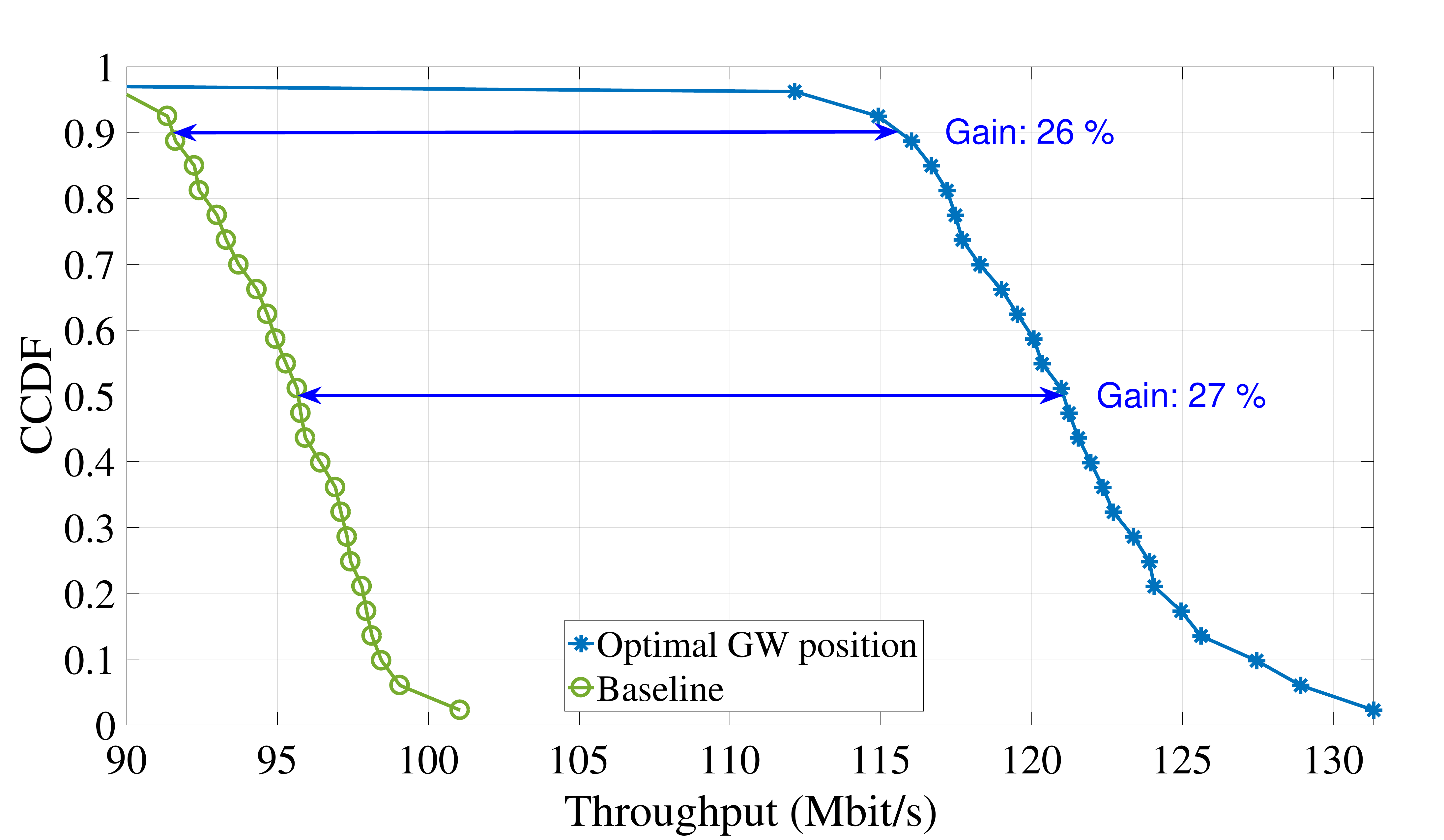}
		\label{GWPAlgorithmEvaluation-Figure: Complex Scenario Throughput 10-90}
	}
	\hfill
	\subfloat[End-to-end delay CDF.]{
		\includegraphics[width=0.47\linewidth]{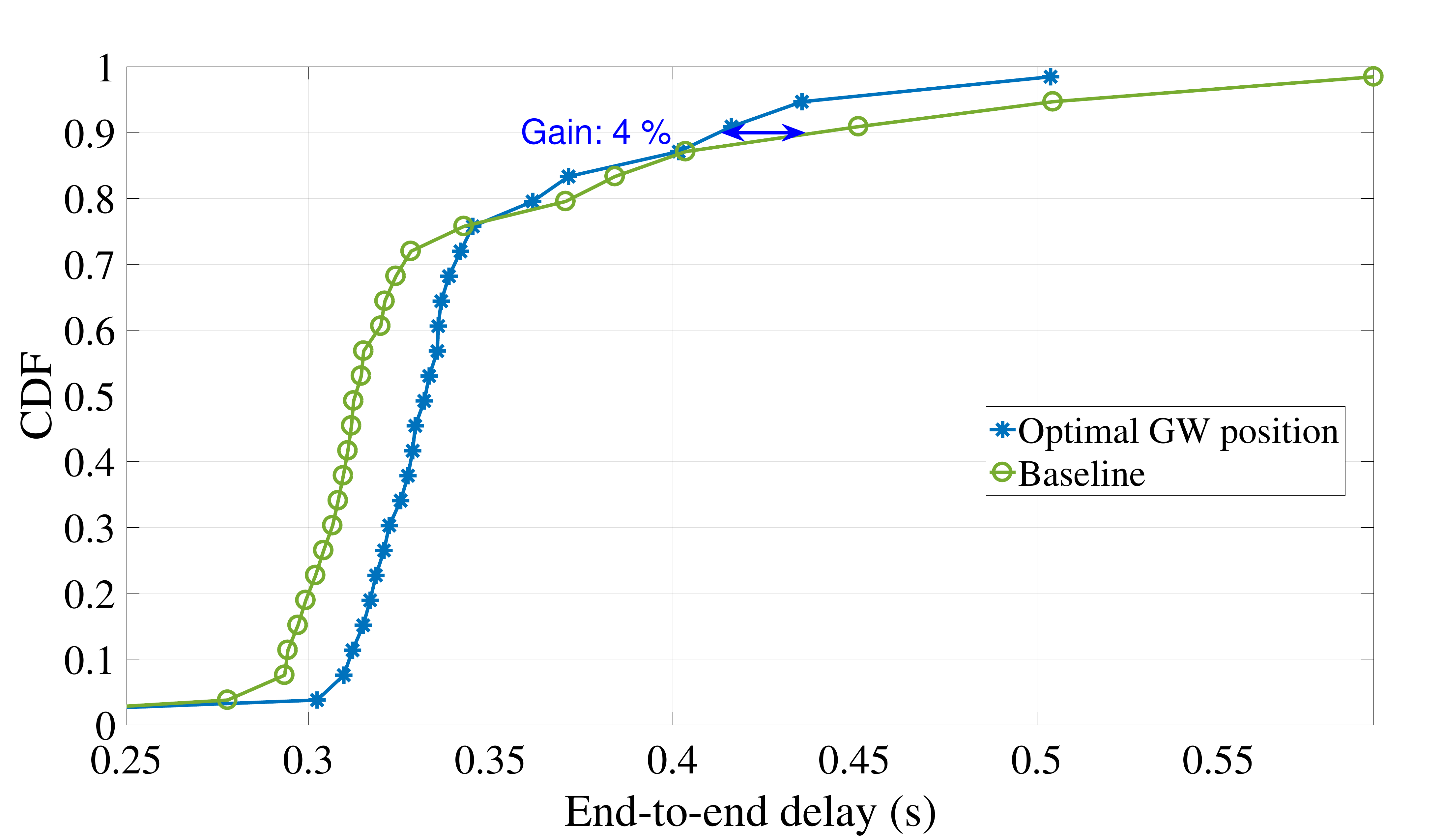}
		\label{GWPAlgorithmEvaluation-Figure: Complex Scenario Delay 10-90}
	}
	\caption{Scenario B - Aggregate Throughput (R) and End-to-end delay results measured in the GW for $\lambda_1 $ and $\lambda_2$ equal to 10\% and 90\% of the channel capacity, respectively.}
	\label{GWPAlgorithmEvaluation-Figure: Complex Scenario Results 10-90}
\end{figure*}

\begin{figure*}[t]
	\centering
	\subfloat[Throughput (R) CCDF.]{
		\includegraphics[width=0.47\linewidth]{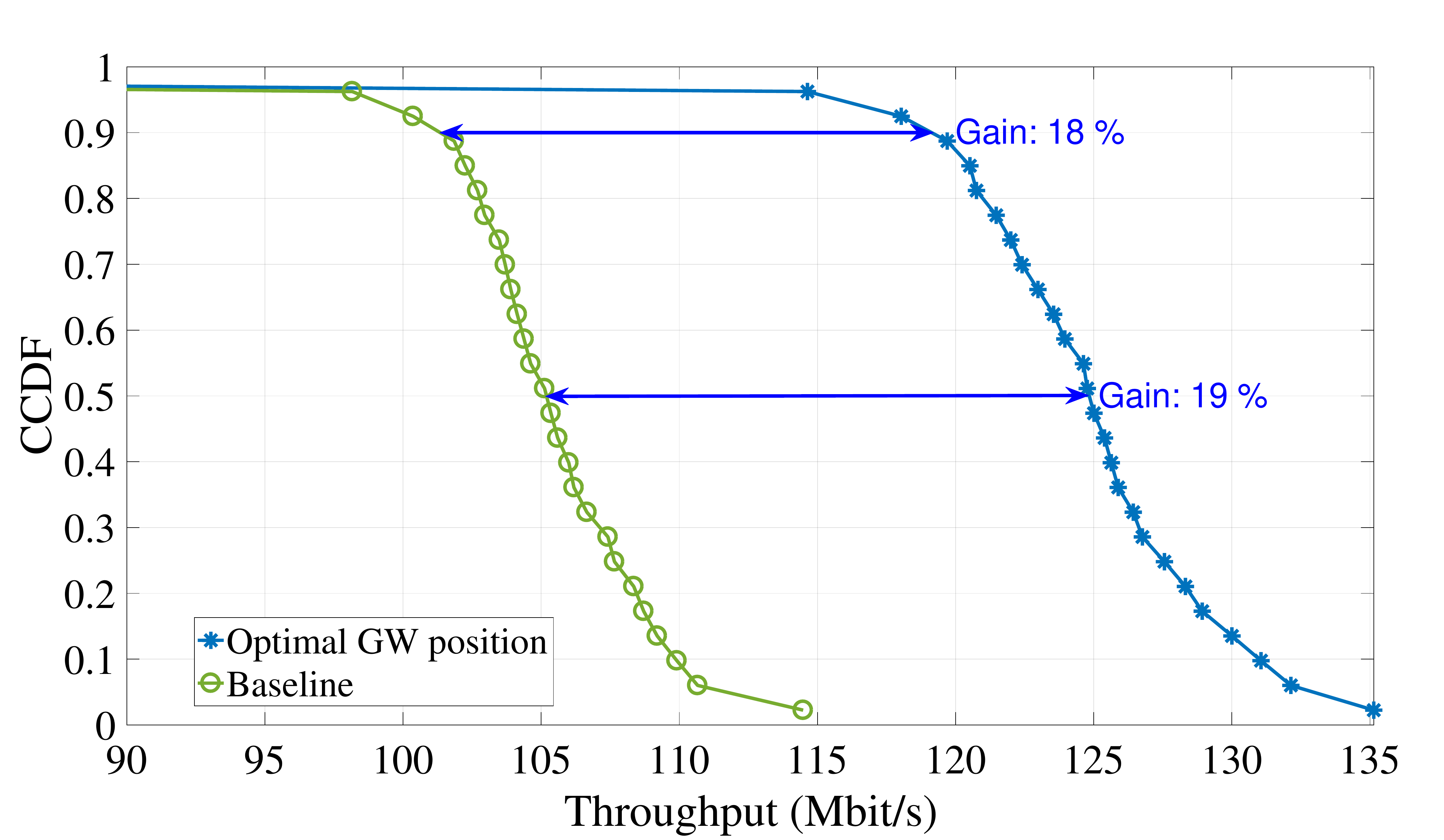}
		\label{GWPAlgorithmEvaluation-Figure: Complex Scenario Throughput 27-75}
	}
	\hfill
	\subfloat[End-to-end delay CDF.]{
		\includegraphics[width=0.47\linewidth]{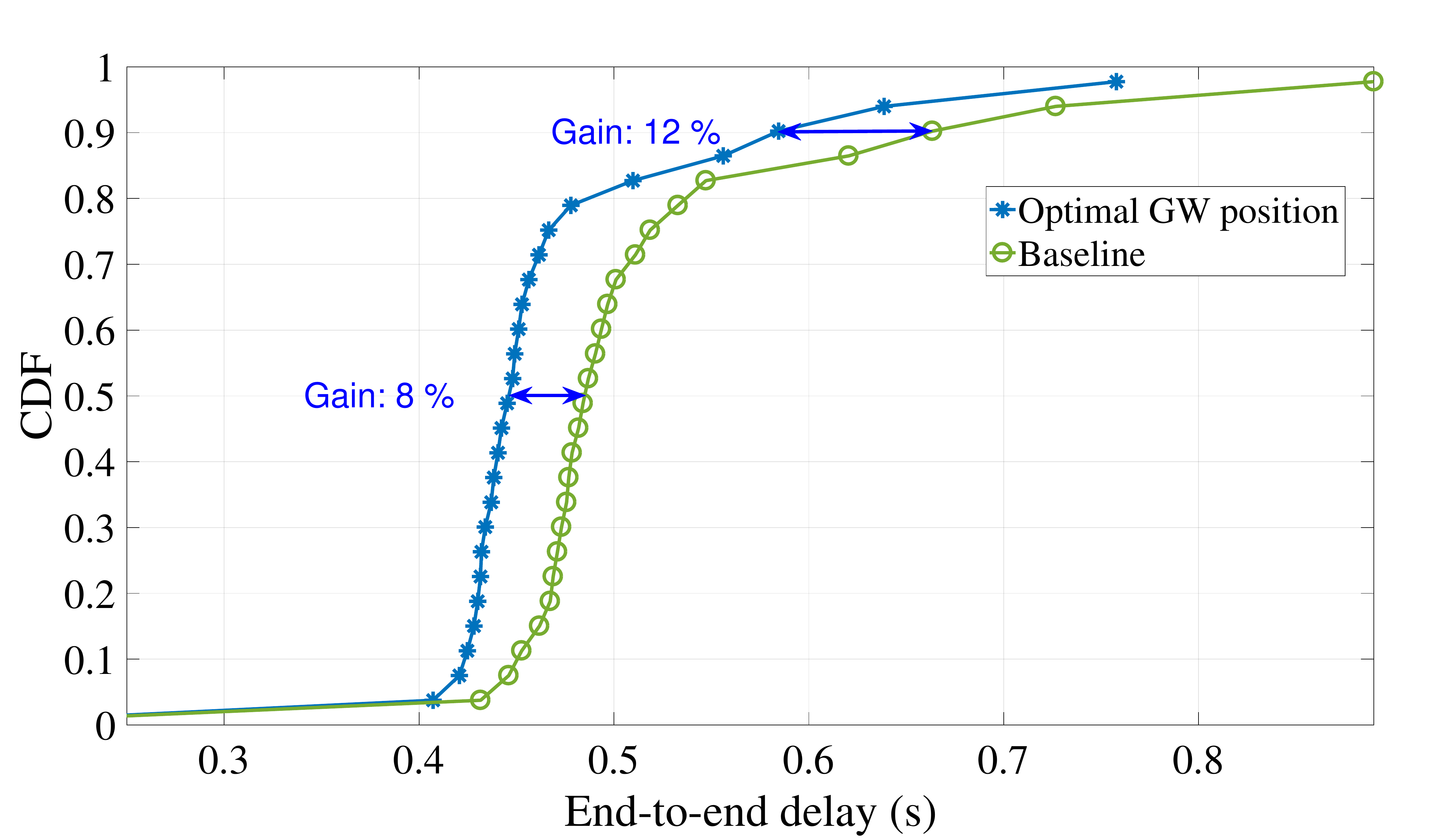}
		\label{GWPAlgorithmEvaluation-Figure: Complex Scenario Delay 25-75}
	}
	\caption{Scenario B - Aggregate Throughput (R) and End-to-end delay results measured in the GW for $\lambda_1 $ and $\lambda_2$ equal to 25\% and 75\% of the channel capacity, respectively.}
	\label{GWPAlgorithmEvaluation-Figure: Complex Scenario Results 25-75}
\end{figure*}

The performance of the GWP algorithm was evaluated considering two performance metrics:

\begin{itemize}
    \item Aggregate Throughput (R): The mean number of bits received per second by the GW.
    \item End-to-end delay: The mean time taken by the packets to reach the application layer of the GW since the instant they were generated by the FMAPs, including queuing, transmission, and propagation delays.
\end{itemize}

\subsection{Simulation Results} \label{GWPAlgorithmEvaluation-Section: Simulation Results}

The results were obtained after 20 simulation runs for each traffic demand combinations that were considered (cf. \cref{GWPAlgorithmEvaluation-Section: Simulation Scenarios}), under the same networking conditions, using $RngSeed = 10$ and $RngRun = \{1, ..., 20\}$. The results are expressed using mean values and they are represented using the CDF for the end-to-end delay and by the CCDF for the aggregate throughput. The CCDF $F'(x)$ represents the percentage of time for which the mean aggregate throughput was higher than $x$, while the CDF $F(x)$ represents the percentage of time for which the mean end-to-end delay was lower than or equal to to $x$.

Regarding Scenario A, when the GW is placed in the optimal position (Position 8 in \cref{GWPAlgorithmEvaluation-Figure: Baseline Scenario}), the aggregate throughput is improved 24\% for the 90th percentile and 21\% for the 50th percentile (median), with respect to the baseline (i.e., the GW placed in the FMAPs center). In parallel, the end-to-end delay is decreased 26\% for both the 90th and 50th percentiles (cf. \cref{GWPAlgorithmEvaluation-Figure: Baseline Scenario Results}). The similar performance results obtained for Position 2 and Position 8, which are depicted in \cref{GWPAlgorithmEvaluation-Figure: Baseline Scenario Results}, are justified by the closer distance between these positions; note that Position 2 was obtained by chance, while Position 8 resulted from the GWP algorithm. %The PDR results denote the action of CoDel, which discards packets held in the transmission queues longer, in order to avoid bufferbloat due to network congestion; if CoDel is not used, the PDR will be close to 1, but the packets will experience longer delays, which is not compliant with the QoS requirements of some nowadays' applications.
In order to meet the higher traffic demand of the right-side FMAPs, the GWP algorithm places the GW closer to them, in order to improve the SNR of the communications links and enable the selection of higher MCS indexes. This improves the overall TMFN performance and the shared medium usage -- the packets are held in the transmission queues for shorter time, the transmission delay decreases, and the throughput increases. The difference between the traffic demand of the FMAPs and the aggregate throughput in the GW are justified by the action of the auto rate mechanism, which control the MCS indexes being used by the FMAPs over time, as well as by the overhead introduced by the UDP, IP, and MAC packet headers.

With respect to Scenario B, when $\lambda_1 $ and $\lambda_2$ are respectively equal to 10\% and 90\% of $L$, an optimal placement for the GW is $(x_0,y_0,z_0) \approx (6.2, 31.0, 8.8)$ for a transmission power $P_{T} = $ \SI{26}{dBm}; this allows to improve the aggregate throughput up to 27\%, considering the 90th and 50th percentiles, while the end-to-end delay is reduced up to 4\% (cf. \cref{GWPAlgorithmEvaluation-Figure: Complex Scenario Results 10-90}). When $\lambda_1 $ and $\lambda_2$ are respectively equal to 25\% and 75\% of the channel capacity, the GWP algorithm defines $(x_0,y_0,z_0) \approx (9.0, 31.1, 2.8)$ as the optimal GW placement for transmission power $P_{T} = $ \SI{24}{dBm}. This allows to improve the aggregate throughput in 18\% with respect to the 90th percentile and 19\% for the 50th percentile, while the end-to-end delay is reduced 12\% for the 90th percentile and 8\% for the 50th percentile (cf. \cref{GWPAlgorithmEvaluation-Figure: Complex Scenario Results 25-75}). These results validate the effectiveness of the GWP algorithm and corroborate our research hypothesis: the TMFN performance can be improved by dynamically adjusting the position of the GW, considering both the positions and the offered traffic of the FMAPs.

%%%%%%%%%%%%%%%%%%%%%%%%%%%%%%%%%%%%%%%%%%%%%%%%%%%%%%%%%%%%%%%%%
% CONCLUSIONS
%%%%%%%%%%%%%%%%%%%%%%%%%%%%%%%%%%%%%%%%%%%%%%%%%%%%%%%%%%%%%%%%%
\section{Conclusions} \label{Conclusions-Section}

This article proposed an integrated solution to control the Traffic-Aware Multi-Tier Flying Networks (TMFN). The synergy created from the combination of the NetPlan algorithm, the RedeFINE routing solution and the GWP algorithm allows the TMFN to seamlessly update its topology according to the users' traffic demand and minimizing the disruption caused by the movement of the UAVs. Therefore, this solution enables the TMFN to be used as an on-demand network, which is able to provide an improved QoS to the users, even in scenarios with a high-density of users and variable traffic demand.

% RESULTS
Even though the three components were designed to be used simultaneously, they were evaluated individually, in order to obtain results that were not affected by the remaining components. The results obtained showed that each component was able to improve the QoS provided to the users, when compared to baselines and state of the art counterparts in the scenarios tested in this article. Also, the experimental results of NetPlan demonstrated the gains obtained by means of simulation. Furthermore, the evaluation of the air-to-ground and ground-to-air channel propagation models for low altitudes demonstrated that both channels are best modeled by the Friis path loss with Rician fast-fading, which corresponds to the theoretical studies proposed in the literature. Moreover, the asymmetry between the air-to-ground and the ground-to-air channels was also confirmed.

% FUTURE WORK
As future work, we plan to develop a prototype of the proposed solution with the NetPlan algorithm, the RedeFINE routing solution and the GWP algorithm fully integrated. Moreover, the RedeFINE solution can be improved by considering the positions of the gateway UAVs determined by GWP in addition to the FMAPs. Finally, we plan to evaluate the performance of the prototype by means of simulation and experimental scenarios.

%%%%%%%%%%%%%%%%%%%%%%%%%%%%%%%%%%%%%%%%%%%%%%%%%%%%%%%%%%%%%%%%%
\section*{Acknowledgments}
%%%%%%%%%%%%%%%%%%%%%%%%%%%%%%%%%%%%%%%%%%%%%%%%%%%%%%%%%%%%%%%%%

%%% ACK TEKEVER
The authors would like to thank Filipe Rodrigues and Filipe Rosa from Tekever for their help setting up the testbed and piloting the UAVs.
%
%%% ACK WISE PROJECT
This work is financed by the ERDF -- European Regional Development Fund through the Operational Programme for Competitiveness and Internationalisation -- COMPETE 2020 Programme and by National Funds through the Portuguese funding agency, FCT -- Fundação para a Ciência e a Tecnologia within project POCI-01-0145-FEDER-016744.
%
%%% ACK EDUARDO PhD GRANT (FCT)
The first author thanks the funding from FCT under the PhD grant PD/BD/113819/2015.
%
%%% ACK ANDRÉ PhD GRANT (FCT)
The second author thanks the funding from FCT under the PhD grant SFRH/BD/137255/2018.

%%%%%%%%%%%%%%%%%%%%%%%%%%%%%%%%%%%%%%%%%%%%%%%%%%%%%%%%%%%%%%%%%
% REFERENCES
%%%%%%%%%%%%%%%%%%%%%%%%%%%%%%%%%%%%%%%%%%%%%%%%%%%%%%%%%%%%%%%%%
\bibliographystyle{elsarticle-num}
\bibliography{References}

\end{document}